\documentclass[letterpaper,twocolumn,prl,aps,superscriptaddress,floatfix]{revtex4}

\usepackage[latin9]{inputenc}
\usepackage{amsmath}
\usepackage{amssymb}
\usepackage{graphicx}
\usepackage{esint}
\usepackage{times}
\usepackage[markup=underlined]{changes}

\usepackage[unicode=true,
bookmarks=true,bookmarksnumbered=false,bookmarksopen=false,
breaklinks=false,pdfborder={0 0 1},backref=false,colorlinks=true]
{hyperref}
\hypersetup{
    linkcolor=magenta, urlcolor=blue, citecolor=blue, pdfstartview={FitH}, hyperfootnotes=false, unicode=true}

\makeatletter

\pdfpageheight\paperheight
\pdfpagewidth\paperwidth

\@ifundefined{textcolor}{}
{%
    \definecolor{BLACK}{gray}{0}
    \definecolor{WHITE}{gray}{1}
    \definecolor{RED}{rgb}{1,0,0}
    \definecolor{GREEN}{rgb}{0,1,0}
    \definecolor{BLUE}{rgb}{0,0,1}
    \definecolor{CYAN}{cmyk}{1,0,0,0}
    \definecolor{MAGENTA}{cmyk}{0,1,0,0}
    \definecolor{YELLOW}{cmyk}{0,0,1,0}
}

\usepackage{xcolor}
\usepackage{soul}

\definecolor{blue}{rgb}{0,0,1}
\definecolor{red}{rgb}{1,0,0}
\definecolor{green}{rgb}{0,1,0}

\usepackage{soul}

\makeatother

\begin{document}
\title{Quantum-enhanced metrology with large Fock states}

\author{Xiaowei Deng}
\thanks{These authors contributed equally to this work.}
\affiliation{International Quantum Academy, Shenzhen 518048, China}
\affiliation{Shenzhen Institute for Quantum Science and Engineering, Southern University of Science and Technology, Shenzhen 518055, China}
\affiliation{Guangdong Provincial Key Laboratory of Quantum Science and Engineering, Southern University of Science and Technology, Shenzhen 518055, China}

\author{Sai Li}
\thanks{These authors contributed equally to this work.}
\affiliation{International Quantum Academy, Shenzhen 518048, China}
\affiliation{Shenzhen Institute for Quantum Science and Engineering, Southern University of Science and Technology, Shenzhen 518055, China}
\affiliation{Guangdong Provincial Key Laboratory of Quantum Science and Engineering, Southern University of Science and Technology, Shenzhen 518055, China}

\author{Zi-Jie Chen}
\thanks{These authors contributed equally to this work.}
\affiliation{CAS Key Laboratory of Quantum Information, University of Science and Technology of China, Hefei, Anhui 230026, China}

\author{Zhongchu Ni}
\affiliation{International Quantum Academy, Shenzhen 518048, China}
\affiliation{Shenzhen Institute for Quantum Science and Engineering, Southern University of Science and Technology, Shenzhen 518055, China}
\affiliation{Guangdong Provincial Key Laboratory of Quantum Science and Engineering, Southern University of Science and Technology, Shenzhen 518055, China}
\affiliation{Department of Physics, Southern University of Science and Technology, Shenzhen 518055, China}

\author{Yanyan Cai}
\affiliation{International Quantum Academy, Shenzhen 518048, China}
\affiliation{Shenzhen Institute for Quantum Science and Engineering, Southern University of Science and Technology, Shenzhen 518055, China}
\affiliation{Guangdong Provincial Key Laboratory of Quantum Science and Engineering, Southern University of Science and Technology, Shenzhen 518055, China}

\author{Jiasheng Mai}
\affiliation{International Quantum Academy, Shenzhen 518048, China}
\affiliation{Shenzhen Institute for Quantum Science and Engineering, Southern University of Science and Technology, Shenzhen 518055, China}
\affiliation{Guangdong Provincial Key Laboratory of Quantum Science and Engineering, Southern University of Science and Technology, Shenzhen 518055, China}

\author{Libo Zhang}
\affiliation{International Quantum Academy, Shenzhen 518048, China}
\affiliation{Shenzhen Institute for Quantum Science and Engineering, Southern University of Science and Technology, Shenzhen 518055, China}
\affiliation{Guangdong Provincial Key Laboratory of Quantum Science and Engineering, Southern University of Science and Technology, Shenzhen 518055, China}

\author{Pan Zheng}
\affiliation{International Quantum Academy, Shenzhen 518048, China}
\affiliation{Shenzhen Institute for Quantum Science and Engineering, Southern University of Science and Technology, Shenzhen 518055, China}
\affiliation{Guangdong Provincial Key Laboratory of Quantum Science and Engineering, Southern University of Science and Technology, Shenzhen 518055, China}

\author{Haifeng Yu}
\affiliation{Beijing Academy of Quantum Information Sciences, Beijing 100193, China}

\author{Chang-Ling Zou}
\email{clzou321@ustc.edu.cn}
\affiliation{CAS Key Laboratory of Quantum Information, University of Science and Technology of China, Hefei, Anhui 230026, China}
\affiliation{Hefei National Laboratory, Hefei 230088, China.}

\author{Song Liu}
\affiliation{International Quantum Academy, Shenzhen 518048, China}
\affiliation{Shenzhen Institute for Quantum Science and Engineering, Southern University of Science and Technology, Shenzhen 518055, China}
\affiliation{Guangdong Provincial Key Laboratory of Quantum Science and Engineering, Southern University of Science and Technology, Shenzhen 518055, China}
\affiliation{Hefei National Laboratory, Hefei 230088, China.}

\author{Fei Yan}
\email{yanfei@baqis.ac.cn}
\altaffiliation[Present address: ]{Beijing Academy of Quantum Information Sciences, Beijing, China}
\affiliation{International Quantum Academy, Shenzhen 518048, China}
\affiliation{Shenzhen Institute for Quantum Science and Engineering, Southern University of Science and Technology, Shenzhen 518055, China}
\affiliation{Guangdong Provincial Key Laboratory of Quantum Science and Engineering, Southern University of Science and Technology, Shenzhen 518055, China}

\author{Yuan Xu}
\email{xuyuan@iqasz.cn}
\affiliation{International Quantum Academy, Shenzhen 518048, China}
\affiliation{Shenzhen Institute for Quantum Science and Engineering, Southern University of Science and Technology, Shenzhen 518055, China}
\affiliation{Guangdong Provincial Key Laboratory of Quantum Science and Engineering, Southern University of Science and Technology, Shenzhen 518055, China}
\affiliation{Hefei National Laboratory, Hefei 230088, China.}

\author{Dapeng Yu}
\email{yudp@sustech.edu.cn}
\affiliation{International Quantum Academy, Shenzhen 518048, China}
\affiliation{Shenzhen Institute for Quantum Science and Engineering, Southern University of Science and Technology, Shenzhen 518055, China}
\affiliation{Guangdong Provincial Key Laboratory of Quantum Science and Engineering, Southern University of Science and Technology, Shenzhen 518055, China}
\affiliation{Department of Physics, Southern University of Science and Technology, Shenzhen 518055, China}
\affiliation{Hefei National Laboratory, Hefei 230088, China.}

\begin{abstract}
Quantum metrology uses non-classical states, such as Fock states with a specific number of photons, to achieve an advantage over classical sensing methods. Typically, quantum metrological performance can be enhanced by increasing the involved excitation numbers, for example by using large photon-number Fock states. However, manipulating these states and demonstrating a quantum metrological advantage is experimentally challenging. Here, we present an efficient method for generating large Fock states approaching 100 photons within a superconducting microwave cavity through the development of a programmable photon number filter. Using these states in displacement and phase measurements, we demonstrate quantum-enhanced metrology approaching the Heisenberg scaling for 40-photon Fock states and achieve a maximum metrological gain of up to 14.8 decibels, highlighting the metrological advantages of large Fock states. Our study could be readily extended to mechanical and optical systems, promising potential applications in weak force detection and dark matter searches.

\end{abstract}
\maketitle
\vskip 0.5cm

Precision measurement and metrology are crucial for a broad range of research, as they enable the accurate measurement of physical quantities~\cite{Tiesinga2021}, the discovery of new phenomena~\cite{Safronova2018}, the validation of physics theories~\cite{Karr2020}, and the development of new technologies~\cite{Ye2008,Wu2019}. 
Quantum metrology~\cite{PARIS2009,degen2017,pirandola2018}, which leverages quantum-mechanical principles, can achieve a measurement precision bounded by the Heisenberg limit (HL)~\cite{braunstein1994,giovannetti2004}, surpassing the standard quantum limit (SQL) in conventional measurements by a factor of $1/\sqrt{N}$, with $N$ being the number of particles or excitation number of probes. 
The quantum enhancement can be achieved by utilizing entanglement as a resource. This is usually realized by preparing the probe system into certain entangled states, such as the Greenberger-Horne-Zeilinger-type states~\cite{gao2010experimental,monz2011,Omran2019,song2019}, NOON states~\cite{nagata2007,chen2010,zhang2018}, spin squeezing states~\cite{hosten2016,colombo2022,xu2022metrological}, and other entangled states~\cite{penasa2016,gilmore2021,marciniak2022} in multimode interferometry. 
However, manipulating these exotic quantum states in a large system remains a formidable challenge.

Alternatively, one may also achieve quantum-enhanced precision measurements in a hardware-efficient fashion without resorting to multipartite entanglement~\cite{Braun2018}. One archetypal example employs a single bosonic mode~\cite{Duivenvoorden2017} as the probe system. Previous works have successfully demonstrated the advantages of quantum sensing by preparing the bosonic system into highly nontrivial states, including Schr\"odinger cat states~\cite{vlastakis2013}, squeezing states~\cite{backes2021}, maximum variance states~\cite{wang2019,mccormick2019}, as well as energy eigenstates or Fock states involving phonons~\cite{chu2018,wolf2019,Podhora2022} and microwave photons~\cite{hofheinz2008,eickbusch2022}. Nonetheless, the demonstrated metrological advantages have been confined to relatively small scales, with $N$ typically on the order of 10, thereby not fully realizing the potential scaling advantages intrinsic to quantum metrology using a single bosonic mode. 

In this work, we demonstrate quantum-enhanced metrology using Fock states with up to 100 photons in a superconducting microwave cavity, almost an order of magnitude improvement compared to previous works~\cite{chu2018,wolf2019,Podhora2022,hofheinz2008,eickbusch2022}. To generate large Fock states, we develop a programmable photon number filter by leveraging the photon-number-dependent response of an ancilla qubit coupled to the cavity. Such a parameterized state preparation method allows us to obtain arbitrary Fock states in logarithm steps. 
Based on these Fock states, we further demonstrate a close-to-Heisenberg scaling for both displacement and phase sensing and achieve a maximum metrological gain of $14.8\pm 0.2\,\mathrm{dB}$ and $12.3\pm 0.5\,\mathrm{dB}$ in each case, illustrating the metrological power of large Fock states in a single bosonic mode.

\begin{figure*}
    \includegraphics{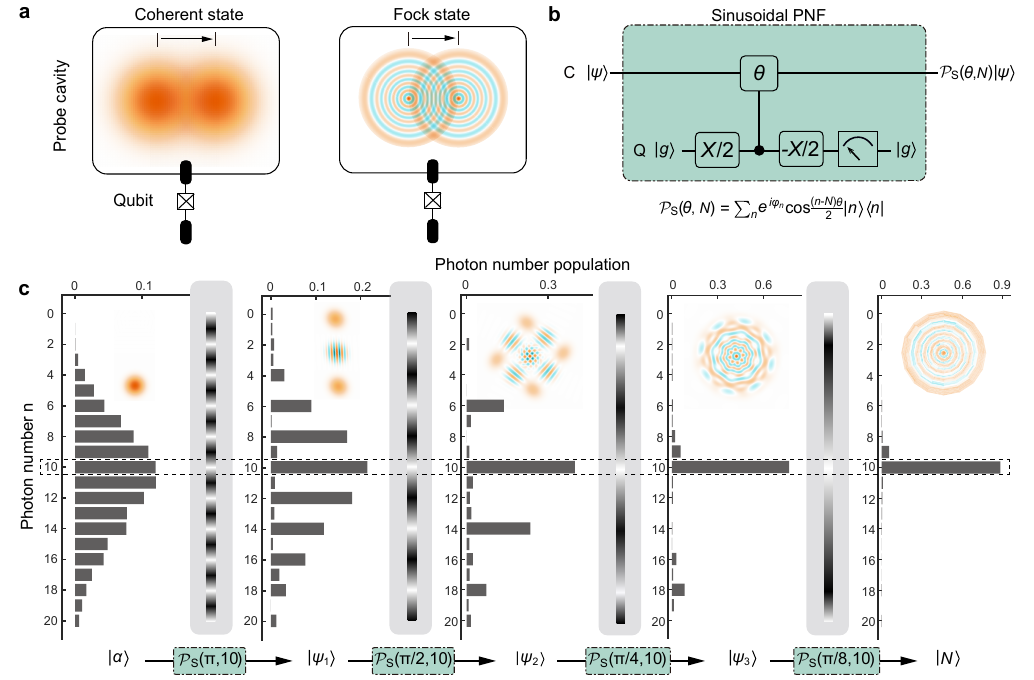}
    \caption{\textbf{Efficient generation of large photon-number Fock states for quantum metrology.}
	\textbf{a} Comparison between the conventional scheme with coherent state (left panel) and the Fock state scheme (right panel) for displacement and phase sensing in a probe cavity with an ancillary qubit. The conventional metrology scheme has a sensitivity limit of $1/2$ when measuring the displaced or rotated coherent states, while the Fock states provide an enhancement factor of $\sqrt{N}$ due to the fine structural features in their Wigner functions in the phase space. 
    \textbf{b} Quantum circuit for the sinusoidal photon-number filter ($\mathcal{P}_\mathrm{S}$) on the bosonic state in the probe cavity (C) by projecting the ancilla qubit (Q) in the ground state. 
    \textbf{c} Measured photon number populations of the cavity state at each stage of four successive sinusoidal PNFs for efficient generation of Fock state $|10\rangle$. Each PNF acts as a grating that periodically blocks certain photon numbers of the cavity state. Insets are simulated Wigner functions of the intermediate states at different stages and measured Wigner function of the final state. }
    \label{fig1}
\end{figure*}

% Explain the Scheme

%% (1) Detection-based Fock state preparation
Employing quantum states with higher excitations in a single quantum system, a bosonic mode in a superconducting quantum circuit provides a hardware-efficient platform for realizing quantum metrology~\cite{cai2021,joshi2021}, as illustrated in Fig.~\ref{fig1}a. In conventional metrology schemes, the bosonic mode serving as the probe is prepared to a quasiclassical coherent state $|\alpha\rangle$. Its Wigner function distribution in the phase space is a Gaussian function, with an uncertainty of $1/2$ and a mean photon number of $N=|\alpha^2|$. The left panel of Fig.~\ref{fig1}a depicts that by either external displacement operation $D(\beta)=e^{\beta(a^\dagger - a)}$ (assuming $\beta$ a positive real number) or phase rotation $e^{-i\phi a^\dagger a}$ of the probe state, the quantum state is displaced by a distance in the phase space. Here, $a^\dagger (a)$ is the creation (annihilation) operator of the bosonic mode. Therefore, the measurement sensitivity is limited by the width of the Gaussian function, which leads to the standard quantum limitations of displacement amplitude sensitivity and phase estimation sensitivity as 
\begin{equation}
\delta \beta_{\mathrm{SQL}} = \frac{1}{2},\,
\delta \phi_{\mathrm{SQL}} =\frac{1}{2\sqrt{N}}.
\label{HL}
\end{equation}

%% (2) Heisenberg limit (both phase/displacement)

Recognized as the most fundamental quantum-mechanical states, Fock states possess an absolute certainty in the photon number of a quantized electromagnetic field, indicating an ultrasensitive response to field perturbations~\cite{oliveira1990}. To go beyond the SQL, nonclassical Fock states was proposed to probe the displacement or phase rotation in Ref.~\cite{wolf2019}. As shown by the right panel of Fig.~\ref{fig1}a, Fock states exhibit sub-Planck phase-space structures of the Wigner functions~\cite{zurek2001}, with their structural features inversely proportional to the photon number, indicating a higher resolution in phase space. By preparing the probe in the Fock state $|N\rangle$ and only performing further displacement and parity measurement operations, the probe can achieve sensitivities
\begin{equation}
\delta \beta_{\mathrm{Fock}} = \frac{1}{2\sqrt{2N+1}},\,
\delta \phi_{\mathrm{Fock}} =\frac{1}{2\sqrt{N(2N+1)}}, 
\label{HL}
\end{equation}
for displacement and phase sensing, respectively (see Methods for details).
For both cases, the scheme based on the Fock state shows a $\sqrt{N}$-enhancement of the measurement sensitivity compared to a coherent state with the same mean photon number, indicating the ability to approach the HL.

\begin{figure*}
    \includegraphics{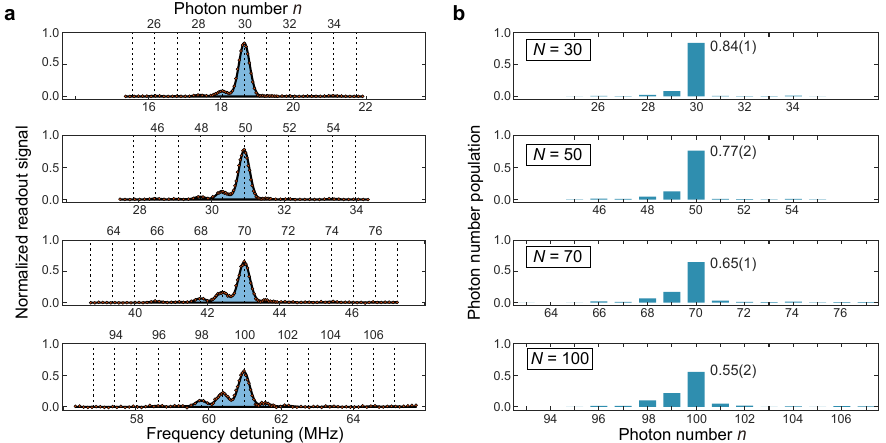}
     \caption{\textbf{Characterization of large photon-number Fock states.}
     \textbf{a} Measured qubit spectrum for various prepared Fock states $|N\rangle$, with $N=30$, 50, 70, and 100. The frequency detuning on the bottom horizontal axis is the drive frequency relative to the qubit frequency in the absence of any photons in the cavity. The solid lines are fits to a sum of Gaussian functions to extract the Fock state populations $P_n$. The total numbers of measurements (the numbers of postselected measurements) are approximately $4\times10^4$ ($2.2\times10^3$),  $5.5\times10^4$ ($2.2\times 10^3$), $7.9\times 10^4$ ($2.5\times10^3$), and $1.0\times10^6$($2.1\times10^3$) with $N=30$, 50, 70, and 100, respectively.
    \textbf{b} Extracted photon number populations $P_n$ of the generated Fock states $|N\rangle$, with $N=30$, 50, 70, and 100. Error bars in the parentheses for the photon number populations $P_N$ are standard errors obtained from the fittings in \textbf{a} and others are not shown for clarity.} 
    \label{fig2}
\end{figure*}

%% (3) The experimental setup + State preparation
The generation of Fock states and their application in Heisenberg-limited quantum metrology are experimentally investigated with a probe mode in a high-Q 3D microwave cavity, which is dispersively coupled to an ancillary superconducting qubit~\cite{blais2021}. With respect to the driving frequencies, the system Hamiltonian reads
\begin{equation}
H/\hbar = \Delta |e\rangle \langle e| -\chi a^\dagger a |e\rangle \langle e|+\epsilon_p(a+a^{\dagger})+\epsilon_q\sigma_x. \label{H0}
\end{equation}
Here, $\sigma_x=|e\rangle \langle g|+|g\rangle \langle e|$ and $|e\rangle (|g\rangle)$ is the excited (ground) state of the superconducting qubit. Employing the dispersive coupling $\chi$ between the probe mode and qubit, and external coherent drives $\epsilon_{p,q}$ on them, efficient quantum control of the composite probe-qubit system can be realized~\cite{vlastakis2013}. 
The displacements of the probe are trivial, and the parity measurement of the probe state can be realized by projectively measuring the qubit~\cite{guerlin2007, sun2014}, while the preparation of the Fock state $|N\rangle$ with $N\gg1$ remains elusive in this platform~\cite{hofheinz2008,eickbusch2022}. 

To tackle this challenge, a projection operation called sinusoidal photon-number filtration (PNF) is developed. The quantum circuit is shown in Fig.~\ref{fig1}b. By choosing the driving detuning on the qubit as $\Delta=N\chi$, the output of the qubit on $|g\rangle$ indicates a projection operation $\mathcal{P}_\mathrm{S}(\theta, N)= \sum_{n}e^{i\varphi_n}\cos{\frac{(n-N) \theta}{2}}|n\rangle \langle n|$, with $\varphi_n$ being an insignificant phase for generating Fock states, and $\theta/\chi$ being the duration of dispersive interaction. By sequentially implementing the sinusoidal PNF with $\theta=\pi/2^{j-1}$ for the $j$-th step ($j=1,2,..,m$), the probe is projected into a space of a general parity of $2^m$, i.e., $\Pi_j\mathcal{P}_\mathrm{S}(\theta_j, N)=\sum_k|N+k2^m\rangle \langle N+k2^m|$, and the parity measurement is a special case of the sinusoidal PNF with $m=1$~\cite{sun2014}. Such a PNF provides a unique and efficient approach for the preparation of Fock states using simple parameterized quantum circuits with a circuit depth of $d=\log_2 \sqrt{N}$ for $N\gg1$, since the photon number uncertainty is $\sqrt{N}$ for an initial coherent state $|\alpha=\sqrt{N}\rangle$. This logarithmic scaling of the circuit depth is more advantageous for efficiently resolving all the photon number Fock states to achieve the quantum-enhanced metrology (see Fig.~\ref{fig4}) than the polynomial scaling using selective number-dependent arbitrary phase gates and displacement operations~\cite{krastanov2015}. In Fig.~\ref{fig1}c, an example of the procedures for preparing Fock $|10\rangle$ by sinusoidal PNF is shown, with the probe initially prepared in a coherent state with $\alpha=\sqrt{10}$. As expected, the measured photon number distributions evolve as they pass through a comb after each PNF. The simulated Wigner functions indicate the preservation of coherence in projected subspaces. The measured Wigner function of the final state verifies the successful preparation of the target Fock state, showing sub-Plank features and rotational symmetry with respect to the origin.

% Experimental state preparation

\begin{figure*}
    \includegraphics{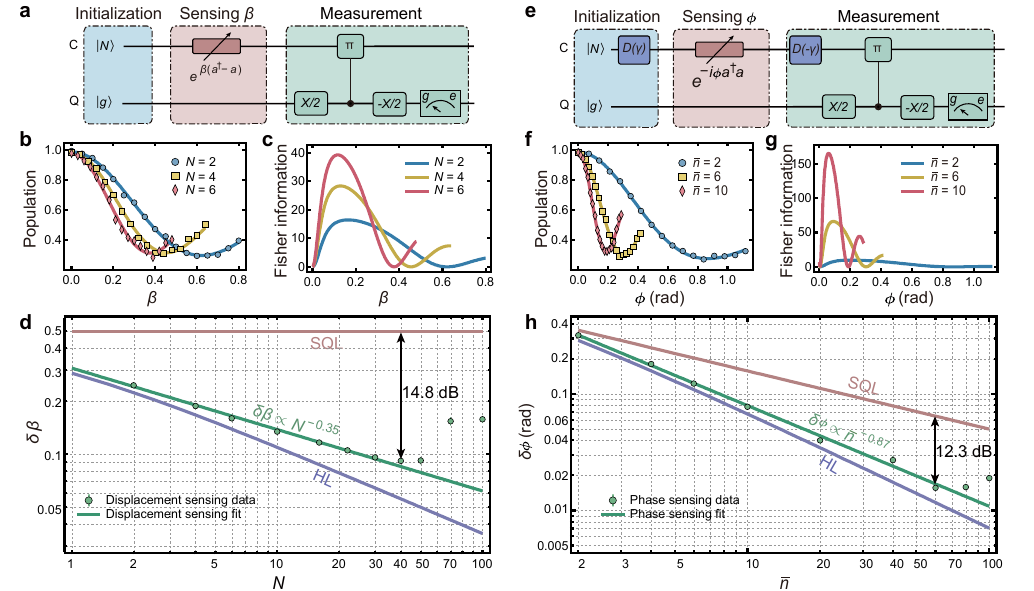}
    \caption{\textbf{Quantum metrology using Fock states.} 
    \textbf{a} Experimental circuit for displacement amplitude sensing. 
    \textbf{b} The measured qubit ground state populations (dots) and corresponding fittings (solid lines) as a function of the displacement amplitude $\beta$ with the probe initially prepared in Fock states $|N\rangle$, with $N = 2,4,6$ as examples. 
    \textbf{c} Fisher information extracted from the fittings in \textbf{b}.
    \textbf{d} The displacement measurement precision $\delta \beta$ against the number of photons $N$ of the initial Fock state, with a $14.8\pm 0.2\,\mathrm{dB}$ enhancement of the precision compared with SQL achieved at $N = 40$. A precision scaling of $N^{-0.35}$ is achieved, as determined from a linear fit in the logarithmic-logarithmic scale. 
    \textbf{e} Experimental circuit for phase sensing. 
    \textbf{f} The measured qubit ground state populations (dots) and corresponding fittings (solid lines) as a function of the phase $\phi$ with the probe initially prepared in various displaced Fock states $D(\gamma=\sqrt{N})|N\rangle$, with $N = 1,3,5$ as examples. 
	\textbf{g} Fisher information extracted from the fittings in \textbf{f}.
     \textbf{h} The phase estimation precision $\delta\phi$ as a function of the average photon number $\bar{n} = 2N$ of the initial displaced Fock state. The precision scales with $\bar{n}^{-0.87}$ and a precision enhancement of $12.3\pm 0.5\,\mathrm{dB}$ surpassing the SQL is achieved at $\bar{n} = 60$. Error bars in \textbf{d} and \textbf{h} are standard errors obtained from error propagation of the fit parameter uncertainties in \textbf{b} and \textbf{f}, respectively. Error bars for other data are smaller than the marker sizes and not shown. }
    \label{fig3}
\end{figure*}

Figure~\ref{fig2} shows the experimental results for the prepared Fock states with $N=30$, 50, 70, and 100. The states are prepared by two PNFs $\mathcal{P}_{\mathrm{S}}(\pi, N)\mathcal{P}_{\mathrm{S}}(\pi/2, N)$, with the assistance of a Gaussian PNF operation (Methods). 
In Fig.~\ref{fig2}a, we characterize the generated Fock states by measuring their photon number distributions through a qubit spectroscopy experiment. By applying a selective Gaussian $\pi$ pulse on the qubit while sweeping the drive frequency detuning, we measure the readout signal spectrum and fit it to a sum of Gaussian functions to extract the height of each peak (Methods). These peaks correspond to the photon number populations $P_n$ of each Fock state $|n\rangle$ after normalizing the Gaussian amplitudes such that $\sum_n{P_n}=1$. 
The extracted photon number populations $P_n$ for the generated large Fock state $|N\rangle$ with $N=30$, 50, 70, and 100 are shown in Fig.~\ref{fig2}b. Comparing the detected photon number populations for different $N$, there are increasing probabilities for $N-1$ and $N-2$ with increasing $N$ because the Fock state decay rate increases with $N$. These measured photon number amplitudes $P_N$ for the Fock state $|N\rangle$ are consistent with those estimated from our error analysis (see Supplementary Fig.~S9). The prepared Fock states are further characterized by Ramsey experiments on the ancilla qubit, with the Ramsey interference oscillating $N$ times faster in the presence of $N$ photons in the cavity (see Supplementary Fig.~S5). 
All these results confirm the successful generation of large photon-number Fock states in the cavity.

\begin{figure*}
    \includegraphics{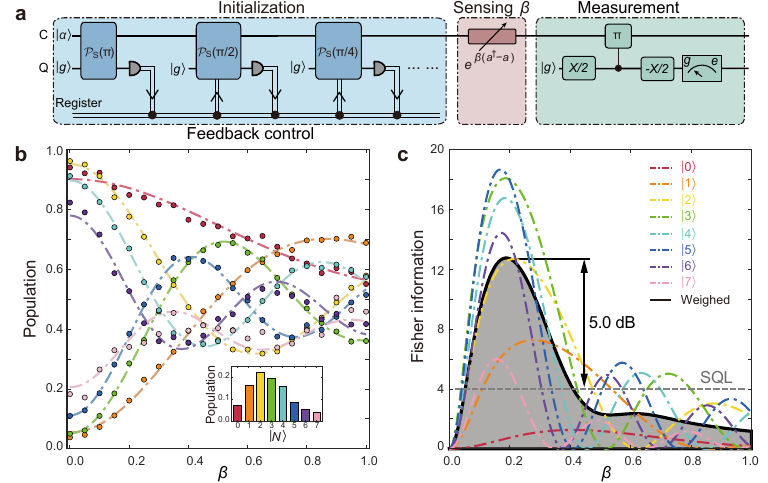}
    \caption{\textbf{Photon-number-resolved quantum metrology scheme for displacement amplitude sensing.}
	\textbf{a} Quantum circuit illustrating the displacement amplitude sensing using coherent initial probe state ($|\alpha=\sqrt{3}\rangle$) and consequent photon-number-resolved measurements, which is implemented by three sinusoidal PNFs for simultaneously resolving 0-7 photons.
    \textbf{b} Measured qubit populations as a function of the displacement amplitude $\beta$ for various traces of the photon-number-resolved Fock states $|N\rangle$. Inset shows the probabilities of traces corresponding to various $|N\rangle$.
	\textbf{c} The Fisher information (dashed lines) extracted from \textbf{b} as a function of the displacement amplitude $\beta$ for different traces, as well as the total Fisher information (solid black line) by weighting all eight traces. }
    \label{fig4}
\end{figure*}

% Metrology gain

Quantum-enhanced metrology using large photon-number Fock states is investigated for both displacement and phase sensing. Using the quantum circuit shown in Fig.~\ref{fig3}a,  the displacement amplitude $|\beta|$ is measured by directly applying a displacement operation to the prepared Fock state $|N\rangle$ and detected by mapping the parity information to the ancilla qubit~($\mathcal{P}_{\mathrm{S}}(\pi)$). The measured qubit ground state populations $P_g$ are plotted as a function of the displacement amplitude $|\beta|$ for various initial Fock states, as shown in Fig.~\ref{fig3}b. The data are fitted to the function $P_g = A \exp{\left(-2|\beta|^2\right)} L_N\left(4|\beta|^2 \right)+B$, where $L_N$ represents the $N$-th order Laguerre polynomials, and $A$ and $B$ are the fitting parameters to account for the measurement imperfections (see Supplementary section II-A). From the population oscillations, we found that the oscillation period, corresponding to the range of the estimated parameter, scales as $1/\sqrt{N}$ (see Supplementary section II-C). The Fisher information serves as a figure of merit for quantifying the average information learned about the parameter $|\beta|$ and is calculated by $F_N(\beta) = \frac{1}{P_g (1-P_g)} \left(\frac{dP_g}{d\beta}  \right)^2$~\cite{braunstein1994,PARIS2009}, as shown in Fig.~\ref{fig3}c. The maximal achievable sensitivity of displacement amplitude sensing can then be estimated using $\delta \beta = 1/\sqrt{F_m}$ with $F_m=\mathrm{max}_\beta(F_N)$ representing the maximal achieved Fisher information. We note that although the Fisher information achieves its maximum at different displacement amplitudes $|\beta|$ for different $N$, the maximal sensitivity to small displacement can be achieved experimentally by initially biasing to the specific $|\beta|$ with a displacement operation. The corresponding extracted $\delta \beta$ as a function of the photon numbers $N$ of the initial Fock state are shown in Fig.~\ref{fig3}d on a logarithmic-logarithmic scale. The results demonstrate that the displacement amplitude sensing with Fock states surpasses the SQL with a maximum metrological gain of $20\log{(\delta \beta_\mathrm{SQL}/\delta \beta)} = 14.8\pm 0.2$~dB achieved at $N=40$. Additionally, a linear fit in the logarithmic-logarithmic scale reveals that the enhancement of the sensitivity scales with the photon number as $N^{-0.35}$ for $N\leq40$, approaching the Heisenberg scaling $N^{-0.5}$. To further enhance the metrological gain, improvements in the quality of the probe cavity and ancilla qubit are required, primarily focusing on suppressing decay and dephasing errors (see Supplementary section II-C).

For phase sensing, the direct application of the Fock state is hindered by the rotation-symmetry structures of the Wigner distributions in the phase space. However, this issue can be addressed by displacing the state away from the origin, as illustrated by the quantum circuit in Fig.~\ref{fig3}e. By initially preparing the probe in a displaced Fock state $D(\gamma) |N\rangle$ with $\gamma=\sqrt{N}$, the phase rotation in the phase space is reminiscent of a displacement of $D(-i\gamma \phi)$. Here, the sensing resources are accounted for as the mean photon number of the initial state $\bar{n}=2N$. Consequently, this displacement yields a similar response curve with respect to $\phi$, as shown in Figs.~\ref{fig3}f and \ref{fig3}g. The resulting phase estimation sensitivity exhibits a maximum metrological gain of $12.3\pm 0.5\,\mathrm{dB}$ for $\bar{n}=60$, and a sensitivity scaling of $\bar{n}^{-0.87}$ for $\bar{n}\leq60$, approaching the Heisenberg scaling $\bar{n}^{-1}$, as depicted in Fig.~\ref{fig3}h.

% Quantum jump Tracking

The experimental results validate the quantum-enhanced metrology approaching the HL for large photon number Fock states. The PNFs generate a target Fock state from a coherent state in a probabilistic manner with the success probability scaling with the photon number as $1/\sqrt{N}$ for $N\gg1$. However, we can still achieve a scaling enhancement of measurement precision with $N^{1/4}$ compared to the classical limit (see Methods). Therefore, it indicates a quantum metrology advantage even with the nondeterministic preparation of the initial probe state. Alternatively, it is unnecessary to select a specific Fock state of a given photon number $N$ while discarding other Fock states with different photon numbers in practical metrology applications. Instead, by adaptively performing parity measurements, we can deterministically collapse the probe state to a photon number and obtain the corresponding Fock state $|n\rangle$. Then, the sensing interrogation and the final measurement can be performed according to $n$, which promises an enhancement of precision scaling with $N^{1/2}$ ($N=\bar{n}$) approaching the HL. Therefore, the advantages of the hardware-efficient quantum metrology platform based on bosonic modes can be further explored based on an idea similar to the quantum jump tracking method~\cite{wang2022}. As a proof-of-principle demonstration, we implement an optimal displacement amplitude sensing scheme by recording all possible traces of the projection measurements on Fock $|N\rangle$ for an initial coherent state of the probe and processing the data according to the measured $N$. The quantum circuit of the quantum jump tracking scheme is depicted in Fig.~\ref{fig4}a, where a sequence of sinusoidal PNFs is applied and the outcomes are recorded to determine the $N$ of the initial probe state. For an initial coherent state with $\alpha = \sqrt{3}$, the experimental results for traces of various $N$ are summarized in Fig.~\ref{fig4}b, with the inset showing the probabilities of these traces. The corresponding Fisher information for each trace [$F\left(|n\rangle\right)$] is represented in Fig.~\ref{fig4}c. 
The total Fisher information is upper bounded by the weighted Fisher information, given by $\sum_{n=0}^{7}{P_n F\left(|n\rangle\right)}$, due to its convexity~\cite{fujiwara2001}.
The experimental results indicate a maximum Fisher information gain of $10\log{(F_\mathrm{weight}/F_\mathrm{SQL})} = 5.04\pm 0.06$~dB for the initial coherent state with $\alpha = \sqrt{3}$, where $F_\mathrm{SQL} = 4$ (dashed line in Fig.~\ref{fig4}c) is the Fisher information of the SQL for displacement amplitude sensing.

% Discussion

In conclusion, we have generated large Fock states with up to 100 photons in a high-quality superconducting microwave cavity and demonstrated quantum-enhanced displacement and phase sensing with these Fock states. The large photon-number Fock states are efficiently generated with programmable parameterized quantum circuits, which provide an alternative way to inspire the future design of control protocols in bosonic systems. By utilizing these highly nonclassical Fock states, we have demonstrated a hardware-efficient approach for quantum-enhanced metrology, with measurement sensitivities approaching the Heisenberg scaling for both displacement and phase sensing. Notably, a metrological gain of $14.8\,\mathrm{dB}$ has been achieved with $N$ approaching $100$. We anticipate that even higher metrological gain is possible with $N$ exceeding one thousand by further optimizing the device's performance~\cite{Milul2023}. In addition, we show that our approach---though depending on postselection---still allows a $N^{1/4}$ quantum-enhanced scaling, and could further achieve a $N^{1/2}$-scaling enhancement through adaptive control.

Our hardware-efficient quantum metrology approach with large photon-number Fock states can readily enable the detection of dark photons with a significantly enhanced signal rate~\cite{Agrawal2023}, since the action of the dark matter wave can be modeled as a classical displacement drive on the cavity states. One prominent advantage of utilizing Fock states is that generating these highly nonclassical states with large photon numbers is less experimentally challenging compared to other nonclassical states, such as the squeezing states~\cite{backes2021} and the GKP states~\cite{Duivenvoorden2017}. Additionally, our control scheme for generating large Fock states can also be applied to trapped ion setups~\cite{leibfried2003} to achieve a high metrological gain, in which platform Fock states with 100 phonons have been achieved and their metrological advantages still await further exploration~\cite{mccormick2019}. Furthermore, our study can be extended to mechanical and optical sensors with the assistance of high-efficiency quantum transducers~\cite{fan2018superconducting,higginbotham2018harnessing,vonlupke2022}, opening up a promising avenue for practical quantum metrology with bosonic modes~\cite{wolf2019,Podhora2022,wang2022,Agrawal2023}.

\smallskip{}

\smallskip{}

\noindent \textbf{\large{}Acknowledgments}{\large\par}

\noindent This work was supported by the Key-Area Research and Development Program of Guangdong Province (Grant No. 2018B030326001), the National Natural Science Foundation of China (Grants No. 12274198, No. 92265210, No. 12061131011, No. 11890704, No. 12322413), the Guangdong Basic and Applied Basic Research Foundation (Grants No. 2024B1515020013, No. 2022A1515010324), the Shenzhen Science and Technology Program (Grant No. RCYX20210706092103021), the Guangdong Provincial Key Laboratory (Grant No. 2019B121203002), the Shenzhen-Hong Kong cooperation zone for technology and innovation (Contract No. HZQB-KCZYB-2020050), the Innovation Program for Quantum Science and Technology (Grant No. 2021ZD0301703), and the Natural Science Foundation of Beijing (Grant No. Z190012). CLZ also acknowledged supports from the Fundamental Research Funds for the Central Universities, the USTC Center for Micro and Nanoscale Research and Fabrication, and USTC Research Funds of the Double First-Class Initiative.

\smallskip{}

\noindent \textbf{\large{}Author contributions}{\large\par}

\noindent Y.X. and D.Y. supervised the project. Y.X. conceived and designed the experiment. X.D. and Sa.L. performed the experiment, analyzed the data, and carried out the numerical simulations under the supervision of Y.X. Z.N. and Sa.L. developed the feedback control technique. Z.-J.C. and C.-L.Z. provided theoretical support for the quantum metrology schemes and data analysis. F.Y. proposed the approach for efficiently generating Fock states with large photon numbers. L.Z., H.Y. and So.L. support the device fabrication. Y.C., J.M., and P.Z. contributed to the experimental and theoretical support. X.D., Sa.L., Z.-J.C., C.-L.Z., F.Y., and Y.X. wrote the manuscript with feedback from all authors.

\smallskip{}

\noindent \textbf{\large{}Competing interests}{\large\par}

\noindent The authors declare no competing interests.

\clearpage{}
%\bibliographystyle{Zou}
%\bibliography{refs}

\begin{thebibliography}{59}%
\makeatletter
\providecommand \@ifxundefined [1]{%
 \@ifx{#1\undefined}
}%
\providecommand \@ifnum [1]{%
 \ifnum #1\expandafter \@firstoftwo
 \else \expandafter \@secondoftwo
 \fi
}%
\providecommand \@ifx [1]{%
 \ifx #1\expandafter \@firstoftwo
 \else \expandafter \@secondoftwo
 \fi
}%
\providecommand \natexlab [1]{#1}%
\providecommand \enquote  [1]{``#1''}%
\providecommand \bibnamefont  [1]{#1}%
\providecommand \bibfnamefont [1]{#1}%
\providecommand \citenamefont [1]{#1}%
\providecommand \href@noop [0]{\@secondoftwo}%
\providecommand \href [0]{\begingroup \@sanitize@url \@href}%
\providecommand \@href[1]{\@@startlink{#1}\@@href}%
\providecommand \@@href[1]{\endgroup#1\@@endlink}%
\providecommand \@sanitize@url [0]{\catcode `\\12\catcode `\$12\catcode
  `\&12\catcode `\#12\catcode `\^12\catcode `\_12\catcode `\%12\relax}%
\providecommand \@@startlink[1]{}%
\providecommand \@@endlink[0]{}%
\providecommand \url  [0]{\begingroup\@sanitize@url \@url }%
\providecommand \@url [1]{\endgroup\@href {#1}{\urlprefix }}%
\providecommand \urlprefix  [0]{URL }%
\providecommand \Eprint [0]{\href }%
\providecommand \doibase [0]{http://dx.doi.org/}%
\providecommand \selectlanguage [0]{\@gobble}%
\providecommand \bibinfo  [0]{\@secondoftwo}%
\providecommand \bibfield  [0]{\@secondoftwo}%
\providecommand \translation [1]{[#1]}%
\providecommand \BibitemOpen [0]{}%
\providecommand \bibitemStop [0]{}%
\providecommand \bibitemNoStop [0]{.\EOS\space}%
\providecommand \EOS [0]{\spacefactor3000\relax}%
\providecommand \BibitemShut  [1]{\csname bibitem#1\endcsname}%
\let\auto@bib@innerbib\@empty
%</preamble>
\bibitem [{\citenamefont {Tiesinga}\ \emph {et~al.}(2021)\citenamefont
  {Tiesinga}, \citenamefont {Mohr}, \citenamefont {Newell},\ and\ \citenamefont
  {Taylor}}]{Tiesinga2021}%
  \BibitemOpen
  \bibfield  {author} {\bibinfo {author} {\bibfnamefont {E.}~\bibnamefont
  {Tiesinga}}, \bibinfo {author} {\bibfnamefont {P.~J.}\ \bibnamefont {Mohr}},
  \bibinfo {author} {\bibfnamefont {D.~B.}\ \bibnamefont {Newell}}, \ and\
  \bibinfo {author} {\bibfnamefont {B.~N.}\ \bibnamefont {Taylor}},\ }\bibfield
   {title} {\enquote {\bibinfo {title} {{CODATA Recommended Values of the
  Fundamental Physical Constants: 2018}},}\ }\href {\doibase 10.1063/5.0064853}
  {\bibfield  {journal} {\bibinfo  {journal} {J. Phys. Chem. Ref. Data}\
  }\textbf {\bibinfo {volume} {50}},\ \bibinfo {pages} {033105} (\bibinfo
  {year} {2021})}\BibitemShut {NoStop}%
\bibitem [{\citenamefont {Safronova}\ \emph {et~al.}(2018)\citenamefont
  {Safronova}, \citenamefont {Budker}, \citenamefont {DeMille}, \citenamefont
  {Kimball}, \citenamefont {Derevianko},\ and\ \citenamefont
  {Clark}}]{Safronova2018}%
  \BibitemOpen
  \bibfield  {author} {\bibinfo {author} {\bibfnamefont {M.~S.}\ \bibnamefont
  {Safronova}}, \bibinfo {author} {\bibfnamefont {D.}~\bibnamefont {Budker}},
  \bibinfo {author} {\bibfnamefont {D.}~\bibnamefont {DeMille}}, \bibinfo
  {author} {\bibfnamefont {D.~F.~J.}\ \bibnamefont {Kimball}}, \bibinfo
  {author} {\bibfnamefont {A.}~\bibnamefont {Derevianko}}, \ and\ \bibinfo
  {author} {\bibfnamefont {C.~W.}\ \bibnamefont {Clark}},\ }\bibfield  {title}
  {\enquote {\bibinfo {title} {{Search for new physics with atoms and
  molecules}},}\ }\href {\doibase 10.1103/RevModPhys.90.025008} {\bibfield
  {journal} {\bibinfo  {journal} {Rev. Mod. Phys.}\ }\textbf {\bibinfo {volume}
  {90}},\ \bibinfo {pages} {025008} (\bibinfo {year} {2018})}\BibitemShut
  {NoStop}%
\bibitem [{\citenamefont {Karr}\ \emph {et~al.}(2020)\citenamefont {Karr},
  \citenamefont {Marchand},\ and\ \citenamefont {Voutier}}]{Karr2020}%
  \BibitemOpen
  \bibfield  {author} {\bibinfo {author} {\bibfnamefont {J.-P.}\ \bibnamefont
  {Karr}}, \bibinfo {author} {\bibfnamefont {D.}~\bibnamefont {Marchand}}, \
  and\ \bibinfo {author} {\bibfnamefont {E.}~\bibnamefont {Voutier}},\
  }\bibfield  {title} {\enquote {\bibinfo {title} {{The proton size}},}\ }\href
  {\doibase 10.1038/s42254-020-0229-x} {\bibfield  {journal} {\bibinfo
  {journal} {Nat. Rev. Phys.}\ }\textbf {\bibinfo {volume} {2}},\ \bibinfo
  {pages} {601} (\bibinfo {year} {2020})}\BibitemShut {NoStop}%
\bibitem [{\citenamefont {Ye}\ \emph {et~al.}(2008)\citenamefont {Ye},
  \citenamefont {Kimble},\ and\ \citenamefont {Katori}}]{Ye2008}%
  \BibitemOpen
  \bibfield  {author} {\bibinfo {author} {\bibfnamefont {J.}~\bibnamefont
  {Ye}}, \bibinfo {author} {\bibfnamefont {H.~J.}\ \bibnamefont {Kimble}}, \
  and\ \bibinfo {author} {\bibfnamefont {H.}~\bibnamefont {Katori}},\
  }\bibfield  {title} {\enquote {\bibinfo {title} {{Quantum State Engineering
  and Precision Metrology Using State-Insensitive Light Traps}},}\ }\href
  {\doibase 10.1126/science.1148259} {\bibfield  {journal} {\bibinfo  {journal}
  {Science}\ }\textbf {\bibinfo {volume} {320}},\ \bibinfo {pages} {1734}
  (\bibinfo {year} {2008})}\BibitemShut {NoStop}%
\bibitem [{\citenamefont {Wu}\ \emph {et~al.}(2019)\citenamefont {Wu},
  \citenamefont {Pagel}, \citenamefont {Malek}, \citenamefont {Nguyen},
  \citenamefont {Zi}, \citenamefont {Scheirer},\ and\ \citenamefont
  {M{\"{u}}ller}}]{Wu2019}%
  \BibitemOpen
  \bibfield  {author} {\bibinfo {author} {\bibfnamefont {X.}~\bibnamefont
  {Wu}}, \bibinfo {author} {\bibfnamefont {Z.}~\bibnamefont {Pagel}}, \bibinfo
  {author} {\bibfnamefont {B.~S.}\ \bibnamefont {Malek}}, \bibinfo {author}
  {\bibfnamefont {T.~H.}\ \bibnamefont {Nguyen}}, \bibinfo {author}
  {\bibfnamefont {F.}~\bibnamefont {Zi}}, \bibinfo {author} {\bibfnamefont
  {D.~S.}\ \bibnamefont {Scheirer}}, \ and\ \bibinfo {author} {\bibfnamefont
  {H.}~\bibnamefont {M{\"{u}}ller}},\ }\bibfield  {title} {\enquote {\bibinfo
  {title} {{Gravity surveys using a mobile atom interferometer}},}\ }\href
  {\doibase 10.1126/sciadv.aax0800} {\bibfield  {journal} {\bibinfo  {journal}
  {Sci. Adv.}\ }\textbf {\bibinfo {volume} {5}},\ \bibinfo {pages} {eaax0800}
  (\bibinfo {year} {2019})}\BibitemShut {NoStop}%
\bibitem [{\citenamefont {PARIS}(2009)}]{PARIS2009}%
  \BibitemOpen
  \bibfield  {author} {\bibinfo {author} {\bibfnamefont {M.~G.~A.}\
  \bibnamefont {PARIS}},\ }\bibfield  {title} {\enquote {\bibinfo {title}
  {{Quantum estimation for quantum technology}},}\ }\href {\doibase
  10.1142/S0219749909004839} {\bibfield  {journal} {\bibinfo  {journal} {Int.
  J. Quantum Inf.}\ }\textbf {\bibinfo {volume} {07}},\ \bibinfo {pages} {125}
  (\bibinfo {year} {2009})}\BibitemShut {NoStop}%
\bibitem [{\citenamefont {Degen}\ \emph {et~al.}(2017)\citenamefont {Degen},
  \citenamefont {Reinhard},\ and\ \citenamefont {Cappellaro}}]{degen2017}%
  \BibitemOpen
  \bibfield  {author} {\bibinfo {author} {\bibfnamefont {C.~L.}\ \bibnamefont
  {Degen}}, \bibinfo {author} {\bibfnamefont {F.}~\bibnamefont {Reinhard}}, \
  and\ \bibinfo {author} {\bibfnamefont {P.}~\bibnamefont {Cappellaro}},\
  }\bibfield  {title} {\enquote {\bibinfo {title} {{Quantum sensing}},}\ }\href
  {\doibase 10.1103/RevModPhys.89.035002} {\bibfield  {journal} {\bibinfo
  {journal} {Rev. Mod. Phys.}\ }\textbf {\bibinfo {volume} {89}},\ \bibinfo
  {pages} {035002} (\bibinfo {year} {2017})}\BibitemShut {NoStop}%
\bibitem [{\citenamefont {Pirandola}\ \emph {et~al.}(2018)\citenamefont
  {Pirandola}, \citenamefont {Bardhan}, \citenamefont {Gehring}, \citenamefont
  {Weedbrook},\ and\ \citenamefont {Lloyd}}]{pirandola2018}%
  \BibitemOpen
  \bibfield  {author} {\bibinfo {author} {\bibfnamefont {S.}~\bibnamefont
  {Pirandola}}, \bibinfo {author} {\bibfnamefont {B.~R.}\ \bibnamefont
  {Bardhan}}, \bibinfo {author} {\bibfnamefont {T.}~\bibnamefont {Gehring}},
  \bibinfo {author} {\bibfnamefont {C.}~\bibnamefont {Weedbrook}}, \ and\
  \bibinfo {author} {\bibfnamefont {S.}~\bibnamefont {Lloyd}},\ }\bibfield
  {title} {\enquote {\bibinfo {title} {Advances in photonic quantum sensing},}\
  }\href {\doibase 10.1038/s41566-018-0301-6} {\bibfield  {journal} {\bibinfo
  {journal} {Nat. Photon.}\ }\textbf {\bibinfo {volume} {12}},\ \bibinfo
  {pages} {724} (\bibinfo {year} {2018})}\BibitemShut {NoStop}%
\bibitem [{\citenamefont {Braunstein}\ and\ \citenamefont
  {Caves}(1994)}]{braunstein1994}%
  \BibitemOpen
  \bibfield  {author} {\bibinfo {author} {\bibfnamefont {S.~L.}\ \bibnamefont
  {Braunstein}}\ and\ \bibinfo {author} {\bibfnamefont {C.~M.}\ \bibnamefont
  {Caves}},\ }\bibfield  {title} {\enquote {\bibinfo {title} {Statistical
  distance and the geometry of quantum states},}\ }\href {\doibase
  10.1103/PhysRevLett.72.3439} {\bibfield  {journal} {\bibinfo  {journal}
  {Phys. Rev. Lett.}\ }\textbf {\bibinfo {volume} {72}},\ \bibinfo {pages}
  {3439} (\bibinfo {year} {1994})}\BibitemShut {NoStop}%
\bibitem [{\citenamefont {Giovannetti}\ \emph {et~al.}(2004)\citenamefont
  {Giovannetti}, \citenamefont {Lloyd},\ and\ \citenamefont
  {Maccone}}]{giovannetti2004}%
  \BibitemOpen
  \bibfield  {author} {\bibinfo {author} {\bibfnamefont {V.}~\bibnamefont
  {Giovannetti}}, \bibinfo {author} {\bibfnamefont {S.}~\bibnamefont {Lloyd}},
  \ and\ \bibinfo {author} {\bibfnamefont {L.}~\bibnamefont {Maccone}},\
  }\bibfield  {title} {\enquote {\bibinfo {title} {Quantum-{{Enhanced
  Measurements}}: {{Beating}} the {{Standard Quantum Limit}}},}\ }\href
  {\doibase 10.1126/science.1104149} {\bibfield  {journal} {\bibinfo  {journal}
  {Science}\ }\textbf {\bibinfo {volume} {306}},\ \bibinfo {pages} {1330}
  (\bibinfo {year} {2004})}\BibitemShut {NoStop}%
\bibitem [{\citenamefont {Gao}\ \emph {et~al.}(2010)\citenamefont {Gao},
  \citenamefont {Lu}, \citenamefont {Yao}, \citenamefont {Xu}, \citenamefont
  {G{\"u}hne}, \citenamefont {Goebel}, \citenamefont {Chen}, \citenamefont
  {Peng}, \citenamefont {Chen},\ and\ \citenamefont
  {Pan}}]{gao2010experimental}%
  \BibitemOpen
  \bibfield  {author} {\bibinfo {author} {\bibfnamefont {W.-B.}\ \bibnamefont
  {Gao}}, \bibinfo {author} {\bibfnamefont {C.-Y.}\ \bibnamefont {Lu}},
  \bibinfo {author} {\bibfnamefont {X.-C.}\ \bibnamefont {Yao}}, \bibinfo
  {author} {\bibfnamefont {P.}~\bibnamefont {Xu}}, \bibinfo {author}
  {\bibfnamefont {O.}~\bibnamefont {G{\"u}hne}}, \bibinfo {author}
  {\bibfnamefont {A.}~\bibnamefont {Goebel}}, \bibinfo {author} {\bibfnamefont
  {Y.-A.}\ \bibnamefont {Chen}}, \bibinfo {author} {\bibfnamefont {C.-Z.}\
  \bibnamefont {Peng}}, \bibinfo {author} {\bibfnamefont {Z.-B.}\ \bibnamefont
  {Chen}}, \ and\ \bibinfo {author} {\bibfnamefont {J.-W.}\ \bibnamefont
  {Pan}},\ }\bibfield  {title} {\enquote {\bibinfo {title} {Experimental
  demonstration of a hyper-entangled ten-qubit {Schr{\"o}dinger} cat state},}\
  }\href {\doibase 10.1038/nphys1603} {\bibfield  {journal} {\bibinfo
  {journal} {Nat. Phys.}\ }\textbf {\bibinfo {volume} {6}},\ \bibinfo {pages}
  {331} (\bibinfo {year} {2010})}\BibitemShut {NoStop}%
\bibitem [{\citenamefont {Monz}\ \emph {et~al.}(2011)\citenamefont {Monz},
  \citenamefont {Schindler}, \citenamefont {Barreiro}, \citenamefont {Chwalla},
  \citenamefont {Nigg}, \citenamefont {Coish}, \citenamefont {Harlander},
  \citenamefont {H{\"a}nsel}, \citenamefont {Hennrich},\ and\ \citenamefont
  {Blatt}}]{monz2011}%
  \BibitemOpen
  \bibfield  {author} {\bibinfo {author} {\bibfnamefont {T.}~\bibnamefont
  {Monz}}, \bibinfo {author} {\bibfnamefont {P.}~\bibnamefont {Schindler}},
  \bibinfo {author} {\bibfnamefont {J.~T.}\ \bibnamefont {Barreiro}}, \bibinfo
  {author} {\bibfnamefont {M.}~\bibnamefont {Chwalla}}, \bibinfo {author}
  {\bibfnamefont {D.}~\bibnamefont {Nigg}}, \bibinfo {author} {\bibfnamefont
  {W.~A.}\ \bibnamefont {Coish}}, \bibinfo {author} {\bibfnamefont
  {M.}~\bibnamefont {Harlander}}, \bibinfo {author} {\bibfnamefont
  {W.}~\bibnamefont {H{\"a}nsel}}, \bibinfo {author} {\bibfnamefont
  {M.}~\bibnamefont {Hennrich}}, \ and\ \bibinfo {author} {\bibfnamefont
  {R.}~\bibnamefont {Blatt}},\ }\bibfield  {title} {\enquote {\bibinfo {title}
  {{14-Qubit Entanglement: Creation and Coherence}},}\ }\href {\doibase
  10.1103/PhysRevLett.106.130506} {\bibfield  {journal} {\bibinfo  {journal}
  {Phys. Rev. Lett.}\ }\textbf {\bibinfo {volume} {106}},\ \bibinfo {pages}
  {130506} (\bibinfo {year} {2011})}\BibitemShut {NoStop}%
\bibitem [{\citenamefont {Omran}\ \emph {et~al.}(2019)\citenamefont {Omran},
  \citenamefont {Levine}, \citenamefont {Keesling}, \citenamefont {Semeghini},
  \citenamefont {Wang}, \citenamefont {Ebadi}, \citenamefont {Bernien},
  \citenamefont {Zibrov}, \citenamefont {Pichler}, \citenamefont {Choi},
  \citenamefont {Cui}, \citenamefont {Rossignolo}, \citenamefont {Rembold},
  \citenamefont {Montangero}, \citenamefont {Calarco}, \citenamefont {Endres},
  \citenamefont {Greiner}, \citenamefont {Vuleti{\'c}},\ and\ \citenamefont
  {Lukin}}]{Omran2019}%
  \BibitemOpen
  \bibfield  {author} {\bibinfo {author} {\bibfnamefont {A.}~\bibnamefont
  {Omran}}, \bibinfo {author} {\bibfnamefont {H.}~\bibnamefont {Levine}},
  \bibinfo {author} {\bibfnamefont {A.}~\bibnamefont {Keesling}}, \bibinfo
  {author} {\bibfnamefont {G.}~\bibnamefont {Semeghini}}, \bibinfo {author}
  {\bibfnamefont {T.~T.}\ \bibnamefont {Wang}}, \bibinfo {author}
  {\bibfnamefont {S.}~\bibnamefont {Ebadi}}, \bibinfo {author} {\bibfnamefont
  {H.}~\bibnamefont {Bernien}}, \bibinfo {author} {\bibfnamefont {A.~S.}\
  \bibnamefont {Zibrov}}, \bibinfo {author} {\bibfnamefont {H.}~\bibnamefont
  {Pichler}}, \bibinfo {author} {\bibfnamefont {S.}~\bibnamefont {Choi}},
  \bibinfo {author} {\bibfnamefont {J.}~\bibnamefont {Cui}}, \bibinfo {author}
  {\bibfnamefont {M.}~\bibnamefont {Rossignolo}}, \bibinfo {author}
  {\bibfnamefont {P.}~\bibnamefont {Rembold}}, \bibinfo {author} {\bibfnamefont
  {S.}~\bibnamefont {Montangero}}, \bibinfo {author} {\bibfnamefont
  {T.}~\bibnamefont {Calarco}}, \bibinfo {author} {\bibfnamefont
  {M.}~\bibnamefont {Endres}}, \bibinfo {author} {\bibfnamefont
  {M.}~\bibnamefont {Greiner}}, \bibinfo {author} {\bibfnamefont
  {V.}~\bibnamefont {Vuleti{\'c}}}, \ and\ \bibinfo {author} {\bibfnamefont
  {M.~D.}\ \bibnamefont {Lukin}},\ }\bibfield  {title} {\enquote {\bibinfo
  {title} {Generation and manipulation of {{Schr\"odinger}} cat states in
  {Rydberg} atom arrays},}\ }\href {\doibase 10.1126/science.aax9743}
  {\bibfield  {journal} {\bibinfo  {journal} {Science}\ }\textbf {\bibinfo
  {volume} {365}},\ \bibinfo {pages} {570} (\bibinfo {year}
  {2019})}\BibitemShut {NoStop}%
\bibitem [{\citenamefont {Song}\ \emph {et~al.}(2019)\citenamefont {Song},
  \citenamefont {Xu}, \citenamefont {Li}, \citenamefont {Zhang}, \citenamefont
  {Zhang}, \citenamefont {Liu}, \citenamefont {Guo}, \citenamefont {Wang},
  \citenamefont {Ren}, \citenamefont {Hao}, \citenamefont {Feng}, \citenamefont
  {Fan}, \citenamefont {Zheng}, \citenamefont {Wang}, \citenamefont {Wang},\
  and\ \citenamefont {Zhu}}]{song2019}%
  \BibitemOpen
  \bibfield  {author} {\bibinfo {author} {\bibfnamefont {C.}~\bibnamefont
  {Song}}, \bibinfo {author} {\bibfnamefont {K.}~\bibnamefont {Xu}}, \bibinfo
  {author} {\bibfnamefont {H.}~\bibnamefont {Li}}, \bibinfo {author}
  {\bibfnamefont {Y.-R.}\ \bibnamefont {Zhang}}, \bibinfo {author}
  {\bibfnamefont {X.}~\bibnamefont {Zhang}}, \bibinfo {author} {\bibfnamefont
  {W.}~\bibnamefont {Liu}}, \bibinfo {author} {\bibfnamefont {Q.}~\bibnamefont
  {Guo}}, \bibinfo {author} {\bibfnamefont {Z.}~\bibnamefont {Wang}}, \bibinfo
  {author} {\bibfnamefont {W.}~\bibnamefont {Ren}}, \bibinfo {author}
  {\bibfnamefont {J.}~\bibnamefont {Hao}}, \bibinfo {author} {\bibfnamefont
  {H.}~\bibnamefont {Feng}}, \bibinfo {author} {\bibfnamefont {H.}~\bibnamefont
  {Fan}}, \bibinfo {author} {\bibfnamefont {D.}~\bibnamefont {Zheng}}, \bibinfo
  {author} {\bibfnamefont {D.-W.}\ \bibnamefont {Wang}}, \bibinfo {author}
  {\bibfnamefont {H.}~\bibnamefont {Wang}}, \ and\ \bibinfo {author}
  {\bibfnamefont {S.-Y.}\ \bibnamefont {Zhu}},\ }\bibfield  {title} {\enquote
  {\bibinfo {title} {Generation of multicomponent atomic {{Schr\"odinger}} cat
  states of up to 20 qubits},}\ }\href {\doibase 10.1126/science.aay0600}
  {\bibfield  {journal} {\bibinfo  {journal} {Science}\ }\textbf {\bibinfo
  {volume} {365}},\ \bibinfo {pages} {574} (\bibinfo {year}
  {2019})}\BibitemShut {NoStop}%
\bibitem [{\citenamefont {Nagata}\ \emph {et~al.}(2007)\citenamefont {Nagata},
  \citenamefont {Okamoto}, \citenamefont {O'Brien}, \citenamefont {Sasaki},\
  and\ \citenamefont {{Shigeki Takeuchi}}}]{nagata2007}%
  \BibitemOpen
  \bibfield  {author} {\bibinfo {author} {\bibfnamefont {T.}~\bibnamefont
  {Nagata}}, \bibinfo {author} {\bibfnamefont {R.}~\bibnamefont {Okamoto}},
  \bibinfo {author} {\bibfnamefont {J.~L.}\ \bibnamefont {O'Brien}}, \bibinfo
  {author} {\bibfnamefont {K.}~\bibnamefont {Sasaki}}, \ and\ \bibinfo {author}
  {\bibnamefont {{Shigeki Takeuchi}}},\ }\bibfield  {title} {\enquote {\bibinfo
  {title} {{Beating the Standard Quantum Limit with Four-Entangled Photons}},}\
  }\href {\doibase 10.1126/science.1138007} {\bibfield  {journal} {\bibinfo
  {journal} {Science}\ }\textbf {\bibinfo {volume} {316}},\ \bibinfo {pages}
  {726} (\bibinfo {year} {2007})}\BibitemShut {NoStop}%
\bibitem [{\citenamefont {Chen}\ \emph {et~al.}(2010)\citenamefont {Chen},
  \citenamefont {Bao}, \citenamefont {Yuan}, \citenamefont {Chen},
  \citenamefont {Zhao},\ and\ \citenamefont {Pan}}]{chen2010}%
  \BibitemOpen
  \bibfield  {author} {\bibinfo {author} {\bibfnamefont {Y.-A.}\ \bibnamefont
  {Chen}}, \bibinfo {author} {\bibfnamefont {X.-H.}\ \bibnamefont {Bao}},
  \bibinfo {author} {\bibfnamefont {Z.-S.}\ \bibnamefont {Yuan}}, \bibinfo
  {author} {\bibfnamefont {S.}~\bibnamefont {Chen}}, \bibinfo {author}
  {\bibfnamefont {B.}~\bibnamefont {Zhao}}, \ and\ \bibinfo {author}
  {\bibfnamefont {J.-W.}\ \bibnamefont {Pan}},\ }\bibfield  {title} {\enquote
  {\bibinfo {title} {{Heralded Generation of an Atomic NOON State}},}\ }\href
  {\doibase 10.1103/PhysRevLett.104.043601} {\bibfield  {journal} {\bibinfo
  {journal} {Phys. Rev. Lett.}\ }\textbf {\bibinfo {volume} {104}},\ \bibinfo
  {pages} {043601} (\bibinfo {year} {2010})}\BibitemShut {NoStop}%
\bibitem [{\citenamefont {Zhang}\ \emph {et~al.}(2018)\citenamefont {Zhang},
  \citenamefont {Um}, \citenamefont {Lv}, \citenamefont {Zhang}, \citenamefont
  {Duan},\ and\ \citenamefont {Kim}}]{zhang2018}%
  \BibitemOpen
  \bibfield  {author} {\bibinfo {author} {\bibfnamefont {J.}~\bibnamefont
  {Zhang}}, \bibinfo {author} {\bibfnamefont {M.}~\bibnamefont {Um}}, \bibinfo
  {author} {\bibfnamefont {D.}~\bibnamefont {Lv}}, \bibinfo {author}
  {\bibfnamefont {J.-N.}\ \bibnamefont {Zhang}}, \bibinfo {author}
  {\bibfnamefont {L.-M.}\ \bibnamefont {Duan}}, \ and\ \bibinfo {author}
  {\bibfnamefont {K.}~\bibnamefont {Kim}},\ }\bibfield  {title} {\enquote
  {\bibinfo {title} {{{NOON States}} of {{Nine Quantized Vibrations}} in {{Two
  Radial Modes}} of a {{Trapped Ion}}},}\ }\href {\doibase
  10.1103/PhysRevLett.121.160502} {\bibfield  {journal} {\bibinfo  {journal}
  {Phys. Rev. Lett.}\ }\textbf {\bibinfo {volume} {121}},\ \bibinfo {pages}
  {160502} (\bibinfo {year} {2018})}\BibitemShut {NoStop}%
\bibitem [{\citenamefont {Hosten}\ \emph {et~al.}(2016)\citenamefont {Hosten},
  \citenamefont {Engelsen}, \citenamefont {Krishnakumar},\ and\ \citenamefont
  {Kasevich}}]{hosten2016}%
  \BibitemOpen
  \bibfield  {author} {\bibinfo {author} {\bibfnamefont {O.}~\bibnamefont
  {Hosten}}, \bibinfo {author} {\bibfnamefont {N.~J.}\ \bibnamefont
  {Engelsen}}, \bibinfo {author} {\bibfnamefont {R.}~\bibnamefont
  {Krishnakumar}}, \ and\ \bibinfo {author} {\bibfnamefont {M.~A.}\
  \bibnamefont {Kasevich}},\ }\bibfield  {title} {\enquote {\bibinfo {title}
  {Measurement noise 100 times lower than the quantum-projection limit using
  entangled atoms},}\ }\href {\doibase 10.1038/nature16176} {\bibfield
  {journal} {\bibinfo  {journal} {Nature}\ }\textbf {\bibinfo {volume} {529}},\
  \bibinfo {pages} {505} (\bibinfo {year} {2016})}\BibitemShut {NoStop}%
\bibitem [{\citenamefont {Colombo}\ \emph {et~al.}(2022)\citenamefont
  {Colombo}, \citenamefont {{Pedrozo-Pe{\~n}afiel}}, \citenamefont
  {Adiyatullin}, \citenamefont {Li}, \citenamefont {Mendez}, \citenamefont
  {Shu},\ and\ \citenamefont {Vuleti{\'c}}}]{colombo2022}%
  \BibitemOpen
  \bibfield  {author} {\bibinfo {author} {\bibfnamefont {S.}~\bibnamefont
  {Colombo}}, \bibinfo {author} {\bibfnamefont {E.}~\bibnamefont
  {{Pedrozo-Pe{\~n}afiel}}}, \bibinfo {author} {\bibfnamefont {A.~F.}\
  \bibnamefont {Adiyatullin}}, \bibinfo {author} {\bibfnamefont
  {Z.}~\bibnamefont {Li}}, \bibinfo {author} {\bibfnamefont {E.}~\bibnamefont
  {Mendez}}, \bibinfo {author} {\bibfnamefont {C.}~\bibnamefont {Shu}}, \ and\
  \bibinfo {author} {\bibfnamefont {V.}~\bibnamefont {Vuleti{\'c}}},\
  }\bibfield  {title} {\enquote {\bibinfo {title} {Time-reversal-based quantum
  metrology with many-body entangled states},}\ }\href {\doibase
  10.1038/s41567-022-01653-5} {\bibfield  {journal} {\bibinfo  {journal} {Nat.
  Phys.}\ }\textbf {\bibinfo {volume} {18}},\ \bibinfo {pages} {925} (\bibinfo
  {year} {2022})}\BibitemShut {NoStop}%
\bibitem [{\citenamefont {Xu}\ \emph {et~al.}(2022)\citenamefont {Xu},
  \citenamefont {Zhang}, \citenamefont {Sun}, \citenamefont {Li}, \citenamefont
  {Song}, \citenamefont {Xiang}, \citenamefont {Huang}, \citenamefont {Li},
  \citenamefont {Shi}, \citenamefont {Chen}, \citenamefont {Song},
  \citenamefont {Zheng}, \citenamefont {Nori}, \citenamefont {Wang},\ and\
  \citenamefont {Fan}}]{xu2022metrological}%
  \BibitemOpen
  \bibfield  {author} {\bibinfo {author} {\bibfnamefont {K.}~\bibnamefont
  {Xu}}, \bibinfo {author} {\bibfnamefont {Y.-R.}\ \bibnamefont {Zhang}},
  \bibinfo {author} {\bibfnamefont {Z.-H.}\ \bibnamefont {Sun}}, \bibinfo
  {author} {\bibfnamefont {H.}~\bibnamefont {Li}}, \bibinfo {author}
  {\bibfnamefont {P.}~\bibnamefont {Song}}, \bibinfo {author} {\bibfnamefont
  {Z.}~\bibnamefont {Xiang}}, \bibinfo {author} {\bibfnamefont
  {K.}~\bibnamefont {Huang}}, \bibinfo {author} {\bibfnamefont
  {H.}~\bibnamefont {Li}}, \bibinfo {author} {\bibfnamefont {Y.-H.}\
  \bibnamefont {Shi}}, \bibinfo {author} {\bibfnamefont {C.-T.}\ \bibnamefont
  {Chen}}, \bibinfo {author} {\bibfnamefont {X.}~\bibnamefont {Song}}, \bibinfo
  {author} {\bibfnamefont {D.}~\bibnamefont {Zheng}}, \bibinfo {author}
  {\bibfnamefont {F.}~\bibnamefont {Nori}}, \bibinfo {author} {\bibfnamefont
  {H.}~\bibnamefont {Wang}}, \ and\ \bibinfo {author} {\bibfnamefont
  {H.}~\bibnamefont {Fan}},\ }\bibfield  {title} {\enquote {\bibinfo {title}
  {{Metrological Characterization of Non-Gaussian Entangled States of
  Superconducting Qubits}},}\ }\href {\doibase 10.1103/PhysRevLett.128.150501}
  {\bibfield  {journal} {\bibinfo  {journal} {Phys. Rev. Lett.}\ }\textbf
  {\bibinfo {volume} {128}},\ \bibinfo {pages} {150501} (\bibinfo {year}
  {2022})}\BibitemShut {NoStop}%
\bibitem [{\citenamefont {Penasa}\ \emph {et~al.}(2016)\citenamefont {Penasa},
  \citenamefont {Gerlich}, \citenamefont {Rybarczyk}, \citenamefont
  {M{\'e}tillon}, \citenamefont {Brune}, \citenamefont {Raimond}, \citenamefont
  {Haroche}, \citenamefont {Davidovich},\ and\ \citenamefont
  {Dotsenko}}]{penasa2016}%
  \BibitemOpen
  \bibfield  {author} {\bibinfo {author} {\bibfnamefont {M.}~\bibnamefont
  {Penasa}}, \bibinfo {author} {\bibfnamefont {S.}~\bibnamefont {Gerlich}},
  \bibinfo {author} {\bibfnamefont {T.}~\bibnamefont {Rybarczyk}}, \bibinfo
  {author} {\bibfnamefont {V.}~\bibnamefont {M{\'e}tillon}}, \bibinfo {author}
  {\bibfnamefont {M.}~\bibnamefont {Brune}}, \bibinfo {author} {\bibfnamefont
  {J.~M.}\ \bibnamefont {Raimond}}, \bibinfo {author} {\bibfnamefont
  {S.}~\bibnamefont {Haroche}}, \bibinfo {author} {\bibfnamefont
  {L.}~\bibnamefont {Davidovich}}, \ and\ \bibinfo {author} {\bibfnamefont
  {I.}~\bibnamefont {Dotsenko}},\ }\bibfield  {title} {\enquote {\bibinfo
  {title} {Measurement of a microwave field amplitude beyond the standard
  quantum limit},}\ }\href {\doibase 10.1103/PhysRevA.94.022313} {\bibfield
  {journal} {\bibinfo  {journal} {Phys. Rev. A}\ }\textbf {\bibinfo {volume}
  {94}},\ \bibinfo {pages} {022313} (\bibinfo {year} {2016})}\BibitemShut
  {NoStop}%
\bibitem [{\citenamefont {Gilmore}\ \emph {et~al.}(2021)\citenamefont
  {Gilmore}, \citenamefont {Affolter}, \citenamefont {{Lewis-Swan}},
  \citenamefont {Barberena}, \citenamefont {Jordan}, \citenamefont {Rey},\ and\
  \citenamefont {Bollinger}}]{gilmore2021}%
  \BibitemOpen
  \bibfield  {author} {\bibinfo {author} {\bibfnamefont {K.~A.}\ \bibnamefont
  {Gilmore}}, \bibinfo {author} {\bibfnamefont {M.}~\bibnamefont {Affolter}},
  \bibinfo {author} {\bibfnamefont {R.~J.}\ \bibnamefont {{Lewis-Swan}}},
  \bibinfo {author} {\bibfnamefont {D.}~\bibnamefont {Barberena}}, \bibinfo
  {author} {\bibfnamefont {E.}~\bibnamefont {Jordan}}, \bibinfo {author}
  {\bibfnamefont {A.~M.}\ \bibnamefont {Rey}}, \ and\ \bibinfo {author}
  {\bibfnamefont {J.~J.}\ \bibnamefont {Bollinger}},\ }\bibfield  {title}
  {\enquote {\bibinfo {title} {Quantum-enhanced sensing of displacements and
  electric fields with two-dimensional trapped-ion crystals},}\ }\href
  {\doibase 10.1126/science.abi5226} {\bibfield  {journal} {\bibinfo  {journal}
  {Science}\ }\textbf {\bibinfo {volume} {373}},\ \bibinfo {pages} {673}
  (\bibinfo {year} {2021})}\BibitemShut {NoStop}%
\bibitem [{\citenamefont {Marciniak}\ \emph {et~al.}(2022)\citenamefont
  {Marciniak}, \citenamefont {Feldker}, \citenamefont {Pogorelov},
  \citenamefont {Kaubruegger}, \citenamefont {Vasilyev}, \citenamefont {{van
  Bijnen}}, \citenamefont {Schindler}, \citenamefont {Zoller}, \citenamefont
  {Blatt},\ and\ \citenamefont {Monz}}]{marciniak2022}%
  \BibitemOpen
  \bibfield  {author} {\bibinfo {author} {\bibfnamefont {C.~D.}\ \bibnamefont
  {Marciniak}}, \bibinfo {author} {\bibfnamefont {T.}~\bibnamefont {Feldker}},
  \bibinfo {author} {\bibfnamefont {I.}~\bibnamefont {Pogorelov}}, \bibinfo
  {author} {\bibfnamefont {R.}~\bibnamefont {Kaubruegger}}, \bibinfo {author}
  {\bibfnamefont {D.~V.}\ \bibnamefont {Vasilyev}}, \bibinfo {author}
  {\bibfnamefont {R.}~\bibnamefont {{van Bijnen}}}, \bibinfo {author}
  {\bibfnamefont {P.}~\bibnamefont {Schindler}}, \bibinfo {author}
  {\bibfnamefont {P.}~\bibnamefont {Zoller}}, \bibinfo {author} {\bibfnamefont
  {R.}~\bibnamefont {Blatt}}, \ and\ \bibinfo {author} {\bibfnamefont
  {T.}~\bibnamefont {Monz}},\ }\bibfield  {title} {\enquote {\bibinfo {title}
  {Optimal metrology with programmable quantum sensors},}\ }\href {\doibase
  10.1038/s41586-022-04435-4} {\bibfield  {journal} {\bibinfo  {journal}
  {Nature}\ }\textbf {\bibinfo {volume} {603}},\ \bibinfo {pages} {604}
  (\bibinfo {year} {2022})}\BibitemShut {NoStop}%
\bibitem [{\citenamefont {Braun}\ \emph {et~al.}(2018)\citenamefont {Braun},
  \citenamefont {Adesso}, \citenamefont {Benatti}, \citenamefont {Floreanini},
  \citenamefont {Marzolino}, \citenamefont {Mitchell},\ and\ \citenamefont
  {Pirandola}}]{Braun2018}%
  \BibitemOpen
  \bibfield  {author} {\bibinfo {author} {\bibfnamefont {D.}~\bibnamefont
  {Braun}}, \bibinfo {author} {\bibfnamefont {G.}~\bibnamefont {Adesso}},
  \bibinfo {author} {\bibfnamefont {F.}~\bibnamefont {Benatti}}, \bibinfo
  {author} {\bibfnamefont {R.}~\bibnamefont {Floreanini}}, \bibinfo {author}
  {\bibfnamefont {U.}~\bibnamefont {Marzolino}}, \bibinfo {author}
  {\bibfnamefont {M.~W.}\ \bibnamefont {Mitchell}}, \ and\ \bibinfo {author}
  {\bibfnamefont {S.}~\bibnamefont {Pirandola}},\ }\bibfield  {title} {\enquote
  {\bibinfo {title} {{Quantum-enhanced measurements without entanglement}},}\
  }\href {\doibase 10.1103/RevModPhys.90.035006} {\bibfield  {journal}
  {\bibinfo  {journal} {Rev. Mod. Phys.}\ }\textbf {\bibinfo {volume} {90}},\
  \bibinfo {pages} {35006} (\bibinfo {year} {2018})}\BibitemShut {NoStop}%
\bibitem [{\citenamefont {Duivenvoorden}\ \emph {et~al.}(2017)\citenamefont
  {Duivenvoorden}, \citenamefont {Terhal},\ and\ \citenamefont
  {Weigand}}]{Duivenvoorden2017}%
  \BibitemOpen
  \bibfield  {author} {\bibinfo {author} {\bibfnamefont {K.}~\bibnamefont
  {Duivenvoorden}}, \bibinfo {author} {\bibfnamefont {B.~M.}\ \bibnamefont
  {Terhal}}, \ and\ \bibinfo {author} {\bibfnamefont {D.}~\bibnamefont
  {Weigand}},\ }\bibfield  {title} {\enquote {\bibinfo {title} {{Single-mode
  displacement sensor}},}\ }\href {\doibase 10.1103/PhysRevA.95.012305}
  {\bibfield  {journal} {\bibinfo  {journal} {Phys. Rev. A}\ }\textbf {\bibinfo
  {volume} {95}},\ \bibinfo {pages} {012305} (\bibinfo {year}
  {2017})}\BibitemShut {NoStop}%
\bibitem [{\citenamefont {Vlastakis}\ \emph {et~al.}(2013)\citenamefont
  {Vlastakis}, \citenamefont {Kirchmair}, \citenamefont {Leghtas},
  \citenamefont {Nigg}, \citenamefont {Frunzio}, \citenamefont {Girvin},
  \citenamefont {Mirrahimi}, \citenamefont {Devoret},\ and\ \citenamefont
  {Schoelkopf}}]{vlastakis2013}%
  \BibitemOpen
  \bibfield  {author} {\bibinfo {author} {\bibfnamefont {B.}~\bibnamefont
  {Vlastakis}}, \bibinfo {author} {\bibfnamefont {G.}~\bibnamefont
  {Kirchmair}}, \bibinfo {author} {\bibfnamefont {Z.}~\bibnamefont {Leghtas}},
  \bibinfo {author} {\bibfnamefont {S.~E.}\ \bibnamefont {Nigg}}, \bibinfo
  {author} {\bibfnamefont {L.}~\bibnamefont {Frunzio}}, \bibinfo {author}
  {\bibfnamefont {S.~M.}\ \bibnamefont {Girvin}}, \bibinfo {author}
  {\bibfnamefont {M.}~\bibnamefont {Mirrahimi}}, \bibinfo {author}
  {\bibfnamefont {M.~H.}\ \bibnamefont {Devoret}}, \ and\ \bibinfo {author}
  {\bibfnamefont {R.~J.}\ \bibnamefont {Schoelkopf}},\ }\bibfield  {title}
  {\enquote {\bibinfo {title} {Deterministically {{Encoding Quantum Information
  Using}} 100-{{Photon Schr\"odinger Cat States}}},}\ }\href {\doibase
  10.1126/science.1243289} {\bibfield  {journal} {\bibinfo  {journal}
  {Science}\ }\textbf {\bibinfo {volume} {342}},\ \bibinfo {pages} {607}
  (\bibinfo {year} {2013})}\BibitemShut {NoStop}%
\bibitem [{\citenamefont {Backes}\ \emph {et~al.}(2021)\citenamefont {Backes},
  \citenamefont {Palken}, \citenamefont {Kenany}, \citenamefont {Brubaker},
  \citenamefont {Cahn},\ and\ \citenamefont {et~al.}}]{backes2021}%
  \BibitemOpen
  \bibfield  {author} {\bibinfo {author} {\bibfnamefont {K.~M.}\ \bibnamefont
  {Backes}}, \bibinfo {author} {\bibfnamefont {D.~A.}\ \bibnamefont {Palken}},
  \bibinfo {author} {\bibfnamefont {S.~A.}\ \bibnamefont {Kenany}}, \bibinfo
  {author} {\bibfnamefont {B.~M.}\ \bibnamefont {Brubaker}}, \bibinfo {author}
  {\bibfnamefont {S.~B.}\ \bibnamefont {Cahn}}, \ and\ \bibinfo {author}
  {\bibnamefont {et~al.}},\ }\bibfield  {title} {\enquote {\bibinfo {title} {A
  quantum enhanced search for dark matter axions},}\ }\href {\doibase
  10.1038/s41586-021-03226-7} {\bibfield  {journal} {\bibinfo  {journal}
  {Nature}\ }\textbf {\bibinfo {volume} {590}},\ \bibinfo {pages} {238}
  (\bibinfo {year} {2021})}\BibitemShut {NoStop}%
\bibitem [{\citenamefont {Wang}\ \emph {et~al.}(2019)\citenamefont {Wang},
  \citenamefont {Wu}, \citenamefont {Ma}, \citenamefont {Cai}, \citenamefont
  {Hu}, \citenamefont {Mu}, \citenamefont {Xu}, \citenamefont {Chen},
  \citenamefont {Wang}, \citenamefont {Song}, \citenamefont {Yuan},
  \citenamefont {Zou}, \citenamefont {Duan},\ and\ \citenamefont
  {Sun}}]{wang2019}%
  \BibitemOpen
  \bibfield  {author} {\bibinfo {author} {\bibfnamefont {W.}~\bibnamefont
  {Wang}}, \bibinfo {author} {\bibfnamefont {Y.}~\bibnamefont {Wu}}, \bibinfo
  {author} {\bibfnamefont {Y.}~\bibnamefont {Ma}}, \bibinfo {author}
  {\bibfnamefont {W.}~\bibnamefont {Cai}}, \bibinfo {author} {\bibfnamefont
  {L.}~\bibnamefont {Hu}}, \bibinfo {author} {\bibfnamefont {X.}~\bibnamefont
  {Mu}}, \bibinfo {author} {\bibfnamefont {Y.}~\bibnamefont {Xu}}, \bibinfo
  {author} {\bibfnamefont {Z.-J.}\ \bibnamefont {Chen}}, \bibinfo {author}
  {\bibfnamefont {H.}~\bibnamefont {Wang}}, \bibinfo {author} {\bibfnamefont
  {Y.~P.}\ \bibnamefont {Song}}, \bibinfo {author} {\bibfnamefont
  {H.}~\bibnamefont {Yuan}}, \bibinfo {author} {\bibfnamefont {C.-L.}\
  \bibnamefont {Zou}}, \bibinfo {author} {\bibfnamefont {L.-M.}\ \bibnamefont
  {Duan}}, \ and\ \bibinfo {author} {\bibfnamefont {L.}~\bibnamefont {Sun}},\
  }\bibfield  {title} {\enquote {\bibinfo {title} {Heisenberg-limited
  single-mode quantum metrology in a superconducting circuit},}\ }\href
  {\doibase 10.1038/s41467-019-12290-7} {\bibfield  {journal} {\bibinfo
  {journal} {Nat. Commun.}\ }\textbf {\bibinfo {volume} {10}},\ \bibinfo
  {pages} {4382} (\bibinfo {year} {2019})}\BibitemShut {NoStop}%
\bibitem [{\citenamefont {McCormick}\ \emph {et~al.}(2019)\citenamefont
  {McCormick}, \citenamefont {Keller}, \citenamefont {Burd}, \citenamefont
  {Wineland}, \citenamefont {Wilson},\ and\ \citenamefont
  {Leibfried}}]{mccormick2019}%
  \BibitemOpen
  \bibfield  {author} {\bibinfo {author} {\bibfnamefont {K.~C.}\ \bibnamefont
  {McCormick}}, \bibinfo {author} {\bibfnamefont {J.}~\bibnamefont {Keller}},
  \bibinfo {author} {\bibfnamefont {S.~C.}\ \bibnamefont {Burd}}, \bibinfo
  {author} {\bibfnamefont {D.~J.}\ \bibnamefont {Wineland}}, \bibinfo {author}
  {\bibfnamefont {A.~C.}\ \bibnamefont {Wilson}}, \ and\ \bibinfo {author}
  {\bibfnamefont {D.}~\bibnamefont {Leibfried}},\ }\bibfield  {title} {\enquote
  {\bibinfo {title} {Quantum-enhanced sensing of a single-ion mechanical
  oscillator},}\ }\href {\doibase 10.1038/s41586-019-1421-y} {\bibfield
  {journal} {\bibinfo  {journal} {Nature}\ }\textbf {\bibinfo {volume} {572}},\
  \bibinfo {pages} {86} (\bibinfo {year} {2019})}\BibitemShut {NoStop}%
\bibitem [{\citenamefont {Chu}\ \emph {et~al.}(2018)\citenamefont {Chu},
  \citenamefont {Kharel}, \citenamefont {Yoon}, \citenamefont {Frunzio},
  \citenamefont {Rakich},\ and\ \citenamefont {Schoelkopf}}]{chu2018}%
  \BibitemOpen
  \bibfield  {author} {\bibinfo {author} {\bibfnamefont {Y.}~\bibnamefont
  {Chu}}, \bibinfo {author} {\bibfnamefont {P.}~\bibnamefont {Kharel}},
  \bibinfo {author} {\bibfnamefont {T.}~\bibnamefont {Yoon}}, \bibinfo {author}
  {\bibfnamefont {L.}~\bibnamefont {Frunzio}}, \bibinfo {author} {\bibfnamefont
  {P.~T.}\ \bibnamefont {Rakich}}, \ and\ \bibinfo {author} {\bibfnamefont
  {R.~J.}\ \bibnamefont {Schoelkopf}},\ }\bibfield  {title} {\enquote {\bibinfo
  {title} {Creation and control of multi-phonon {{Fock}} states in a bulk
  acoustic-wave resonator},}\ }\href {\doibase 10.1038/s41586-018-0717-7}
  {\bibfield  {journal} {\bibinfo  {journal} {Nature}\ }\textbf {\bibinfo
  {volume} {563}},\ \bibinfo {pages} {666} (\bibinfo {year}
  {2018})}\BibitemShut {NoStop}%
\bibitem [{\citenamefont {Wolf}\ \emph {et~al.}(2019)\citenamefont {Wolf},
  \citenamefont {Shi}, \citenamefont {Heip}, \citenamefont {Gessner},
  \citenamefont {Pezz{\`e}}, \citenamefont {Smerzi}, \citenamefont {Schulte},
  \citenamefont {Hammerer},\ and\ \citenamefont {Schmidt}}]{wolf2019}%
  \BibitemOpen
  \bibfield  {author} {\bibinfo {author} {\bibfnamefont {F.}~\bibnamefont
  {Wolf}}, \bibinfo {author} {\bibfnamefont {C.}~\bibnamefont {Shi}}, \bibinfo
  {author} {\bibfnamefont {J.~C.}\ \bibnamefont {Heip}}, \bibinfo {author}
  {\bibfnamefont {M.}~\bibnamefont {Gessner}}, \bibinfo {author} {\bibfnamefont
  {L.}~\bibnamefont {Pezz{\`e}}}, \bibinfo {author} {\bibfnamefont
  {A.}~\bibnamefont {Smerzi}}, \bibinfo {author} {\bibfnamefont
  {M.}~\bibnamefont {Schulte}}, \bibinfo {author} {\bibfnamefont
  {K.}~\bibnamefont {Hammerer}}, \ and\ \bibinfo {author} {\bibfnamefont
  {P.~O.}\ \bibnamefont {Schmidt}},\ }\bibfield  {title} {\enquote {\bibinfo
  {title} {Motional {{Fock}} states for quantum-enhanced amplitude and phase
  measurements with trapped ions},}\ }\href {\doibase
  10.1038/s41467-019-10576-4} {\bibfield  {journal} {\bibinfo  {journal} {Nat.
  Commun.}\ }\textbf {\bibinfo {volume} {10}},\ \bibinfo {pages} {2929}
  (\bibinfo {year} {2019})}\BibitemShut {NoStop}%
\bibitem [{\citenamefont {Podhora}\ \emph {et~al.}(2022)\citenamefont
  {Podhora}, \citenamefont {Lachman}, \citenamefont {Pham}, \citenamefont
  {Le\ifmmode~\check{s}\else \v{s}\fi{}und\'ak}, \citenamefont
  {\ifmmode~\check{C}\else \v{C}\fi{}\'{\i}p}, \citenamefont
  {Slodi\ifmmode~\check{c}\else \v{c}\fi{}ka},\ and\ \citenamefont
  {Filip}}]{Podhora2022}%
  \BibitemOpen
  \bibfield  {author} {\bibinfo {author} {\bibfnamefont {L.}~\bibnamefont
  {Podhora}}, \bibinfo {author} {\bibfnamefont {L.}~\bibnamefont {Lachman}},
  \bibinfo {author} {\bibfnamefont {T.}~\bibnamefont {Pham}}, \bibinfo {author}
  {\bibfnamefont {A.}~\bibnamefont {Le\ifmmode~\check{s}\else
  \v{s}\fi{}und\'ak}}, \bibinfo {author} {\bibfnamefont {O.}~\bibnamefont
  {\ifmmode~\check{C}\else \v{C}\fi{}\'{\i}p}}, \bibinfo {author}
  {\bibfnamefont {L.}~\bibnamefont {Slodi\ifmmode~\check{c}\else
  \v{c}\fi{}ka}}, \ and\ \bibinfo {author} {\bibfnamefont {R.}~\bibnamefont
  {Filip}},\ }\bibfield  {title} {\enquote {\bibinfo {title} {Quantum
  non-gaussianity of multiphonon states of a single atom},}\ }\href {\doibase
  10.1103/PhysRevLett.129.013602} {\bibfield  {journal} {\bibinfo  {journal}
  {Phys. Rev. Lett.}\ }\textbf {\bibinfo {volume} {129}},\ \bibinfo {pages}
  {013602} (\bibinfo {year} {2022})}\BibitemShut {NoStop}%
\bibitem [{\citenamefont {Hofheinz}\ \emph {et~al.}(2008)\citenamefont
  {Hofheinz}, \citenamefont {Weig}, \citenamefont {Ansmann}, \citenamefont
  {Bialczak}, \citenamefont {Lucero}, \citenamefont {Neeley}, \citenamefont
  {O'Connell}, \citenamefont {Wang}, \citenamefont {Martinis},\ and\
  \citenamefont {Cleland}}]{hofheinz2008}%
  \BibitemOpen
  \bibfield  {author} {\bibinfo {author} {\bibfnamefont {M.}~\bibnamefont
  {Hofheinz}}, \bibinfo {author} {\bibfnamefont {E.~M.}\ \bibnamefont {Weig}},
  \bibinfo {author} {\bibfnamefont {M.}~\bibnamefont {Ansmann}}, \bibinfo
  {author} {\bibfnamefont {R.~C.}\ \bibnamefont {Bialczak}}, \bibinfo {author}
  {\bibfnamefont {E.}~\bibnamefont {Lucero}}, \bibinfo {author} {\bibfnamefont
  {M.}~\bibnamefont {Neeley}}, \bibinfo {author} {\bibfnamefont {A.~D.}\
  \bibnamefont {O'Connell}}, \bibinfo {author} {\bibfnamefont {H.}~\bibnamefont
  {Wang}}, \bibinfo {author} {\bibfnamefont {J.~M.}\ \bibnamefont {Martinis}},
  \ and\ \bibinfo {author} {\bibfnamefont {A.~N.}\ \bibnamefont {Cleland}},\
  }\bibfield  {title} {\enquote {\bibinfo {title} {Generation of {{Fock}}
  states in a superconducting quantum circuit},}\ }\href {\doibase
  10.1038/nature07136} {\bibfield  {journal} {\bibinfo  {journal} {Nature}\
  }\textbf {\bibinfo {volume} {454}},\ \bibinfo {pages} {310} (\bibinfo {year}
  {2008})}\BibitemShut {NoStop}%
\bibitem [{\citenamefont {Eickbusch}\ \emph {et~al.}(2022)\citenamefont
  {Eickbusch}, \citenamefont {Sivak}, \citenamefont {Ding}, \citenamefont
  {Elder}, \citenamefont {Jha}, \citenamefont {Venkatraman}, \citenamefont
  {Royer}, \citenamefont {Girvin}, \citenamefont {Schoelkopf},\ and\
  \citenamefont {Devoret}}]{eickbusch2022}%
  \BibitemOpen
  \bibfield  {author} {\bibinfo {author} {\bibfnamefont {A.}~\bibnamefont
  {Eickbusch}}, \bibinfo {author} {\bibfnamefont {V.}~\bibnamefont {Sivak}},
  \bibinfo {author} {\bibfnamefont {A.~Z.}\ \bibnamefont {Ding}}, \bibinfo
  {author} {\bibfnamefont {S.~S.}\ \bibnamefont {Elder}}, \bibinfo {author}
  {\bibfnamefont {S.~R.}\ \bibnamefont {Jha}}, \bibinfo {author} {\bibfnamefont
  {J.}~\bibnamefont {Venkatraman}}, \bibinfo {author} {\bibfnamefont
  {B.}~\bibnamefont {Royer}}, \bibinfo {author} {\bibfnamefont {S.~M.}\
  \bibnamefont {Girvin}}, \bibinfo {author} {\bibfnamefont {R.~J.}\
  \bibnamefont {Schoelkopf}}, \ and\ \bibinfo {author} {\bibfnamefont {M.~H.}\
  \bibnamefont {Devoret}},\ }\bibfield  {title} {\enquote {\bibinfo {title}
  {Fast universal control of an oscillator with weak dispersive coupling to a
  qubit},}\ }\href {\doibase 10.1038/s41567-022-01776-9} {\bibfield  {journal}
  {\bibinfo  {journal} {Nat. Phys.}\ }\textbf {\bibinfo {volume} {18}},\
  \bibinfo {pages} {1464} (\bibinfo {year} {2022})}\BibitemShut {NoStop}%
\bibitem [{\citenamefont {Cai}\ \emph {et~al.}(2021)\citenamefont {Cai},
  \citenamefont {Ma}, \citenamefont {Wang}, \citenamefont {Zou},\ and\
  \citenamefont {Sun}}]{cai2021}%
  \BibitemOpen
  \bibfield  {author} {\bibinfo {author} {\bibfnamefont {W.}~\bibnamefont
  {Cai}}, \bibinfo {author} {\bibfnamefont {Y.}~\bibnamefont {Ma}}, \bibinfo
  {author} {\bibfnamefont {W.}~\bibnamefont {Wang}}, \bibinfo {author}
  {\bibfnamefont {C.-L.}\ \bibnamefont {Zou}}, \ and\ \bibinfo {author}
  {\bibfnamefont {L.}~\bibnamefont {Sun}},\ }\bibfield  {title} {\enquote
  {\bibinfo {title} {{Bosonic Quantum Error Correction Codes in Superconducting
  Quantum Circuits}},}\ }\href {\doibase 10.1016/j.fmre.2020.12.006} {\bibfield
   {journal} {\bibinfo  {journal} {Fundam. Res.}\ }\textbf {\bibinfo {volume}
  {1}},\ \bibinfo {pages} {50} (\bibinfo {year} {2021})}\BibitemShut {NoStop}%
\bibitem [{\citenamefont {Joshi}\ \emph {et~al.}(2021)\citenamefont {Joshi},
  \citenamefont {Noh},\ and\ \citenamefont {Gao}}]{joshi2021}%
  \BibitemOpen
  \bibfield  {author} {\bibinfo {author} {\bibfnamefont {A.}~\bibnamefont
  {Joshi}}, \bibinfo {author} {\bibfnamefont {K.}~\bibnamefont {Noh}}, \ and\
  \bibinfo {author} {\bibfnamefont {Y.~Y.}\ \bibnamefont {Gao}},\ }\bibfield
  {title} {\enquote {\bibinfo {title} {Quantum information processing with
  bosonic qubits in circuit {{QED}}},}\ }\href {\doibase
  10.1088/2058-9565/abe989} {\bibfield  {journal} {\bibinfo  {journal} {Quantum
  Sci. and Technol.}\ }\textbf {\bibinfo {volume} {6}},\ \bibinfo {pages}
  {033001} (\bibinfo {year} {2021})}\BibitemShut {NoStop}%
\bibitem [{\citenamefont {de~Oliveira}\ \emph {et~al.}(1990)\citenamefont
  {de~Oliveira}, \citenamefont {Kim}, \citenamefont {Knight},\ and\
  \citenamefont {Buek}}]{oliveira1990}%
  \BibitemOpen
  \bibfield  {author} {\bibinfo {author} {\bibfnamefont {F.~A.~M.}\
  \bibnamefont {de~Oliveira}}, \bibinfo {author} {\bibfnamefont {M.~S.}\
  \bibnamefont {Kim}}, \bibinfo {author} {\bibfnamefont {P.~L.}\ \bibnamefont
  {Knight}}, \ and\ \bibinfo {author} {\bibfnamefont {V.}~\bibnamefont
  {Buek}},\ }\bibfield  {title} {\enquote {\bibinfo {title} {Properties of
  displaced number states},}\ }\href {\doibase 10.1103/PhysRevA.41.2645}
  {\bibfield  {journal} {\bibinfo  {journal} {Phys. Rev. A}\ }\textbf {\bibinfo
  {volume} {41}},\ \bibinfo {pages} {2645} (\bibinfo {year}
  {1990})}\BibitemShut {NoStop}%
\bibitem [{\citenamefont {Zurek}(2001)}]{zurek2001}%
  \BibitemOpen
  \bibfield  {author} {\bibinfo {author} {\bibfnamefont {W.~H.}\ \bibnamefont
  {Zurek}},\ }\bibfield  {title} {\enquote {\bibinfo {title} {Sub-{{Planck}}
  structure in phase space and its relevance for quantum decoherence},}\ }\href
  {\doibase 10.1038/35089017} {\bibfield  {journal} {\bibinfo  {journal}
  {Nature}\ }\textbf {\bibinfo {volume} {412}},\ \bibinfo {pages} {712}
  (\bibinfo {year} {2001})}\BibitemShut {NoStop}%
\bibitem [{\citenamefont {Blais}\ \emph {et~al.}(2021)\citenamefont {Blais},
  \citenamefont {Grimsmo}, \citenamefont {Girvin},\ and\ \citenamefont
  {Wallraff}}]{blais2021}%
  \BibitemOpen
  \bibfield  {author} {\bibinfo {author} {\bibfnamefont {A.}~\bibnamefont
  {Blais}}, \bibinfo {author} {\bibfnamefont {A.~L.}\ \bibnamefont {Grimsmo}},
  \bibinfo {author} {\bibfnamefont {S.~M.}\ \bibnamefont {Girvin}}, \ and\
  \bibinfo {author} {\bibfnamefont {A.}~\bibnamefont {Wallraff}},\ }\bibfield
  {title} {\enquote {\bibinfo {title} {Circuit quantum electrodynamics},}\
  }\href {\doibase 10.1103/RevModPhys.93.025005} {\bibfield  {journal}
  {\bibinfo  {journal} {Rev. Mod. Phys.}\ }\textbf {\bibinfo {volume} {93}},\
  \bibinfo {pages} {025005} (\bibinfo {year} {2021})}\BibitemShut {NoStop}%
\bibitem [{\citenamefont {Guerlin}\ \emph {et~al.}(2007)\citenamefont
  {Guerlin}, \citenamefont {Bernu}, \citenamefont {Del{\'e}glise},
  \citenamefont {Sayrin}, \citenamefont {Gleyzes}, \citenamefont {Kuhr},
  \citenamefont {Brune}, \citenamefont {Raimond},\ and\ \citenamefont
  {Haroche}}]{guerlin2007}%
  \BibitemOpen
  \bibfield  {author} {\bibinfo {author} {\bibfnamefont {C.}~\bibnamefont
  {Guerlin}}, \bibinfo {author} {\bibfnamefont {J.}~\bibnamefont {Bernu}},
  \bibinfo {author} {\bibfnamefont {S.}~\bibnamefont {Del{\'e}glise}}, \bibinfo
  {author} {\bibfnamefont {C.}~\bibnamefont {Sayrin}}, \bibinfo {author}
  {\bibfnamefont {S.}~\bibnamefont {Gleyzes}}, \bibinfo {author} {\bibfnamefont
  {S.}~\bibnamefont {Kuhr}}, \bibinfo {author} {\bibfnamefont {M.}~\bibnamefont
  {Brune}}, \bibinfo {author} {\bibfnamefont {J.-M.}\ \bibnamefont {Raimond}},
  \ and\ \bibinfo {author} {\bibfnamefont {S.}~\bibnamefont {Haroche}},\
  }\bibfield  {title} {\enquote {\bibinfo {title} {Progressive field-state
  collapse and quantum non-demolition photon counting},}\ }\href {\doibase
  10.1038/nature06057} {\bibfield  {journal} {\bibinfo  {journal} {Nature}\
  }\textbf {\bibinfo {volume} {448}},\ \bibinfo {pages} {889} (\bibinfo {year}
  {2007})}\BibitemShut {NoStop}%
\bibitem [{\citenamefont {Sun}\ \emph {et~al.}(2014)\citenamefont {Sun},
  \citenamefont {Petrenko}, \citenamefont {Leghtas}, \citenamefont {Vlastakis},
  \citenamefont {Kirchmair}, \citenamefont {Sliwa}, \citenamefont {Narla},
  \citenamefont {Hatridge}, \citenamefont {Shankar}, \citenamefont {Blumoff},
  \citenamefont {Frunzio}, \citenamefont {Mirrahimi}, \citenamefont {Devoret},\
  and\ \citenamefont {Schoelkopf}}]{sun2014}%
  \BibitemOpen
  \bibfield  {author} {\bibinfo {author} {\bibfnamefont {L.}~\bibnamefont
  {Sun}}, \bibinfo {author} {\bibfnamefont {A.}~\bibnamefont {Petrenko}},
  \bibinfo {author} {\bibfnamefont {Z.}~\bibnamefont {Leghtas}}, \bibinfo
  {author} {\bibfnamefont {B.}~\bibnamefont {Vlastakis}}, \bibinfo {author}
  {\bibfnamefont {G.}~\bibnamefont {Kirchmair}}, \bibinfo {author}
  {\bibfnamefont {K.~M.}\ \bibnamefont {Sliwa}}, \bibinfo {author}
  {\bibfnamefont {A.}~\bibnamefont {Narla}}, \bibinfo {author} {\bibfnamefont
  {M.}~\bibnamefont {Hatridge}}, \bibinfo {author} {\bibfnamefont
  {S.}~\bibnamefont {Shankar}}, \bibinfo {author} {\bibfnamefont
  {J.}~\bibnamefont {Blumoff}}, \bibinfo {author} {\bibfnamefont
  {L.}~\bibnamefont {Frunzio}}, \bibinfo {author} {\bibfnamefont
  {M.}~\bibnamefont {Mirrahimi}}, \bibinfo {author} {\bibfnamefont {M.~H.}\
  \bibnamefont {Devoret}}, \ and\ \bibinfo {author} {\bibfnamefont {R.~J.}\
  \bibnamefont {Schoelkopf}},\ }\bibfield  {title} {\enquote {\bibinfo {title}
  {Tracking photon jumps with repeated quantum non-demolition parity
  measurements},}\ }\href {\doibase 10.1038/nature13436} {\bibfield  {journal}
  {\bibinfo  {journal} {Nature}\ }\textbf {\bibinfo {volume} {511}},\ \bibinfo
  {pages} {444} (\bibinfo {year} {2014})}\BibitemShut {NoStop}%
\bibitem [{\citenamefont {Krastanov}\ \emph {et~al.}(2015)\citenamefont
  {Krastanov}, \citenamefont {Albert}, \citenamefont {Shen}, \citenamefont
  {Zou}, \citenamefont {Heeres}, \citenamefont {Vlastakis}, \citenamefont
  {Schoelkopf},\ and\ \citenamefont {Jiang}}]{krastanov2015}%
  \BibitemOpen
  \bibfield  {author} {\bibinfo {author} {\bibfnamefont {S.}~\bibnamefont
  {Krastanov}}, \bibinfo {author} {\bibfnamefont {V.~V.}\ \bibnamefont
  {Albert}}, \bibinfo {author} {\bibfnamefont {C.}~\bibnamefont {Shen}},
  \bibinfo {author} {\bibfnamefont {C.-L.}\ \bibnamefont {Zou}}, \bibinfo
  {author} {\bibfnamefont {R.~W.}\ \bibnamefont {Heeres}}, \bibinfo {author}
  {\bibfnamefont {B.}~\bibnamefont {Vlastakis}}, \bibinfo {author}
  {\bibfnamefont {R.~J.}\ \bibnamefont {Schoelkopf}}, \ and\ \bibinfo {author}
  {\bibfnamefont {L.}~\bibnamefont {Jiang}},\ }\bibfield  {title} {\enquote
  {\bibinfo {title} {Universal control of an oscillator with dispersive
  coupling to a qubit},}\ }\href {\doibase 10.1103/PhysRevA.92.040303}
  {\bibfield  {journal} {\bibinfo  {journal} {Phys. Rev. A}\ }\textbf {\bibinfo
  {volume} {92}},\ \bibinfo {pages} {040303} (\bibinfo {year}
  {2015})}\BibitemShut {NoStop}%
\bibitem [{\citenamefont {Wang}\ \emph
  {et~al.}(2022{\natexlab{a}})\citenamefont {Wang}, \citenamefont {Chen},
  \citenamefont {Liu}, \citenamefont {Cai}, \citenamefont {Ma}, \citenamefont
  {Mu}, \citenamefont {Pan}, \citenamefont {Hua}, \citenamefont {Hu},
  \citenamefont {Xu}, \citenamefont {Wang}, \citenamefont {Song}, \citenamefont
  {Zou}, \citenamefont {Zou},\ and\ \citenamefont {Sun}}]{wang2022}%
  \BibitemOpen
  \bibfield  {author} {\bibinfo {author} {\bibfnamefont {W.}~\bibnamefont
  {Wang}}, \bibinfo {author} {\bibfnamefont {Z.-J.}\ \bibnamefont {Chen}},
  \bibinfo {author} {\bibfnamefont {X.}~\bibnamefont {Liu}}, \bibinfo {author}
  {\bibfnamefont {W.}~\bibnamefont {Cai}}, \bibinfo {author} {\bibfnamefont
  {Y.}~\bibnamefont {Ma}}, \bibinfo {author} {\bibfnamefont {X.}~\bibnamefont
  {Mu}}, \bibinfo {author} {\bibfnamefont {X.}~\bibnamefont {Pan}}, \bibinfo
  {author} {\bibfnamefont {Z.}~\bibnamefont {Hua}}, \bibinfo {author}
  {\bibfnamefont {L.}~\bibnamefont {Hu}}, \bibinfo {author} {\bibfnamefont
  {Y.}~\bibnamefont {Xu}}, \bibinfo {author} {\bibfnamefont {H.}~\bibnamefont
  {Wang}}, \bibinfo {author} {\bibfnamefont {Y.~P.}\ \bibnamefont {Song}},
  \bibinfo {author} {\bibfnamefont {X.-B.}\ \bibnamefont {Zou}}, \bibinfo
  {author} {\bibfnamefont {C.-L.}\ \bibnamefont {Zou}}, \ and\ \bibinfo
  {author} {\bibfnamefont {L.}~\bibnamefont {Sun}},\ }\bibfield  {title}
  {\enquote {\bibinfo {title} {Quantum-enhanced radiometry via approximate
  quantum error correction},}\ }\href {\doibase 10.1038/s41467-022-30410-8}
  {\bibfield  {journal} {\bibinfo  {journal} {Nat. Commun.}\ }\textbf {\bibinfo
  {volume} {13}},\ \bibinfo {pages} {3214} (\bibinfo {year}
  {2022}{\natexlab{a}})}\BibitemShut {NoStop}%
\bibitem [{\citenamefont {Fujiwara}(2001)}]{fujiwara2001}%
  \BibitemOpen
  \bibfield  {author} {\bibinfo {author} {\bibfnamefont {A.}~\bibnamefont
  {Fujiwara}},\ }\bibfield  {title} {\enquote {\bibinfo {title} {Quantum
  channel identification problem},}\ }\href {\doibase
  10.1103/PhysRevA.63.042304} {\bibfield  {journal} {\bibinfo  {journal} {Phys.
  Rev. A}\ }\textbf {\bibinfo {volume} {63}},\ \bibinfo {pages} {042304}
  (\bibinfo {year} {2001})}\BibitemShut {NoStop}%
\bibitem [{\citenamefont {Milul}\ \emph {et~al.}(2023)\citenamefont {Milul},
  \citenamefont {Guttel}, \citenamefont {Goldblatt}, \citenamefont {Hazanov},
  \citenamefont {Joshi}, \citenamefont {Chausovsky}, \citenamefont {Kahn},
  \citenamefont {\ifmmode~\mbox{\c{C}}\else \c{C}\fi{}ifty\"urek},
  \citenamefont {Lafont},\ and\ \citenamefont {Rosenblum}}]{Milul2023}%
  \BibitemOpen
  \bibfield  {author} {\bibinfo {author} {\bibfnamefont {O.}~\bibnamefont
  {Milul}}, \bibinfo {author} {\bibfnamefont {B.}~\bibnamefont {Guttel}},
  \bibinfo {author} {\bibfnamefont {U.}~\bibnamefont {Goldblatt}}, \bibinfo
  {author} {\bibfnamefont {S.}~\bibnamefont {Hazanov}}, \bibinfo {author}
  {\bibfnamefont {L.~M.}\ \bibnamefont {Joshi}}, \bibinfo {author}
  {\bibfnamefont {D.}~\bibnamefont {Chausovsky}}, \bibinfo {author}
  {\bibfnamefont {N.}~\bibnamefont {Kahn}}, \bibinfo {author} {\bibfnamefont
  {E.}~\bibnamefont {\ifmmode~\mbox{\c{C}}\else \c{C}\fi{}ifty\"urek}},
  \bibinfo {author} {\bibfnamefont {F.}~\bibnamefont {Lafont}}, \ and\ \bibinfo
  {author} {\bibfnamefont {S.}~\bibnamefont {Rosenblum}},\ }\bibfield  {title}
  {\enquote {\bibinfo {title} {{Superconducting Cavity Qubit with Tens of
  Milliseconds Single-Photon Coherence Time}},}\ }\href {\doibase
  10.1103/PRXQuantum.4.030336} {\bibfield  {journal} {\bibinfo  {journal} {PRX
  Quantum}\ }\textbf {\bibinfo {volume} {4}},\ \bibinfo {pages} {030336}
  (\bibinfo {year} {2023})}\BibitemShut {NoStop}%
\bibitem [{\citenamefont {Agrawal}\ \emph {et~al.}(2024)\citenamefont
  {Agrawal}, \citenamefont {Dixit}, \citenamefont {Roy}, \citenamefont
  {Chakram}, \citenamefont {He}, \citenamefont {Naik}, \citenamefont
  {Schuster},\ and\ \citenamefont {Chou}}]{Agrawal2023}%
  \BibitemOpen
  \bibfield  {author} {\bibinfo {author} {\bibfnamefont {A.}~\bibnamefont
  {Agrawal}}, \bibinfo {author} {\bibfnamefont {A.~V.}\ \bibnamefont {Dixit}},
  \bibinfo {author} {\bibfnamefont {T.}~\bibnamefont {Roy}}, \bibinfo {author}
  {\bibfnamefont {S.}~\bibnamefont {Chakram}}, \bibinfo {author} {\bibfnamefont
  {K.}~\bibnamefont {He}}, \bibinfo {author} {\bibfnamefont {R.~K.}\
  \bibnamefont {Naik}}, \bibinfo {author} {\bibfnamefont {D.~I.}\ \bibnamefont
  {Schuster}}, \ and\ \bibinfo {author} {\bibfnamefont {A.}~\bibnamefont
  {Chou}},\ }\bibfield  {title} {\enquote {\bibinfo {title} {{Stimulated
  Emission of Signal Photons from Dark Matter Waves}},}\ }\href {\doibase
  10.1103/PhysRevLett.132.140801} {\bibfield  {journal} {\bibinfo  {journal}
  {Phys. Rev. Lett.}\ }\textbf {\bibinfo {volume} {132}},\ \bibinfo {pages}
  {140801} (\bibinfo {year} {2024})}\BibitemShut {NoStop}%
\bibitem [{\citenamefont {Leibfried}\ \emph {et~al.}(2003)\citenamefont
  {Leibfried}, \citenamefont {Blatt}, \citenamefont {Monroe},\ and\
  \citenamefont {Wineland}}]{leibfried2003}%
  \BibitemOpen
  \bibfield  {author} {\bibinfo {author} {\bibfnamefont {D.}~\bibnamefont
  {Leibfried}}, \bibinfo {author} {\bibfnamefont {R.}~\bibnamefont {Blatt}},
  \bibinfo {author} {\bibfnamefont {C.}~\bibnamefont {Monroe}}, \ and\ \bibinfo
  {author} {\bibfnamefont {D.}~\bibnamefont {Wineland}},\ }\bibfield  {title}
  {\enquote {\bibinfo {title} {{Quantum Dynamics of Single Trapped Ions}},}\
  }\href {\doibase 10.1103/RevModPhys.75.281} {\bibfield  {journal} {\bibinfo
  {journal} {Rev. Mod. Phys.}\ }\textbf {\bibinfo {volume} {75}},\ \bibinfo
  {pages} {281} (\bibinfo {year} {2003})}\BibitemShut {NoStop}%
\bibitem [{\citenamefont {Fan}\ \emph {et~al.}(2018)\citenamefont {Fan},
  \citenamefont {Zou}, \citenamefont {Cheng}, \citenamefont {Guo},
  \citenamefont {Han}, \citenamefont {Gong}, \citenamefont {Wang},\ and\
  \citenamefont {Tang}}]{fan2018superconducting}%
  \BibitemOpen
  \bibfield  {author} {\bibinfo {author} {\bibfnamefont {L.}~\bibnamefont
  {Fan}}, \bibinfo {author} {\bibfnamefont {C.-L.}\ \bibnamefont {Zou}},
  \bibinfo {author} {\bibfnamefont {R.}~\bibnamefont {Cheng}}, \bibinfo
  {author} {\bibfnamefont {X.}~\bibnamefont {Guo}}, \bibinfo {author}
  {\bibfnamefont {X.}~\bibnamefont {Han}}, \bibinfo {author} {\bibfnamefont
  {Z.}~\bibnamefont {Gong}}, \bibinfo {author} {\bibfnamefont {S.}~\bibnamefont
  {Wang}}, \ and\ \bibinfo {author} {\bibfnamefont {H.~X.}\ \bibnamefont
  {Tang}},\ }\bibfield  {title} {\enquote {\bibinfo {title} {Superconducting
  cavity electro-optics: A platform for coherent photon conversion between
  superconducting and photonic circuits},}\ }\href {\doibase
  10.1126/sciadv.aar4994} {\bibfield  {journal} {\bibinfo  {journal} {Sci.
  Adv.}\ }\textbf {\bibinfo {volume} {4}},\ \bibinfo {pages} {eaar4994}
  (\bibinfo {year} {2018})}\BibitemShut {NoStop}%
\bibitem [{\citenamefont {Higginbotham}\ \emph {et~al.}(2018)\citenamefont
  {Higginbotham}, \citenamefont {Burns}, \citenamefont {Urmey}, \citenamefont
  {Peterson}, \citenamefont {Kampel}, \citenamefont {Brubaker}, \citenamefont
  {Smith}, \citenamefont {Lehnert},\ and\ \citenamefont
  {Regal}}]{higginbotham2018harnessing}%
  \BibitemOpen
  \bibfield  {author} {\bibinfo {author} {\bibfnamefont {A.~P.}\ \bibnamefont
  {Higginbotham}}, \bibinfo {author} {\bibfnamefont {P.}~\bibnamefont {Burns}},
  \bibinfo {author} {\bibfnamefont {M.}~\bibnamefont {Urmey}}, \bibinfo
  {author} {\bibfnamefont {R.}~\bibnamefont {Peterson}}, \bibinfo {author}
  {\bibfnamefont {N.}~\bibnamefont {Kampel}}, \bibinfo {author} {\bibfnamefont
  {B.}~\bibnamefont {Brubaker}}, \bibinfo {author} {\bibfnamefont
  {G.}~\bibnamefont {Smith}}, \bibinfo {author} {\bibfnamefont
  {K.}~\bibnamefont {Lehnert}}, \ and\ \bibinfo {author} {\bibfnamefont
  {C.}~\bibnamefont {Regal}},\ }\bibfield  {title} {\enquote {\bibinfo {title}
  {Harnessing electro-optic correlations in an efficient mechanical
  converter},}\ }\href {\doibase 10.1038/s41567-018-0210-0} {\bibfield
  {journal} {\bibinfo  {journal} {Nat. Phys.}\ }\textbf {\bibinfo {volume}
  {14}},\ \bibinfo {pages} {1038} (\bibinfo {year} {2018})}\BibitemShut
  {NoStop}%
\bibitem [{\citenamefont {{von L{\"u}pke}}\ \emph {et~al.}(2022)\citenamefont
  {{von L{\"u}pke}}, \citenamefont {Yang}, \citenamefont {Bild}, \citenamefont
  {Michaud}, \citenamefont {Fadel},\ and\ \citenamefont {Chu}}]{vonlupke2022}%
  \BibitemOpen
  \bibfield  {author} {\bibinfo {author} {\bibfnamefont {U.}~\bibnamefont {{von
  L{\"u}pke}}}, \bibinfo {author} {\bibfnamefont {Y.}~\bibnamefont {Yang}},
  \bibinfo {author} {\bibfnamefont {M.}~\bibnamefont {Bild}}, \bibinfo {author}
  {\bibfnamefont {L.}~\bibnamefont {Michaud}}, \bibinfo {author} {\bibfnamefont
  {M.}~\bibnamefont {Fadel}}, \ and\ \bibinfo {author} {\bibfnamefont
  {Y.}~\bibnamefont {Chu}},\ }\bibfield  {title} {\enquote {\bibinfo {title}
  {Parity measurement in the strong dispersive regime of circuit quantum
  acoustodynamics},}\ }\href {\doibase 10.1038/s41567-022-01591-2} {\bibfield
  {journal} {\bibinfo  {journal} {Nat. Phys.}\ }\textbf {\bibinfo {volume}
  {18}},\ \bibinfo {pages} {794} (\bibinfo {year} {2022})}\BibitemShut
  {NoStop}%
\end{thebibliography}

\begin{thebibliography}{59}%
\makeatletter
\providecommand \@ifxundefined [1]{%
 \@ifx{#1\undefined}
}%
\providecommand \@ifnum [1]{%
 \ifnum #1\expandafter \@firstoftwo
 \else \expandafter \@secondoftwo
 \fi
}%
\providecommand \@ifx [1]{%
 \ifx #1\expandafter \@firstoftwo
 \else \expandafter \@secondoftwo
 \fi
}%
\providecommand \natexlab [1]{#1}%
\providecommand \enquote  [1]{``#1''}%
\providecommand \bibnamefont  [1]{#1}%
\providecommand \bibfnamefont [1]{#1}%
\providecommand \citenamefont [1]{#1}%
\providecommand \href@noop [0]{\@secondoftwo}%
\providecommand \href [0]{\begingroup \@sanitize@url \@href}%
\providecommand \@href[1]{\@@startlink{#1}\@@href}%
\providecommand \@@href[1]{\endgroup#1\@@endlink}%
\providecommand \@sanitize@url [0]{\catcode `\\12\catcode `\$12\catcode
  `\&12\catcode `\#12\catcode `\^12\catcode `\_12\catcode `\%12\relax}%
\providecommand \@@startlink[1]{}%
\providecommand \@@endlink[0]{}%
\providecommand \url  [0]{\begingroup\@sanitize@url \@url }%
\providecommand \@url [1]{\endgroup\@href {#1}{\urlprefix }}%
\providecommand \urlprefix  [0]{URL }%
\providecommand \Eprint [0]{\href }%
\providecommand \doibase [0]{http://dx.doi.org/}%
\providecommand \selectlanguage [0]{\@gobble}%
\providecommand \bibinfo  [0]{\@secondoftwo}%
\providecommand \bibfield  [0]{\@secondoftwo}%
\providecommand \translation [1]{[#1]}%
\providecommand \BibitemOpen [0]{}%
\providecommand \bibitemStop [0]{}%
\providecommand \bibitemNoStop [0]{.\EOS\space}%
\providecommand \EOS [0]{\spacefactor3000\relax}%
\providecommand \BibitemShut  [1]{\csname bibitem#1\endcsname}%
\let\auto@bib@innerbib\@empty
%</preamble>
\bibitem [{\citenamefont {Ni}\ \emph {et~al.}(2023)\citenamefont {Ni},
  \citenamefont {Li}, \citenamefont {Deng}, \citenamefont {Cai}, \citenamefont
  {Zhang}, \citenamefont {Wang}, \citenamefont {Yang}, \citenamefont {Yu},
  \citenamefont {Yan}, \citenamefont {Liu}, \citenamefont {Zou}, \citenamefont
  {Sun}, \citenamefont {Zheng}, \citenamefont {Xu},\ and\ \citenamefont
  {Yu}}]{ni2023}%
  \BibitemOpen
  \bibfield  {author} {\bibinfo {author} {\bibfnamefont {Z.}~\bibnamefont
  {Ni}}, \bibinfo {author} {\bibfnamefont {S.}~\bibnamefont {Li}}, \bibinfo
  {author} {\bibfnamefont {X.}~\bibnamefont {Deng}}, \bibinfo {author}
  {\bibfnamefont {Y.}~\bibnamefont {Cai}}, \bibinfo {author} {\bibfnamefont
  {L.}~\bibnamefont {Zhang}}, \bibinfo {author} {\bibfnamefont
  {W.}~\bibnamefont {Wang}}, \bibinfo {author} {\bibfnamefont {Z.-B.}\
  \bibnamefont {Yang}}, \bibinfo {author} {\bibfnamefont {H.}~\bibnamefont
  {Yu}}, \bibinfo {author} {\bibfnamefont {F.}~\bibnamefont {Yan}}, \bibinfo
  {author} {\bibfnamefont {S.}~\bibnamefont {Liu}}, \bibinfo {author}
  {\bibfnamefont {C.-L.}\ \bibnamefont {Zou}}, \bibinfo {author} {\bibfnamefont
  {L.}~\bibnamefont {Sun}}, \bibinfo {author} {\bibfnamefont {S.-B.}\
  \bibnamefont {Zheng}}, \bibinfo {author} {\bibfnamefont {Y.}~\bibnamefont
  {Xu}}, \ and\ \bibinfo {author} {\bibfnamefont {D.}~\bibnamefont {Yu}},\
  }\bibfield  {title} {\enquote {\bibinfo {title} {Beating the break-even point
  with a discrete-variable-encoded logical qubit},}\ }\href {\doibase
  10.1038/s41586-023-05784-4} {\bibfield  {journal} {\bibinfo  {journal}
  {Nature}\ }\textbf {\bibinfo {volume} {616}},\ \bibinfo {pages} {56}
  (\bibinfo {year} {2023})}\BibitemShut {NoStop}%
\bibitem [{\citenamefont {Koch}\ \emph {et~al.}(2007)\citenamefont {Koch},
  \citenamefont {Yu}, \citenamefont {Gambetta}, \citenamefont {Houck},
  \citenamefont {Schuster}, \citenamefont {Majer}, \citenamefont {Blais},
  \citenamefont {Devoret}, \citenamefont {Girvin},\ and\ \citenamefont
  {Schoelkopf}}]{koch2007}%
  \BibitemOpen
  \bibfield  {author} {\bibinfo {author} {\bibfnamefont {J.}~\bibnamefont
  {Koch}}, \bibinfo {author} {\bibfnamefont {T.~M.}\ \bibnamefont {Yu}},
  \bibinfo {author} {\bibfnamefont {J.}~\bibnamefont {Gambetta}}, \bibinfo
  {author} {\bibfnamefont {A.~A.}\ \bibnamefont {Houck}}, \bibinfo {author}
  {\bibfnamefont {D.~I.}\ \bibnamefont {Schuster}}, \bibinfo {author}
  {\bibfnamefont {J.}~\bibnamefont {Majer}}, \bibinfo {author} {\bibfnamefont
  {A.}~\bibnamefont {Blais}}, \bibinfo {author} {\bibfnamefont {M.~H.}\
  \bibnamefont {Devoret}}, \bibinfo {author} {\bibfnamefont {S.~M.}\
  \bibnamefont {Girvin}}, \ and\ \bibinfo {author} {\bibfnamefont {R.~J.}\
  \bibnamefont {Schoelkopf}},\ }\bibfield  {title} {\enquote {\bibinfo {title}
  {Charge-insensitive qubit design derived from the {{Cooper}} pair box},}\
  }\href {\doibase 10.1103/PhysRevA.76.042319} {\bibfield  {journal} {\bibinfo
  {journal} {Phys. Rev. A}\ }\textbf {\bibinfo {volume} {76}},\ \bibinfo
  {pages} {042319} (\bibinfo {year} {2007})}\BibitemShut {NoStop}%
\bibitem [{\citenamefont {Axline}\ \emph {et~al.}(2016)\citenamefont {Axline},
  \citenamefont {Reagor}, \citenamefont {Heeres}, \citenamefont {Reinhold},
  \citenamefont {Wang}, \citenamefont {Shain}, \citenamefont {Pfaff},
  \citenamefont {Chu}, \citenamefont {Frunzio},\ and\ \citenamefont
  {Schoelkopf}}]{axline2016}%
  \BibitemOpen
  \bibfield  {author} {\bibinfo {author} {\bibfnamefont {C.}~\bibnamefont
  {Axline}}, \bibinfo {author} {\bibfnamefont {M.}~\bibnamefont {Reagor}},
  \bibinfo {author} {\bibfnamefont {R.}~\bibnamefont {Heeres}}, \bibinfo
  {author} {\bibfnamefont {P.}~\bibnamefont {Reinhold}}, \bibinfo {author}
  {\bibfnamefont {C.}~\bibnamefont {Wang}}, \bibinfo {author} {\bibfnamefont
  {K.}~\bibnamefont {Shain}}, \bibinfo {author} {\bibfnamefont
  {W.}~\bibnamefont {Pfaff}}, \bibinfo {author} {\bibfnamefont
  {Y.}~\bibnamefont {Chu}}, \bibinfo {author} {\bibfnamefont {L.}~\bibnamefont
  {Frunzio}}, \ and\ \bibinfo {author} {\bibfnamefont {R.~J.}\ \bibnamefont
  {Schoelkopf}},\ }\bibfield  {title} {\enquote {\bibinfo {title} {An
  architecture for integrating planar and {{3D cQED}} devices},}\ }\href
  {\doibase 10.1063/1.4959241} {\bibfield  {journal} {\bibinfo  {journal}
  {Appl. Phys. Lett.}\ }\textbf {\bibinfo {volume} {109}},\ \bibinfo {pages}
  {042601} (\bibinfo {year} {2016})}\BibitemShut {NoStop}%
\bibitem [{\citenamefont {Reagor}\ \emph {et~al.}(2016)\citenamefont {Reagor},
  \citenamefont {Pfaff}, \citenamefont {Axline}, \citenamefont {Heeres},
  \citenamefont {Ofek}, \citenamefont {Sliwa}, \citenamefont {Holland},
  \citenamefont {Wang}, \citenamefont {Blumoff}, \citenamefont {Chou},
  \citenamefont {Hatridge}, \citenamefont {Frunzio}, \citenamefont {Devoret},
  \citenamefont {Jiang},\ and\ \citenamefont {Schoelkopf}}]{reagor2016}%
  \BibitemOpen
  \bibfield  {author} {\bibinfo {author} {\bibfnamefont {M.}~\bibnamefont
  {Reagor}}, \bibinfo {author} {\bibfnamefont {W.}~\bibnamefont {Pfaff}},
  \bibinfo {author} {\bibfnamefont {C.}~\bibnamefont {Axline}}, \bibinfo
  {author} {\bibfnamefont {R.~W.}\ \bibnamefont {Heeres}}, \bibinfo {author}
  {\bibfnamefont {N.}~\bibnamefont {Ofek}}, \bibinfo {author} {\bibfnamefont
  {K.}~\bibnamefont {Sliwa}}, \bibinfo {author} {\bibfnamefont
  {E.}~\bibnamefont {Holland}}, \bibinfo {author} {\bibfnamefont
  {C.}~\bibnamefont {Wang}}, \bibinfo {author} {\bibfnamefont {J.}~\bibnamefont
  {Blumoff}}, \bibinfo {author} {\bibfnamefont {K.}~\bibnamefont {Chou}},
  \bibinfo {author} {\bibfnamefont {M.~J.}\ \bibnamefont {Hatridge}}, \bibinfo
  {author} {\bibfnamefont {L.}~\bibnamefont {Frunzio}}, \bibinfo {author}
  {\bibfnamefont {M.~H.}\ \bibnamefont {Devoret}}, \bibinfo {author}
  {\bibfnamefont {L.}~\bibnamefont {Jiang}}, \ and\ \bibinfo {author}
  {\bibfnamefont {R.~J.}\ \bibnamefont {Schoelkopf}},\ }\bibfield  {title}
  {\enquote {\bibinfo {title} {Quantum memory with millisecond coherence in
  circuit {{QED}}},}\ }\href {\doibase 10.1103/PhysRevB.94.014506} {\bibfield
  {journal} {\bibinfo  {journal} {Phys. Rev. B}\ }\textbf {\bibinfo {volume}
  {94}},\ \bibinfo {pages} {014506} (\bibinfo {year} {2016})}\BibitemShut
  {NoStop}%
\bibitem [{\citenamefont {Reagor}\ \emph {et~al.}(2013)\citenamefont {Reagor},
  \citenamefont {Paik}, \citenamefont {Catelani}, \citenamefont {Sun},
  \citenamefont {Axline}, \citenamefont {Holland}, \citenamefont {Pop},
  \citenamefont {Masluk}, \citenamefont {Brecht}, \citenamefont {Frunzio},
  \citenamefont {Devoret}, \citenamefont {Glazman},\ and\ \citenamefont
  {Schoelkopf}}]{reagor2013}%
  \BibitemOpen
  \bibfield  {author} {\bibinfo {author} {\bibfnamefont {M.}~\bibnamefont
  {Reagor}}, \bibinfo {author} {\bibfnamefont {H.}~\bibnamefont {Paik}},
  \bibinfo {author} {\bibfnamefont {G.}~\bibnamefont {Catelani}}, \bibinfo
  {author} {\bibfnamefont {L.}~\bibnamefont {Sun}}, \bibinfo {author}
  {\bibfnamefont {C.}~\bibnamefont {Axline}}, \bibinfo {author} {\bibfnamefont
  {E.}~\bibnamefont {Holland}}, \bibinfo {author} {\bibfnamefont {I.~M.}\
  \bibnamefont {Pop}}, \bibinfo {author} {\bibfnamefont {N.~A.}\ \bibnamefont
  {Masluk}}, \bibinfo {author} {\bibfnamefont {T.}~\bibnamefont {Brecht}},
  \bibinfo {author} {\bibfnamefont {L.}~\bibnamefont {Frunzio}}, \bibinfo
  {author} {\bibfnamefont {M.~H.}\ \bibnamefont {Devoret}}, \bibinfo {author}
  {\bibfnamefont {L.}~\bibnamefont {Glazman}}, \ and\ \bibinfo {author}
  {\bibfnamefont {R.~J.}\ \bibnamefont {Schoelkopf}},\ }\bibfield  {title}
  {\enquote {\bibinfo {title} {Reaching 10 ms single photon lifetimes for
  superconducting aluminum cavities},}\ }\href {\doibase 10.1063/1.4807015}
  {\bibfield  {journal} {\bibinfo  {journal} {Appl. Phys. Lett.}\ }\textbf
  {\bibinfo {volume} {102}},\ \bibinfo {pages} {192604} (\bibinfo {year}
  {2013})}\BibitemShut {NoStop}%
\bibitem [{\citenamefont {Place}\ \emph {et~al.}(2021)\citenamefont {Place},
  \citenamefont {Rodgers}, \citenamefont {Mundada}, \citenamefont {Smitham},
  \citenamefont {Fitzpatrick}, \citenamefont {Leng}, \citenamefont {Premkumar},
  \citenamefont {Bryon}, \citenamefont {Vrajitoarea}, \citenamefont {Sussman},
  \citenamefont {Cheng}, \citenamefont {Madhavan}, \citenamefont {Babla},
  \citenamefont {Le}, \citenamefont {Gang}, \citenamefont {J{\"a}ck},
  \citenamefont {Gyenis}, \citenamefont {Yao}, \citenamefont {Cava},
  \citenamefont {{de Leon}},\ and\ \citenamefont {Houck}}]{place2021}%
  \BibitemOpen
  \bibfield  {author} {\bibinfo {author} {\bibfnamefont {A.~P.~M.}\
  \bibnamefont {Place}}, \bibinfo {author} {\bibfnamefont {L.~V.~H.}\
  \bibnamefont {Rodgers}}, \bibinfo {author} {\bibfnamefont {P.}~\bibnamefont
  {Mundada}}, \bibinfo {author} {\bibfnamefont {B.~M.}\ \bibnamefont
  {Smitham}}, \bibinfo {author} {\bibfnamefont {M.}~\bibnamefont
  {Fitzpatrick}}, \bibinfo {author} {\bibfnamefont {Z.}~\bibnamefont {Leng}},
  \bibinfo {author} {\bibfnamefont {A.}~\bibnamefont {Premkumar}}, \bibinfo
  {author} {\bibfnamefont {J.}~\bibnamefont {Bryon}}, \bibinfo {author}
  {\bibfnamefont {A.}~\bibnamefont {Vrajitoarea}}, \bibinfo {author}
  {\bibfnamefont {S.}~\bibnamefont {Sussman}}, \bibinfo {author} {\bibfnamefont
  {G.}~\bibnamefont {Cheng}}, \bibinfo {author} {\bibfnamefont
  {T.}~\bibnamefont {Madhavan}}, \bibinfo {author} {\bibfnamefont {H.~K.}\
  \bibnamefont {Babla}}, \bibinfo {author} {\bibfnamefont {X.~H.}\ \bibnamefont
  {Le}}, \bibinfo {author} {\bibfnamefont {Y.}~\bibnamefont {Gang}}, \bibinfo
  {author} {\bibfnamefont {B.}~\bibnamefont {J{\"a}ck}}, \bibinfo {author}
  {\bibfnamefont {A.}~\bibnamefont {Gyenis}}, \bibinfo {author} {\bibfnamefont
  {N.}~\bibnamefont {Yao}}, \bibinfo {author} {\bibfnamefont {R.~J.}\
  \bibnamefont {Cava}}, \bibinfo {author} {\bibfnamefont {N.~P.}\ \bibnamefont
  {{de Leon}}}, \ and\ \bibinfo {author} {\bibfnamefont {A.~A.}\ \bibnamefont
  {Houck}},\ }\bibfield  {title} {\enquote {\bibinfo {title} {New material
  platform for superconducting transmon qubits with coherence times exceeding
  0.3 milliseconds},}\ }\href {\doibase 10.1038/s41467-021-22030-5} {\bibfield
  {journal} {\bibinfo  {journal} {Nat. Commun.}\ }\textbf {\bibinfo {volume}
  {12}},\ \bibinfo {pages} {1779} (\bibinfo {year} {2021})}\BibitemShut
  {NoStop}%
\bibitem [{\citenamefont {Wang}\ \emph
  {et~al.}(2022{\natexlab{b}})\citenamefont {Wang}, \citenamefont {Li},
  \citenamefont {Xu}, \citenamefont {Li}, \citenamefont {Wang}, \citenamefont
  {Yang}, \citenamefont {Mi}, \citenamefont {Liang}, \citenamefont {Su},
  \citenamefont {Yang}, \citenamefont {Wang}, \citenamefont {Wang},
  \citenamefont {Li}, \citenamefont {Chen}, \citenamefont {Li}, \citenamefont
  {Linghu}, \citenamefont {Han}, \citenamefont {Zhang}, \citenamefont {Feng},
  \citenamefont {Song}, \citenamefont {Ma}, \citenamefont {Zhang},
  \citenamefont {Wang}, \citenamefont {Zhao}, \citenamefont {Liu},
  \citenamefont {Xue}, \citenamefont {Jin},\ and\ \citenamefont
  {Yu}}]{wang2022b}%
  \BibitemOpen
  \bibfield  {author} {\bibinfo {author} {\bibfnamefont {C.}~\bibnamefont
  {Wang}}, \bibinfo {author} {\bibfnamefont {X.}~\bibnamefont {Li}}, \bibinfo
  {author} {\bibfnamefont {H.}~\bibnamefont {Xu}}, \bibinfo {author}
  {\bibfnamefont {Z.}~\bibnamefont {Li}}, \bibinfo {author} {\bibfnamefont
  {J.}~\bibnamefont {Wang}}, \bibinfo {author} {\bibfnamefont {Z.}~\bibnamefont
  {Yang}}, \bibinfo {author} {\bibfnamefont {Z.}~\bibnamefont {Mi}}, \bibinfo
  {author} {\bibfnamefont {X.}~\bibnamefont {Liang}}, \bibinfo {author}
  {\bibfnamefont {T.}~\bibnamefont {Su}}, \bibinfo {author} {\bibfnamefont
  {C.}~\bibnamefont {Yang}}, \bibinfo {author} {\bibfnamefont {G.}~\bibnamefont
  {Wang}}, \bibinfo {author} {\bibfnamefont {W.}~\bibnamefont {Wang}}, \bibinfo
  {author} {\bibfnamefont {Y.}~\bibnamefont {Li}}, \bibinfo {author}
  {\bibfnamefont {M.}~\bibnamefont {Chen}}, \bibinfo {author} {\bibfnamefont
  {C.}~\bibnamefont {Li}}, \bibinfo {author} {\bibfnamefont {K.}~\bibnamefont
  {Linghu}}, \bibinfo {author} {\bibfnamefont {J.}~\bibnamefont {Han}},
  \bibinfo {author} {\bibfnamefont {Y.}~\bibnamefont {Zhang}}, \bibinfo
  {author} {\bibfnamefont {Y.}~\bibnamefont {Feng}}, \bibinfo {author}
  {\bibfnamefont {Y.}~\bibnamefont {Song}}, \bibinfo {author} {\bibfnamefont
  {T.}~\bibnamefont {Ma}}, \bibinfo {author} {\bibfnamefont {J.}~\bibnamefont
  {Zhang}}, \bibinfo {author} {\bibfnamefont {R.}~\bibnamefont {Wang}},
  \bibinfo {author} {\bibfnamefont {P.}~\bibnamefont {Zhao}}, \bibinfo {author}
  {\bibfnamefont {W.}~\bibnamefont {Liu}}, \bibinfo {author} {\bibfnamefont
  {G.}~\bibnamefont {Xue}}, \bibinfo {author} {\bibfnamefont {Y.}~\bibnamefont
  {Jin}}, \ and\ \bibinfo {author} {\bibfnamefont {H.}~\bibnamefont {Yu}},\
  }\bibfield  {title} {\enquote {\bibinfo {title} {Towards practical quantum
  computers: Transmon qubit with a lifetime approaching 0.5 milliseconds},}\
  }\href {\doibase 10.1038/s41534-021-00510-2} {\bibfield  {journal} {\bibinfo
  {journal} {npj Quantum Inf.}\ }\textbf {\bibinfo {volume} {8}},\ \bibinfo
  {pages} {3} (\bibinfo {year} {2022}{\natexlab{b}})}\BibitemShut {NoStop}%
\bibitem [{\citenamefont {Heeres}\ \emph {et~al.}(2015)\citenamefont {Heeres},
  \citenamefont {Vlastakis}, \citenamefont {Holland}, \citenamefont
  {Krastanov}, \citenamefont {Albert}, \citenamefont {Frunzio}, \citenamefont
  {Jiang},\ and\ \citenamefont {Schoelkopf}}]{heeres2015}%
  \BibitemOpen
  \bibfield  {author} {\bibinfo {author} {\bibfnamefont {R.~W.}\ \bibnamefont
  {Heeres}}, \bibinfo {author} {\bibfnamefont {B.}~\bibnamefont {Vlastakis}},
  \bibinfo {author} {\bibfnamefont {E.}~\bibnamefont {Holland}}, \bibinfo
  {author} {\bibfnamefont {S.}~\bibnamefont {Krastanov}}, \bibinfo {author}
  {\bibfnamefont {V.~V.}\ \bibnamefont {Albert}}, \bibinfo {author}
  {\bibfnamefont {L.}~\bibnamefont {Frunzio}}, \bibinfo {author} {\bibfnamefont
  {L.}~\bibnamefont {Jiang}}, \ and\ \bibinfo {author} {\bibfnamefont {R.~J.}\
  \bibnamefont {Schoelkopf}},\ }\bibfield  {title} {\enquote {\bibinfo {title}
  {Cavity {{State Manipulation Using Photon-Number Selective Phase Gates}}},}\
  }\href {\doibase 10.1103/PhysRevLett.115.137002} {\bibfield  {journal}
  {\bibinfo  {journal} {Phys. Rev. Lett.}\ }\textbf {\bibinfo {volume} {115}},\
  \bibinfo {pages} {137002} (\bibinfo {year} {2015})}\BibitemShut {NoStop}%
\bibitem [{\citenamefont {Schuster}\ \emph {et~al.}(2007)\citenamefont
  {Schuster}, \citenamefont {Houck}, \citenamefont {Schreier}, \citenamefont
  {Wallraff}, \citenamefont {Gambetta}, \citenamefont {Blais}, \citenamefont
  {Frunzio}, \citenamefont {Majer}, \citenamefont {Johnson}, \citenamefont
  {Devoret}, \citenamefont {Girvin},\ and\ \citenamefont
  {Schoelkopf}}]{Schuster2007}%
  \BibitemOpen
  \bibfield  {author} {\bibinfo {author} {\bibfnamefont {D.~I.}\ \bibnamefont
  {Schuster}}, \bibinfo {author} {\bibfnamefont {A.~A.}\ \bibnamefont {Houck}},
  \bibinfo {author} {\bibfnamefont {J.~A.}\ \bibnamefont {Schreier}}, \bibinfo
  {author} {\bibfnamefont {A.}~\bibnamefont {Wallraff}}, \bibinfo {author}
  {\bibfnamefont {J.~M.}\ \bibnamefont {Gambetta}}, \bibinfo {author}
  {\bibfnamefont {A.}~\bibnamefont {Blais}}, \bibinfo {author} {\bibfnamefont
  {L.}~\bibnamefont {Frunzio}}, \bibinfo {author} {\bibfnamefont
  {J.}~\bibnamefont {Majer}}, \bibinfo {author} {\bibfnamefont
  {B.}~\bibnamefont {Johnson}}, \bibinfo {author} {\bibfnamefont {M.~H.}\
  \bibnamefont {Devoret}}, \bibinfo {author} {\bibfnamefont {S.~M.}\
  \bibnamefont {Girvin}}, \ and\ \bibinfo {author} {\bibfnamefont {R.~J.}\
  \bibnamefont {Schoelkopf}},\ }\bibfield  {title} {\enquote {\bibinfo {title}
  {Resolving photon number states in a superconducting circuit},}\ }\href
  {\doibase 10.1038/nature05461} {\bibfield  {journal} {\bibinfo  {journal}
  {Nature}\ }\textbf {\bibinfo {volume} {445}},\ \bibinfo {pages} {515}
  (\bibinfo {year} {2007})}\BibitemShut {NoStop}%
\end{thebibliography}

%Control: production of eprint (-1) disabled
%

\clearpage{}
\setcounter{figure}{0} 
\noindent \textbf{\large{}Methods}{\large\par}

\noindent \textbf{Device and setup}

\noindent The experimental device, similar to that in Ref.~\cite{ni2023}, consists of a superconducting transmon qubit~\cite{koch2007}, a Purcell-filtered stripline readout resonator~\cite{axline2016}, and a 3D coaxial stub cavity~\cite{reagor2016} serving as the probe cavity for quantum metrology experiments. The probe cavity, machined from a single block of high purity (99.9995\%) aluminum, is chemically etched to improve its quality factor~\cite{reagor2013}. The transmon qubit and readout resonator are made by a thin tantalum film on a sapphire chip~\cite{place2021, wang2022b}, which is fit with the small waveguide tunnel of the 3D cavity. The transmon qubit has a resonance frequency $\omega_q/2\pi=4.878$~GHz, an energy relaxation time $T_1=93~\mu$s, and a pure dephasing time $T_\phi = 445~\mu$s. The high-Q probe cavity has a resonance frequency $\omega_c/2\pi=6.597$~GHz, a single-photon lifetime $T_1=1.2$~ms, and a pure dephasing time $T_\phi=4.0$~ms. The dispersive coupling strength between the qubit and the probe cavity is measured to be $\chi_\mathrm{qc}/2\pi=0.626$~MHz. To achieve a fast high-fidelity qubit readout, the readout resonator is designed with a fast decay rate $\kappa_r/2\pi=1.92$~MHz, matching the dispersive shift $\chi_\mathrm{qr}/2\pi=2.0$~MHz. 
The control of the ancillary transmon qubit, probe cavity, and readout resonator is achieved using Zurich Instruments HDAWG and UHFQA, employing single-sideband IQ modulations. Additionally, the UHFQA acquires the transmitted readout signal, performs demodulation and discrimination, and sends the digitized results to the HDAWG in real-time via a DIO link cable for fast feedback control (see Supplementary section I-A).

\vbox{}

\noindent \textbf{Gaussian photon number filter} 

\noindent In principle, the sinusoidal photon number filter (PNF) $\mathcal{P}_{\mathrm{S}}$ can be applied to the entire Hilbert space without any truncation. However, to concentrate on a subspace centered around the desired Fock state and thus realize a more efficient projection to a target Fock state with fewer operations, we employ an additional Gaussian PNF operation. This is accomplished by applying a qubit flip pulse with a Gaussian temporal envelope~\cite{vlastakis2013,heeres2015} with a standard deviation of $\sigma_t$ and a detuning of $N\chi_\mathrm{qc}$, which leads to the projection operator $\mathcal{P}_G(\sigma, N) \approx \sum_{n}e^{i\varphi_n}e^{- (n-N))^2/4\sigma^2}|n\rangle \langle n|$ with $\varphi_n$ being an insignificant phase similar to the sinusoidal filter. The Gaussian standard deviation of the filter function is approximated as $\sigma\approx 1/\left(\sqrt{2}\chi_\mathrm{qc}\sigma_t \right)$ for the Fock state distribution. Detailed experimental results regarding the Gaussian PNF can be found in the Supplementary Fig.~S3. Therefore, while $\mathcal{P}_{\mathrm{G}}$ selects a finite range of photon numbers, $\mathcal{P}_{\mathrm{S}}$ rules out photon numbers with certain parity. The combination of these two PNFs avoids the complicated optimization of pulses and is robust against decoherence (see Supplementary section I-B).

\vbox{}

\noindent \textbf{Photon number population reconstruction} 

\noindent The generated Fock states with large photon numbers are characterized by measuring their photon number distribution through a qubit spectroscopy experiment. In this experiment, we apply a selective Gaussian $\pi$ pulse on the ancillary qubit while sweeping the drive frequency. The measured readout signal as a function of the qubit drive frequency is fitted to a function of multicomponent Gaussian distributions denoted as $S(f) = \sum_{n=N-N_c}^{n=N+N_c}{A_n\exp{\left(-\frac{(f-f_n)^2}{2\sigma_f^2} \right)}+B}$, since each resoloved peak corresponds to each photon number component with the peak height $A_n$ proportional to the corresponding photon number population $P_n$~\cite{Schuster2007}. 
Here, the fitting parameter $\sigma_f$ is the frequency domain standard deviation of the selective Gaussian pulse and the background $B$ accounts for the measurement imperfections due to residual crosstalk between the probe cavity and readout resonator. These two parameters are constrained to be the same for all $n$, while $f_n$ is the precharacterized qubit resonance frequency with $n$ photons in the probe cavity. The cutoff number of Gaussian components $N_c$ is chosen to be sufficiently large to include all the relevant Fock components. By normalizing the fitted Gaussian amplitudes $A_n$, the photon number populations by $P_n = A_n/\sum_{n}{A_n}$ can be extracted by ensuring the normalization of the photon number population $\sum_n{P_n}=1$. More details can be found in Section I-B of the Supplementary Information.

\vbox{}

\noindent \textbf{Precision limitations of quantum metrology}

\noindent Quantum metrology involves the precise estimation of an unknown parameter $\lambda$ by utilizing quantum resource states, and generally consists of three steps: (i) preparing the probe in a known quantum state; (ii) interrogating the unknown parameter $\lambda$, i.e., encoding the parameter into the probe state, usually using a unitary transformation $U_\lambda = \exp{(-i\lambda h)}$ with $h$ being the interaction Hamiltonian; and (iii) measuring the final state with the encoded parameter $\lambda$. By comparing the input and output states of the probe, the unknown parameter can be inferred, with a precision limitation of estimation $\delta\lambda \ge 1/\sqrt{Q}$ according to the Cramer-Rao bound~\cite{braunstein1994}. Here,  $Q=4(\Delta h)^2=4(\langle h^2 \rangle - \langle h \rangle ^2)$ represents the quantum Fisher information (QFI) which is the supremum of the Fisher information among all possible measurements, and  $\Delta h$ is the standard deviation of the Hamiltonian $h$ under the initial probe state.

For displacement amplitude sensing, where $\lambda=\beta$ (assuming $\beta$ is a real number here), the interrogation interaction leads to the unitary transformation $U_\beta = \exp{(-i\beta h_\beta)}$ with $h_\beta = i(a^\dag - a)$. The standard quantum limit (SQL) is calculated by preparing the probe in a coherent state $|\alpha\rangle$. The QFI is then determined as $Q_\mathrm{SQL} = 4(\langle \alpha | h_\beta^2  |\alpha\rangle - \langle \alpha | h_\beta | \alpha \rangle ^2) = 4$, and the corresponding precision $\delta \beta_\mathrm{SQL} = 1/2$ is independent of the average photon number of the initial coherent state. By utilizing the Fock state $|N\rangle$ as the initial probe state, the QFI is calculated as $Q_\mathrm{Fock} = 4(\langle N | h_\beta^2  |N\rangle - \langle N | h_\beta | N \rangle ^2) = 4(2N+1)$, which leads to an estimation precision $\delta \beta_\mathrm{Fock} = 1/2\sqrt{2N+1}$. Comparing the precision for coherent and Fock states, a $\sqrt{N}$-enhancement of the measurement sensitivity is achievable considering the same input excitation numbers in the probe state, and the results indicate the scheme based on Fock states can approach the Heisenberg limit (HL). It can be further demonstrated that even when considering displacement in arbitrary directions, utilizing Fock states and parity measurements can still saturate the QFI. For more details, see section II-C in the Supplementary Information.

For phase measurement, where $\lambda=\phi$, the interrogation gives $U_\phi = \exp{(-i\phi h_\phi)}$ with $h_\phi = a^\dag a$. The SQL is calculated in a similar way with a coherent state $|\alpha\rangle$ in the probe, as the QFI $Q_\mathrm{SQL} = 4(\langle \alpha | h_\phi^2  |\alpha\rangle - \langle \alpha | h_\phi | \alpha \rangle ^2) = 4N$ is determined and the estimation precision $\delta \phi_\mathrm{SQL} = 1/2\sqrt{N}$ is obtained, with $N=|\alpha|^2$ representing the average photon number of the initial coherent state. By utilizing the displaced Fock state $D(\sqrt{N})|N\rangle$ as the initial probe state, the QFI is calculated as $Q_\mathrm{Fock} = 4(\langle N |D(-\sqrt{N}) h_\phi^2  D(\sqrt{N}) |N\rangle - \langle N |D(-\sqrt{N}) h_\phi D(\sqrt{N})| N \rangle ^2) = 4N(2N+1)$. Thus, $\delta \phi_\mathrm{Fock} = 1/2\sqrt{N(2N+1)}$, indicating a $\sqrt{N}$-enhancement of the measurement sensitivity that approaches the HL. It can also be proven that the parity measurement combined with the displacement operation can saturate the QFI here. (Supplementary Section II-D).

\vbox{}

\noindent \textbf{Extracting the Fisher information from measured data} 

\noindent In the quantum metrology experiments, the classical Fisher information for estimating the parameter $\lambda$ is extracted from the measured probability distributions $P_\mu$ using the equation $F_\lambda = \sum_\mu {\frac{1}{P_\mu} \left( \frac{\partial  P_\mu}{\partial \lambda}   \right)^2 }$~\cite{braunstein1994,PARIS2009}, where a set of measurement projections \{$M_\mu$\} is considered. In our experiment, the measurement performed on the ancillary qubit yields only two outcomes $\mu=g,e$, corresponding to the populations with the qubit in the ground and excited states. Therefore, the Fisher information can be further expressed as $F_\lambda = \frac{1}{P_g(1-P_g)} \left( \frac{d P_g}{d\lambda} \right)^2 $. Finally, the maximum achievable Fisher information $F_m = \mathrm{max}_\lambda (F_\lambda)$ is used to estimate the measurement precision of the parameter $\lambda$ by $\delta \lambda = 1/\sqrt{F_m}$. 

For the displacement amplitude sensing experiment, the measured qubit ground state population $P_g$ is plotted as a function of the displacement amplitude $\beta$ in Fig.~\ref{fig3}b. The populations is fitted to a function of $P_g = Ae^{-2|\beta|^2}L_N(4|\beta|^2)+B$ (Supplementary Section II-A ) with $L_N$ being the $N$-th order Laguerre polynomials, and $A$ and $B$ being the fitting parameters to account for measurement imperfections. With the fitting result, the classical Fisher information $F_\beta$ is calculated according to the equations described above, and plotted in Fig.~\ref{fig3}c. The maximum achieved Fisher information is used to obtain the estimation precision $\delta \beta$ for the parameter $\beta$. 

For the phase measurement, the measured qubit population $P_g$ as a function of $\phi$ shown in Fig.~\ref{fig3}f, is fitted to a function of $P_g = Ae^{-2N(\phi+\phi_0)^2}L_N(4N(\phi+\phi_0)^2)+B$ with $A$ and $B$ being the fitting parameters to account for the measurement imperfections and $\phi_0$ being the fitting parameter to account for the phase rotation due to the self-Kerr interaction during the sensing process. In a similar way, the calculated Fisher information $F_\phi$ shown in Fig.~\ref{fig3}g is employed to obtain the estimation precision $\delta \phi$ for the parameter $\phi$.

\vbox{}

\noindent \textbf{Calculating the success probability and the scaling enhancement} 

\noindent In the quantum metrology experiment, the initial Fock states are generated using PNFs to project the coherent state into the target photon number $N$ with a success probability of $P_N = e^{-N} N^N/N! \approx 1/\sqrt{2\pi N}$ for $N\gg 1$ according to the Stirling's approximation.

With considering the post-selected preparation of the Fock states, the quantum Fisher information for displacement sensing can be calculated as $Q_\mathrm{postFock} = 4(2N+1)P_N \approx \frac{8}{\sqrt{2\pi}}N^{1/2}$ for $N\gg 1$, resulting in a scaling factor of $N^{1/2}$ better than $Q_\mathrm{SQL}=4$ for quasiclassical coherent states. Therefore, the measurement precision of the displacement amplitude $\delta \beta_\mathrm{postFock} = 1/\sqrt{Q_\mathrm{postFock}} \approx \sqrt{\frac{\sqrt{2\pi}}{8}}N^{-1/4}$ can still achieve a scaling enhancement factor of $N^{1/4}$ over the classical limit, highlighting the quantum metrological advantage despite the probabilistic preparation of the initial Fock state.

Similarly, in the phase sensing experiment, the quantum Fisher information is $Q_\mathrm{postFock} \approx \frac{2}{\sqrt{\pi}}(\bar{n}^{3/2}+\bar{n}^{1/2})$ when considering the postselected preparation of the Fock states $|N\rangle$ with $N=\bar{n}/2$ in our scheme. The corresponding measurement precision $\delta \phi_\mathrm{postFock} = 1/\sqrt{Q_\mathrm{postFock}} \approx \sqrt{\frac{\sqrt{\pi}}{2}}\bar{n}^{-3/4}$ can also achieve a scaling enhancement factor of $\bar{n}^{1/4}$ over the classical limit.

\vbox{}

\smallskip{}

\noindent \textbf{\large{}Data availability}{\large\par}

\noindent Source data for all the Figures in the main article are available with the paper. All other data relevant to this study are available from the corresponding authors upon reasonable request.

\smallskip{}

\noindent \textbf{\large{}Code availability}{\large\par}

\noindent The code used for this study is available from the corresponding authors upon reasonable request.

\smallskip{}

\noindent \textbf{\large{}Additional information}{\large\par}

\noindent \textbf{Supplementary information} The online version contains supplementary material.

\noindent \textbf{Correspondence and requests for materials} should be addressed to C.-L.Z., F.Y., Y.X. and D. Y.

\clearpage{}

\end{document}

\title{Supplementary Information for ``Quantum-enhanced metrology with large Fock states"}
\author{Xiaowei Deng}
\thanks{These authors contributed equally to this work.}
\affiliation{International Quantum Academy, Shenzhen 518048, China}
\affiliation{Shenzhen Institute for Quantum Science and Engineering, Southern University of Science and Technology, Shenzhen 518055, China}
\affiliation{Guangdong Provincial Key Laboratory of Quantum Science and Engineering, Southern University of Science and Technology, Shenzhen 518055, China}

\author{Sai Li}
\thanks{These authors contributed equally to this work.}
\affiliation{International Quantum Academy, Shenzhen 518048, China}
\affiliation{Shenzhen Institute for Quantum Science and Engineering, Southern University of Science and Technology, Shenzhen 518055, China}
\affiliation{Guangdong Provincial Key Laboratory of Quantum Science and Engineering, Southern University of Science and Technology, Shenzhen 518055, China}

\author{Zi-Jie Chen}
\thanks{These authors contributed equally to this work.}
\affiliation{CAS Key Laboratory of Quantum Information, University of Science and Technology of China, Hefei, Anhui 230026, China}

\author{Zhongchu Ni}
\affiliation{International Quantum Academy, Shenzhen 518048, China}
\affiliation{Shenzhen Institute for Quantum Science and Engineering, Southern University of Science and Technology, Shenzhen 518055, China}
\affiliation{Guangdong Provincial Key Laboratory of Quantum Science and Engineering, Southern University of Science and Technology, Shenzhen 518055, China}
\affiliation{Department of Physics, Southern University of Science and Technology, Shenzhen 518055, China}

\author{Yanyan Cai}
\affiliation{International Quantum Academy, Shenzhen 518048, China}
\affiliation{Shenzhen Institute for Quantum Science and Engineering, Southern University of Science and Technology, Shenzhen 518055, China}
\affiliation{Guangdong Provincial Key Laboratory of Quantum Science and Engineering, Southern University of Science and Technology, Shenzhen 518055, China}

\author{Jiasheng Mai}
\affiliation{International Quantum Academy, Shenzhen 518048, China}
\affiliation{Shenzhen Institute for Quantum Science and Engineering, Southern University of Science and Technology, Shenzhen 518055, China}
\affiliation{Guangdong Provincial Key Laboratory of Quantum Science and Engineering, Southern University of Science and Technology, Shenzhen 518055, China}

\author{Libo Zhang}
\affiliation{International Quantum Academy, Shenzhen 518048, China}
\affiliation{Shenzhen Institute for Quantum Science and Engineering, Southern University of Science and Technology, Shenzhen 518055, China}
\affiliation{Guangdong Provincial Key Laboratory of Quantum Science and Engineering, Southern University of Science and Technology, Shenzhen 518055, China}

\author{Pan Zheng}
\affiliation{International Quantum Academy, Shenzhen 518048, China}
\affiliation{Shenzhen Institute for Quantum Science and Engineering, Southern University of Science and Technology, Shenzhen 518055, China}
\affiliation{Guangdong Provincial Key Laboratory of Quantum Science and Engineering, Southern University of Science and Technology, Shenzhen 518055, China}

\author{Haifeng Yu}
\affiliation{Beijing Academy of Quantum Information Sciences, Beijing 100193, China}

\author{Chang-Ling Zou}
\email{clzou321@ustc.edu.cn}
\affiliation{CAS Key Laboratory of Quantum Information, University of Science and Technology of China, Hefei, Anhui 230026, China}
\affiliation{Hefei National Laboratory, Hefei 230088, China.}

\author{Song Liu}
\affiliation{International Quantum Academy, Shenzhen 518048, China}
\affiliation{Shenzhen Institute for Quantum Science and Engineering, Southern University of Science and Technology, Shenzhen 518055, China}
\affiliation{Guangdong Provincial Key Laboratory of Quantum Science and Engineering, Southern University of Science and Technology, Shenzhen 518055, China}
\affiliation{Hefei National Laboratory, Hefei 230088, China.}

\author{Fei Yan}
\email{yanfei@baqis.ac.cn}
\altaffiliation[Present address: ]{Beijing Academy of Quantum Information Sciences, Beijing, China}
\affiliation{International Quantum Academy, Shenzhen 518048, China}
\affiliation{Shenzhen Institute for Quantum Science and Engineering, Southern University of Science and Technology, Shenzhen 518055, China}
\affiliation{Guangdong Provincial Key Laboratory of Quantum Science and Engineering, Southern University of Science and Technology, Shenzhen 518055, China}

\author{Yuan Xu}
\email{xuyuan@iqasz.cn}
\affiliation{International Quantum Academy, Shenzhen 518048, China}
\affiliation{Shenzhen Institute for Quantum Science and Engineering, Southern University of Science and Technology, Shenzhen 518055, China}
\affiliation{Guangdong Provincial Key Laboratory of Quantum Science and Engineering, Southern University of Science and Technology, Shenzhen 518055, China}
\affiliation{Hefei National Laboratory, Hefei 230088, China.}

\author{Dapeng Yu}
\email{yudp@sustech.edu.cn}
\affiliation{International Quantum Academy, Shenzhen 518048, China}
\affiliation{Shenzhen Institute for Quantum Science and Engineering, Southern University of Science and Technology, Shenzhen 518055, China}
\affiliation{Guangdong Provincial Key Laboratory of Quantum Science and Engineering, Southern University of Science and Technology, Shenzhen 518055, China}
\affiliation{Department of Physics, Southern University of Science and Technology, Shenzhen 518055, China}
\affiliation{Hefei National Laboratory, Hefei 230088, China.}

\maketitle
% \toc
\tableofcontents

%\clearpage{}
\section{Experimental Details}

\subsection{Device and parameters}
\subsubsection{Experimental device}
The three-dimensional (3D) circuit quantum electrodynamics (QED) device used in this experiment is similar to that in Ref.~\cite{ni2023}, with the schematic depicted in Fig.~\ref{figS1}. The device consists of a superconducting transmon qubit~\cite{koch2007}, a 3D coaxial stub cavity~\cite{reagor2016}, and a Purcell-filtered stripline readout resonator~\cite{axline2016} within a single package. The package is machined from a single block of high purity (5N5) aluminum and chemically etched approximately 100~$\mu$m to improve the cavity\rq{}s surface quality, thereby enhancing its lifetime~\cite{reagor2013}. 

The coaxial stub cavity as the probe cavity is constructed with a center stub conductor with a diameter of 3.2~mm and a cylindrical outer conductor with a diameter of 9.6 mm. The center stub is electrically connected to the ground of the outer conductor at one end and terminated open into a cylindrical waveguide at the other end, forming a 3D $\lambda/4$ transmission line resonator. The resonance frequency of the fundamental mode in the cavity is determined by the center stub height of 9.8 mm. A horizontal tunnel with a diameter of 3.2 mm located at a height near the top of the center stub is designed to accommodate a sapphire chip. 

On the sapphire chip, the transmon qubit and the Purcell-filtered stripline readout resonator are patterned lithographically, with an optical image of the chip shown in Fig.~\ref{figS1}. The transmon qubit is designed with two antenna pads to capacitively couple to the probe cavity mode and the readout resonator mode. The Purcell-filtered readout resonator is constructed with two quasiplanar $\lambda/2$ transmission line resonators, which are formed by the metal wall of the tunnel and two metal strips on the sapphire chip inserted in the tunnel. The two striplines of the readout and filter modes are designed with wiggles to minimize the patterns\rq{} physical footprint on the chip~\cite{chou2018}. These two modes are strongly coupled to a $50~\Omega$ transmission line to enable fast qubit readout. The filter mode is utilized to protect the qubit and probe cavity from spontaneous decay into the outside environment. 

The patterns on the sapphire chip are defined using a standard electron-beam lithography process. Initially, a BCC alpha-phase tantalum film~\cite{place2021, wang2022b} is deposited on the sapphire substrate to pattern the qubit antenna pads and the readout and filter striplines. The trilayer Josephson junction of the qubit, consisting of $\mathrm{Al}/\mathrm{Al}_2\mathrm{O}_3/\mathrm{Al}$, is then created by a double angle evaporation technique, before which ion milling is utilized to remove any oxides on the surface of the tantalum film. After fabrication, the chip diced from a $430~\mu$m-thick c-plane sapphire wafer into 3 mm $\times$ 24 mm dimensions, is tightly clamped at one end and inserted into the tunnel.

\subsubsection{Measurement setup}

\begin{figure}[b]
    \includegraphics{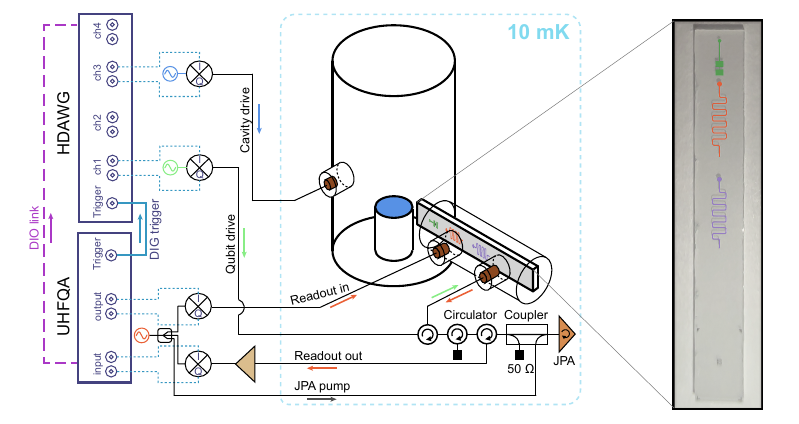}
    \caption{Measurement setup and experimental device schematic.} 
    \label{figS1}
\end{figure}

The experimental device is cooled to a temperature below 10 mK in a dilution refrigerator and controlled using microwave signals generated from room-temperature electronics. The measurement setup is illustrated in Fig.~\ref{figS1}. 

The transmon qubit and probe cavity are both controlled with microwave signals produced by single-sideband in-phase and quadrature (IQ) modulations of continuous-microwave tones generated from respective signal generators. The Zurich Instruments high-density arbitrary waveform generator (HDAWG) is employed to realize qubit drives using a pair of digital-to-analog converter (DAC) channels, while another pair of DAC channels of the same HDAWG are utilized as an IQ pair to generate cavity drive pulses.  

The qubit readout pulse is generated by single-sideband IQ modulation using a pair of DAC channels of the Zurich Instruments ultra-high frequency quantum analyzer (UHFQA) and is sent to the readout resonator. After amplification, the transmitted readout signal is downconverted, digitized, and recorded by the analog-to-digital converter (ADC) of the same UHFQA. To align the timing sequence of the qubit and cavity control pulses with the readout pulse, a digital trigger is used to synchronize the UHFQA and the HDAWG.

To achieve fast feedback control in the quantum metrology experiment, we connect the UHFQA to the HDAWG using a digital input/output (DIO) link cable. This allows the UHFQA to send the digitized readout signals to the HDAWG in real-time through the DIO link, and the HDAWG can selectively play different waveforms based on the received DIO signal. The feedback latency, which is defined as the duration between sending the last point of the readout pulse from the UHFQA and sending out the first point of the feedback pulse from the HDAWG, is approximately 591~ns in our experiment, including the signal traveling time through the experimental circuitry.

\subsubsection{System Hamiltonian and device parameters}
The 3D circuit QED system, containing two nearly linear bosonic modes (probe cavity and readout resonator) and an anharmonic bosonic mode (superconducting qubit), can be described with the dispersive Hamiltonian with relevant terms:
\begin{eqnarray}
\label{H1}
\hat{H}/\hbar &=& \omega_\mathrm{q} \hat{a}_\mathrm{q}^\dag \hat{a}_\mathrm{q} + \omega_\mathrm{c} \hat{a}_\mathrm{c}^\dag \hat{a}_\mathrm{c} +\omega_\mathrm{r} \hat{a}_\mathrm{r}^\dag \hat{a}_\mathrm{r} \notag\\
&-& \frac{K_\mathrm{q}}{2}\hat{a}_\mathrm{q}^{\dag 2}\hat{a}_\mathrm{q}^{2}-\frac{K_\mathrm{c}}{2}\hat{a}_\mathrm{c}^{\dag 2}\hat{a}_\mathrm{c}^{2} \notag\\
&-& \chi_\mathrm{qr}\hat{a}_\mathrm{q}^\dag\hat{a}_\mathrm{q}\hat{a}_\mathrm{r}^\dag\hat{a}_\mathrm{r} - \chi_\mathrm{qc}\hat{a}_\mathrm{q}^\dag\hat{a}_\mathrm{q}\hat{a}_\mathrm{c}^\dag\hat{a}_\mathrm{c} + \frac{\chi'_\mathrm{qc}}{2}\hat{a}_\mathrm{q}^\dag\hat{a}_\mathrm{q}\hat{a}_\mathrm{c}^{\dag 2}\hat{a}_\mathrm{c}^{2}
,
\end{eqnarray}
where $\omega_\mathrm{q,c,r}$, $a_\mathrm{q,c,r}$, and $K_\mathrm{q,c,r}$ are the resonance frequencies, annihilation operators, and self-Kerrs of the ancilla qubit, probe cavity, and readout resonator, respectively; $\chi_\mathrm{qr}$ and $\chi_\mathrm{qc}$ are the cross-Kerrs between the qubit and the readout resonator and probe cavity modes; and $\chi'_\mathrm{qc}$ is the higher-order cross-Kerr between the ancilla qubit and the probe cavity. The measured values of these parameters are summarized and listed in Table~\ref{TableS1}.

\begin{table}[t]
\caption{Hamiltonian parameters and coherence times.}
\centering
\begin{tabular}{p{3cm}<{\centering} p{3cm}<{\centering}  p{1.7cm}<{\centering} p{1.7cm}<{\centering} p{1.7cm}<{\centering} p{1.7cm}<{\centering} p{1.7cm}<{\centering} p{1.7cm}<{\centering}}
  \hline
  \hline
    &&&&&&&\\[-0.9em]
  \multicolumn{2}{c}{Measured parameters} &  \multicolumn{2}{c}{Probe cavity ($C$)} & \multicolumn{2}{c}{Transmon qubit ($Q$)} & \multicolumn{2}{c}{Readout resonator ($R$)} \\\hline
  &&&&&&&\\[-0.9em]
  \multicolumn{2}{c}{Mode frequency $\omega_\mathrm{c,q,r}/2\pi$ } & \multicolumn{2}{c}{6.597~GHz} & \multicolumn{2}{c}{4.878~GHz} & \multicolumn{2}{c}{8.530~GHz}  \\
  &&&&&&&\\[-0.9em]
  \hline
  &&&&&&&\\[-0.9em]
  \multicolumn{2}{c}{Self-Kerr $K_\mathrm{c,q,r}/2\pi$ } & \multicolumn{2}{c}{0.44~kHz} & \multicolumn{2}{c}{214~MHz} & \multicolumn{2}{c}{-}  \\
  &&&&&&&\\[-0.9em]
  \hline
  &&&&&&&\\[-0.9em]
  \multicolumn{2}{c}{Cross-Kerr $(\chi_\mathrm{qc}, \chi_\mathrm{qr})/2\pi$ } & &\multicolumn{2}{c}{0.626~MHz} & \multicolumn{2}{c}{2.0~MHz} & \\
  &&&&&&&\\[-0.9em]
  \hline  
  &&&&&&&\\[-0.9em]
  \multicolumn{2}{c}{Second-order cross-Kerr $\chi'_\mathrm{qc}/2\pi$ } & & \multicolumn{2}{c}{0.328~kHz} &  \multicolumn{2}{c}{-} & \\
  &&&&&&&\\[-0.9em]
  \hline 
  \hline
  &&&&&&&\\[-0.9em]
  \multicolumn{2}{c}{Relaxation time $T_1$ } & \multicolumn{2}{c}{$1.2~$ms} & \multicolumn{2}{c}{$93~\mu$s} & \multicolumn{2}{c}{83~ns}  \\
  &&&&&&&\\[-0.9em]
  \multicolumn{2}{c}{Pure dephasing time $T_\phi$ } & \multicolumn{2}{c}{$4.0~$ms} & \multicolumn{2}{c}{$445~\mu$s} & \multicolumn{2}{c}{-}  \\
  &&&&&&&\\[-0.9em]
  \multicolumn{2}{c}{Thermal population } & \multicolumn{2}{c}{$<$0.1\%} & \multicolumn{2}{c}{2\%} & \multicolumn{2}{c}{-}  \\
  &&&&&&&\\[-0.9em]
  \hline 
  \hline
\end{tabular}
\label{TableS1}
\end{table}

\begin{figure}[b]
    \includegraphics{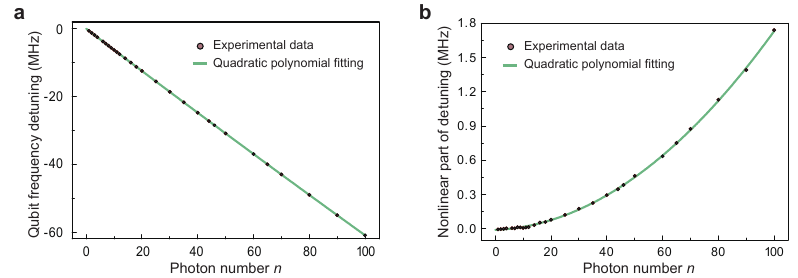}
    \caption{Qubit spectroscopy experiment for determining the dispersive shifts $\chi_\mathrm{qc}$ and $\chi'_\mathrm{qc}$ between the qubit and the probe cavity. (a) Measured qubit transition frequency detuning as a function of the photon number $n$ in the cavity, which is fitted to a second-order polynomial (solid line). (b) The deviation of the qubit transition frequency detuning from the linear part as a function of the photon number $n$ in the cavity, exhibiting a quadratic relationship. }
    \label{figS2}
\end{figure}

The dispersive interaction between the ancillary transmon qubit and the probe cavity will result in a photon-number-dependent qubit frequency shift, which is utilized to implement photon number filter operations to generate Fock states in the cavity. The linear ($\chi_\mathrm{qc}$) and quadratic ($\chi'_\mathrm{qc}$) photon-number-dependent dispersive shifts are the dominant terms in the dispersive Hamiltonian and are obtained from qubit spectroscopy experiments with the cavity initially prepared in different Fock states or coherent states.
In Fig.~\ref{figS2}a, the measured qubit transition frequency detuning relative to the qubit frequency with zero photons in the cavity is plotted as a function of the photon number $n$ in the cavity and is fitted to a low-order polynomial function. In this experiment, a second-order polynomial is sufficient to fit the experimental results to extract the linear dispersive shift $\chi_\mathrm{qc}/2\pi=626$~kHz and the second-order dispersive shift $\chi'_\mathrm{qc}/2\pi=328$~Hz. This is confirmed by the plot of the deviation of the qubit transition frequency from the linear part, as shown in Fig.~\ref{figS2}b, which exhibits an obvious quadratic relationship.

The coherence times of the three modes in this system are characterized using standard circuit QED measurements and summarized in Table~\ref{TableS1}. The transmon qubit has coherence times of approximately $T_{1} = 93~\mu$s and $T_{2}^* = 131~\mu$s, corresponding to a pure dephasing time of $T_{\phi} = 445~\mu$s. The bosonic mode in the probe cavity has single-photon lifetimes of approximately $T_{1} = 1.2~$ms and $T_{2}^* = 1.5~$ms, corresponding to a pure dephasing time of $T_{\phi} = 4.0~$ms. The decay rate of the readout resonator $\kappa_\mathrm{r}/2\pi = 1.92$~MHz (corresponding to a decay time constant of 83~ns) is carefully designed to match the dispersive shift $\chi_\mathrm{qr}/2\pi = 2.0$~MHz, to achieve a high-fidelity qubit readout. The qubit readout is optimized with a pulse duration of approximately 500~ns, resulting in an average fidelity of 0.995 and an average QNDness of 0.991.

\subsection{Fock state generation and characterization}

\subsubsection{Photon number filtration (PNF)}
In quantum metrology experiments, the initial Fock states with large photon numbers are efficiently generated with two kinds of programmable PNFs: Gaussian PNF and sinusoidal PNF, by utilizing the dispersive interaction between the probe cavity and the transmon qubit. 

The Gaussian PNF is realized by first initializing the transmon qubit in the excited state $|e\rangle$, resulting in a joint state of $|e\rangle |\psi_\mathrm{in}\rangle =\sum_n{c_n|e,n\rangle}$. Then, a selective $\pi$ pulse with a Gaussian envelope $\epsilon (t) = A \exp {\left(-t^2/2\sigma_t^2 \right)} $ is applied on the qubit with the drive frequency matching the qubit frequency in the presence of $N$ photons in the cavity~\cite{heeres2015, vlastakis2013}. Here $A$ is the pulse amplitude with $\int{\epsilon(t)dt} = \pi/2$ to achieve a $\pi$ pulse and $\sigma_t$ is the Gaussian standard deviation. Finally, the qubit is postselected in the ground state $|g\rangle$ to project the cavity state. The interaction Hamiltonian of the selective pulse can be written as
\begin{eqnarray}
    \hat{H}_\mathrm{int}/\hbar &=& -\chi_\mathrm{qc}(a^\dagger a - N)|e\rangle \langle e | + \epsilon(t) \sigma_y \\\notag
    &=& \sum_n {\{ -\chi_\mathrm{qc}(n - N)|e\rangle \langle e | + \epsilon(t) \sigma_y \} |n\rangle \langle n | }, 
\end{eqnarray}
where $a$ ($a^\dagger$) is the annihilation (creation) operator of the bosonic mode, and $\sigma_y$ is the y-axis Pauli operator of the qubit. The Hamiltonian of each photon number block can be transformed into a rotating frame as 
\begin{eqnarray}
\tilde{H}_n(t)/\hbar =  \epsilon(t) e^{i\Delta_nt|e\rangle \langle e |} \sigma_y e^{-i\Delta_nt|e\rangle \langle e |},
\end{eqnarray}
with $\Delta_n=-(n-N)\chi_\mathrm{qc}$. For exactly $N$ photons in the probe cavity, the above Hamiltonian flips the qubit state from the initial excited state to the ground state. For other photon number states in the probe cavity $\sum_{n\ne N}{c_n|n\rangle}$, we can approximate the final state as
\begin{eqnarray}
    |\psi_\mathrm{out}\rangle &=& \sum_{n\ne N} {\{ 1 - \frac{i}{\hbar}\int^t{\tilde{H}_n(s) ds} \}c_n |e, n\rangle } \\\notag
    &\approx& \sum_{n\ne N} {c_n \{|e, n\rangle - \hat{\epsilon}(\Delta_n) |g, n\rangle  \} }, 
\end{eqnarray}
with only considering the first order Dyson expansion. Here $\hat{\epsilon}(\Delta_n) \propto e^{-\Delta_n^2 \sigma_t^2/2}$ is the Fourier transform of the pulse amplitude envelop $\epsilon(t)$ at a frequency of $\Delta_n$. After projecting the qubit into ground state, the photon number coefficient $c_n$ of the cavity state has been transformed into $c_n e^{-\Delta_n^2 \sigma_t^2/2} = c_n e^{-{(n-N)^2}/{4\sigma^2}}$ with $\sigma = 1/\left(\sqrt{2}\chi_\mathrm{qc}\sigma_t\right)$. 

\begin{figure}[t]
    \includegraphics{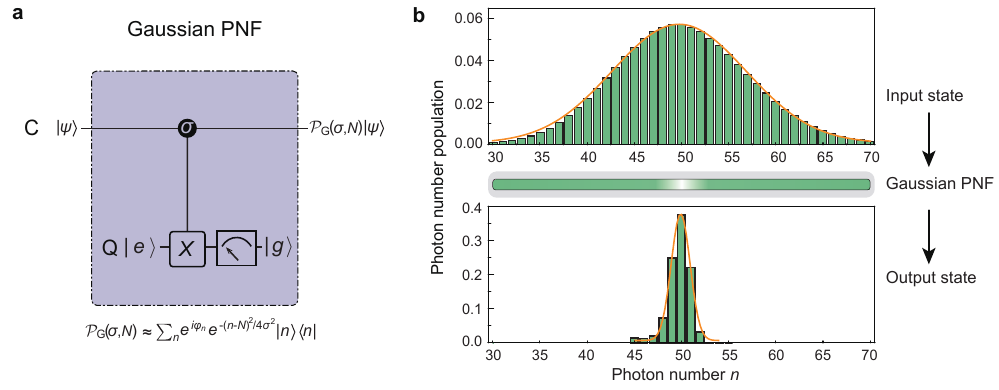}
    \caption{(a) Quantum circuit for implementing the Gaussian PNF on the bosonic state in the probe cavity. The PNF operation involves a controlled gate that flips the qubit state based on the photon numbers in the cavity, as determined by the parameter $\sigma$ of the Gaussian function. (b) Characterization of the Gaussian PNF operation. An ideal photon number population of a coherent state with $\alpha=\sqrt{50}$ (top) is filtered out by a Gaussian PNF $\mathcal{P}_\mathrm{G}(\sigma, 50)$, resulting in the measured photon number populations with a smaller standard deviation (bottom).} 
    \label{figS3}
\end{figure}

Therefore, this process transforms the input bosonic state $\ket{\psi_\mathrm{in}}=\sum_n{c_n|n\rangle}$ to the output state $\ket{\psi_\mathrm{out}} \propto \mathcal{P}_\mathrm{G}(\sigma, N) \ket{\psi_\mathrm{in}}$ (neglecting the normalization factor) through an approximate Gaussian filter function $\mathcal{P}_\mathrm{G}(\sigma, N) \approx \sum_n{e^{-{(n-N)^2}/{4\sigma^2} }e^{i\varphi_n}\ket{n}\bra{n}}$ with $\varphi_n$ being an insignificant phase factor for each Fock component. The quantum circuit of the Gaussian PNF is shown in Fig.~\ref{figS3}(a), where a controlled gate flips the qubit state based on the photon numbers in the cavity, as determined by the parameter $\sigma$ of the Gaussian PNF. As a demonstration, we implement the Gaussian PNF operation on an initial coherent state with an average photon number of $N=50$. The measurement result is shown in Fig.~\ref{figS3}b, where the standard deviation of the photon number population is reduced from $\sqrt{50}$ to approximately 0.9 by choosing $\sigma_t=200$~ns of the selective pulse. In principle, only one Gaussian PNF is sufficient to generate Fock states with arbitrary photon numbers by engineering $\sigma \ll 1$ but at the expense of a long operation time. In our experiment, the parameter $\sigma$ is chosen as a trade-off based on the fact that the larger $\sigma$ is, the more residual photon number populations need to be filtered out further, and the smaller $\sigma$ is, the more decoherence-induced errors due to the long selective $\pi$ pulse. 

%\mathscr{P}

%\subsection{Sinusoidal photon number filter}
The sinusoidal PNF is implemented by inserting a conditional unitary $C_\theta = \ket{g}\bra{g} \otimes I+ \ket{e}\bra{e} \otimes e^{i \theta (a^\dagger a - N) }$ into a Ramsey-type sequence: $X/2 \rightarrow C_\theta \rightarrow -X/2$, where $\theta/\chi_\mathrm{qc}$ is the dispersive interaction duration. This process can be written with a matrix form in the qubit basis as
\begin{eqnarray}
\label{sPNF}
U_\mathrm{S}(\theta) &&= \frac{1}{2}\left(
\begin{array}{cc}
I+e^{i\theta(a^\dagger a-N)},  & -iI+ie^{i\theta(a^\dagger a-N)} \\
iI-ie^{i\theta(a^\dagger a-N)}, & I+e^{i\theta(a^\dagger a-N)}     
\end{array}
\right),
\end{eqnarray}
where $I$ represents an identity operator implemented on the bosonic state in the probe cavity. After projecting the qubit into the ground state, the filter operation transforms the input bosonic state $\ket{\psi_\mathrm{in}}$ to the output state $\ket{\psi_\mathrm{out}}\propto \mathcal{P}_\mathrm{S}(\theta, N) \ket{\psi_\mathrm{in}}$ through a sinusoidal filter function $\mathcal{P}_\mathrm{S}(\theta, N) \approx \sum_{n}{\cos{\frac{(n-N))\theta}{2}} e^{i\varphi_n}\ket{n}\bra{n}}$ with $\varphi_n$ being an insignificant phase factor for generating Fock states. In addition, the sinusoidal PNF can also be generalized to $\mathcal{P}_\mathrm{S}(\theta, \varphi, N) = \sum_n{\sin{\frac{(n-N)\theta - \varphi}{2}} \ket{n}\bra{n}}$ by adding a programmable phase $\varphi$, which represents the angle of the rotation axis relative to the $x$ axis of the second $\pi/2$ pulse in the Ramsey experiment. This generalized sinusoidal PNF, similar to the distillation operation described in Ref.~\cite{teja2023distillation}, is utilized in the photon-number-resolved quantum metrology experiment, as discussed in Section I-C2.

We note that the PNFs demonstrated here are built out of the selective number-dependent arbitrary phase (SNAP) gates~\cite{krastanov2015} and postselection measurements of the qubit modes. The SNAP gate consists of two selective $\pi$ rotations applied on the qubit to acquire a geometric phase conditional on the specific Fock component. Whereas, the PNFs incorporate a single selective rotation followed by a postselection measurement on the ancilla qubit. The selective rotation in the Gaussian PNF is a single selective $\pi$ pulse with Gaussian envelope, which is similar to that of the SNAP gate. Meanwhile, the Ramsey-type sequence in the sinusoidal PNF can also be seen as a selective pulse conditional on all the Fock components with the same photon-number parity. 

\subsubsection{Sequence optimization to generate large Fock states}
\begin{figure}[t]
    \includegraphics{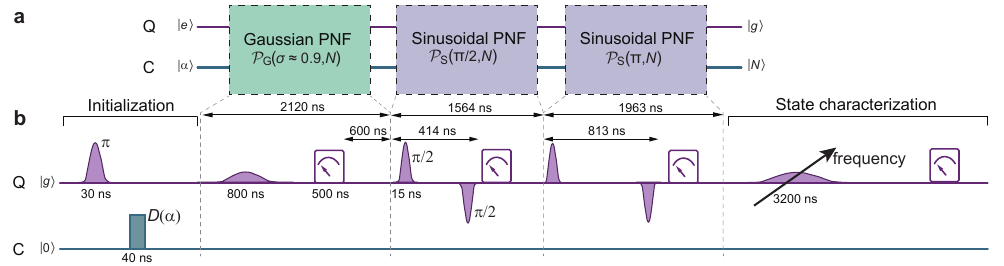}
    \caption{Parameterized quantum circuit with sequential PNFs (a) and experimental sequence (b) for efficiently generating Fock states with large photon numbers.}  
    \label{figS4}
\end{figure}

To efficiently generate Fock states with large photon numbers, we combine the Gaussian and sinusoidal PNFs in a parameterized quantum circuit, as shown in Fig.~\ref{figS4}a.

In the circuit, the ancillary qubit is initialized in the excited state, while the probe cavity is in a coherent state with an average photon number of $N$. We first implement a Gaussian PNF to compress the photon number distribution of the coherent state into a narrow distribution around the target photon number $N$. The chosen parameter $\sigma$ in the Gaussian PNF is a compromise based on the fact that the smaller $\sigma$ is, the more decoherenced-induced errors due to the long operation time, and the larger $\sigma$ is, the more residual photon number populations there are. By choosing $\sigma\approx 0.9$, the Gaussian PNF operation largely reduces the photon number population $P_n\approx 0$ for $|n-N|> 2$. The subsequent two sinusoidal PNFs $\mathcal{P}_\mathrm{S}(\pi/2, N)$ and $\mathcal{P}_\mathrm{S}(\pi, N)$ remove the photon number populations $P_{N\pm2}$ and $P_{N\pm1}$, respectively. The order of these two sinusoidal PNFs is chosen to improve the fidelity of the target Fock state in the presence of the photon decay during PNF operations. With these three steps in a parameterized quantum circuit, the target Fock state $\ket{N}$ with a large photon number $N$ can be efficiently generated. For the experiment shown in Fig. 2 in the main text, the success probabilities of the three PNFs are (16.0\%, 61.7\%, 56.6\%), (12.7\%, 59.3\%, 54.0\%), (10.5\%, 57.5\%, 53.0\%), and (5.9\%, 61.2\%, 58.7\%) for generating Fock states $|30\rangle$, $|50\rangle$, $|70\rangle$ and $|100\rangle$, respectively, corresponding to a total success probability of 5.5\%, 4.0\%, 3.2\%, and 2.1\%.

The pulse sequences for the three PNFs are shown in Fig.~\ref{figS4}b. Owing to the residual cross-Kerr interaction between the probe cavity and the readout resonator, the qubit readout will unavoidably be affected when the probe cavity contains large photon numbers. To mitigate this effect, we measure the single-shot readout clusters corresponding to the qubit ground and excited states for various photon numbers in the cavity, fit the two clusters to a Gaussian distribution to determine the center and standard deviation of the clusters, and postselect the single-shot data within 2.4 times the standard deviation around the center while disregarding others to attain confidence in the measured qubit state. Ultimately, the qubit readout exhibits an average QNDness of 0.98 with a mean photon of 100 in the cavity.

\subsubsection{Fock state characterization}
We experimentally characterize the generated Fock state $\ket{N}$ by measuring its photon number distribution $P_{n}$ with a qubit spectroscopy experiment, with the sequence shown in Fig.~\ref{figS4}b. The measured readout signals for different Fock states are fitted to a function with multicomponent Gaussian distributions: $S(f) = \sum_{n=N-N_c}^{n=N+N_c}{A_n\exp{\left(-\frac{(f-f_n)^2}{2\sigma_f^2} \right)}+B}$, since each resolved peak corresponds to each photon number component with the peak height proportional to the corresponding photon number population~\cite{Schuster2007}. Here $A_n$, $B$, and $\sigma_f$ are the fitting parameters and $f_n$ is the precharacterized qubit resonance frequency with $n$ photons in the probe cavity in Fig.~\ref{figS2}. Here $\sigma_f$ is the frequency domain standard deviation of the selective Gaussian pulse in the state characterization process and the background $B$ accounts for the measurement imperfections due to residual crosstalk between the probe cavity and the readout resonator. These two parameters are both constrained to be the same for all $n$. The cutoff of the number of Gaussian components $N_c$ is chosen to be sufficiently large to include all the relevant Fock components. From this fit, the photon number population $P_n$ can be extracted by normalizing the Gaussian amplitudes by $P_n = A_n/\sum_{n}{A_n}$, which ensures the normalization of the photon number population $\sum_n{P_n}=1$. The normalized readout signal $[S(f)-B]/\sum_{n}{A_n}$ and the extracted photon number populations $P_n$ for each generated Fock state $\ket{N}$ with $N=30$, 50, 70 and 100 are shown in Fig. 2 in the main text.

\begin{figure}[t]
    \includegraphics{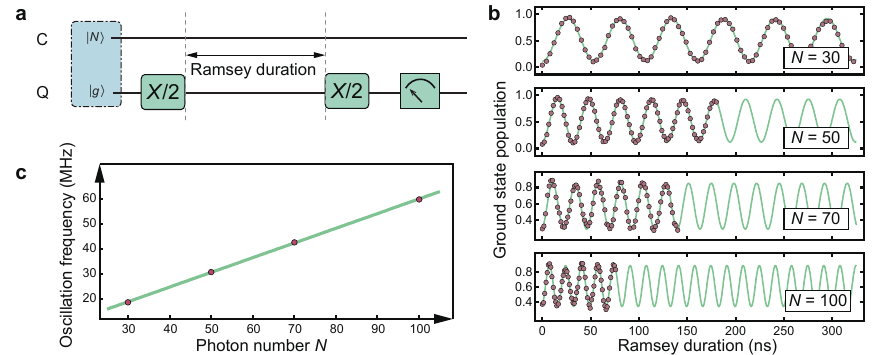}
    \caption{(a) Quantum circuit for performing a Ramsey experiment on the ancilla qubit with the cavity initially prepared in a Fock state $|N\rangle$. (b) Measured (dots) Ramsey interference patterns of ancillary qubit for different Fock states $|N\rangle$ with $N=30$, 50, 70, and 100. Each curve is fitted to a single-component sinusoidal function (solid lines) to extract the dominant oscillation frequency. (c) Oscillation frequencies (dots) extracted in (b) as a function of the photon number $N$ and a linear fit (solid line).}
    \label{figS5}
\end{figure}

%\subsection{Fock state characterize with ancilla qubit Ramsey oscillations}
In addition, the generated Fock states are further characterized by performing a Ramsey experiment on the ancilla qubit with the quantum circuit shown in Fig.~\ref{figS5}a, as the measured qubit ground state probability oscillates with the Ramsey duration. As shown in Fig.~\ref{figS5}b, the qubit Ramsey interference pattern oscillates $N$ times faster in the presence of $N$ photons in the cavity, due to the photon-number-dependent dispersive Hamiltonian. The extracted Ramsey oscillation frequencies from a single-component sinusoidal fitting of Fig.~\ref{figS5}b are plotted in Fig.~\ref{figS5}c, showing excellent agreement with the linear scaling with $N$. 
These experimental results indicate the successful generation of the large photon-number Fock states in the cavity.

\subsection{Experimental metrology schemes}
\subsubsection{Experimental sequence}

\begin{figure}[t]
    \includegraphics{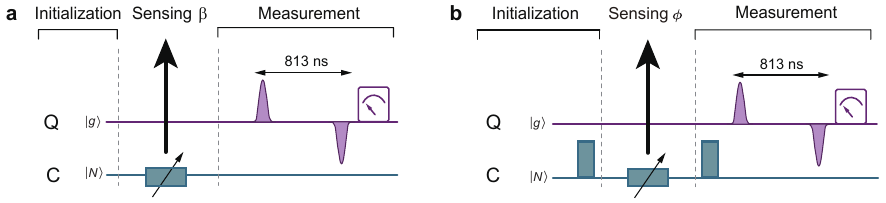}
    \caption{Experimental sequence for displacement amplitude sensing (a) and phase sensing (b).} 
    \label{figS6}
\end{figure}

Fock states with sub-Planck structures in phase space can be utilized to enhance the performance of precision measurements~\cite{Zhang2015,Aaron2020}. The experimental sequences for performing the displacement amplitude and phase sensing with Fock states are shown in Fig.~\ref{figS6}, where we always initialize the transmon qubit to the ground state $\ket{g}$ and the probe cavity to the Fock state $\ket{N}$. The initialization sequence of the Fock state is shown in Fig.~\ref{figS4}, which contains a Gaussian PNF and two sinusoidal PNFs in a parameterized quantum circuit. The final parity measurement is implemented by inserting a controlled-phase gate $C_\pi = \ket{g}\bra{g} \otimes I+ \ket{e}\bra{e} \otimes e^{i \pi a^\dagger a }$ between two $\pi/2$ pulses, which is realized by waiting a duration of $\pi/\chi_\mathrm{qc}$ due to the native dispersive interaction Hamiltonian $\hat{H} = -\chi_\mathrm{qc}a^\dagger a |e\rangle \langle e|$ between the probe cavity and the ancilla qubit.

For displacement amplitude sensing, a displacement drive is applied to encode the parameter $\beta$ to the initial resource state $\ket{N}$. The parameter $\beta$ is finally estimated by mapping the parity information of the cavity to the ancillary qubit state, with the experimental sequence shown in Fig.~\ref{figS6}a.

For the phase sensing experiment, the resource states are prepared by displacing the Fock state $\ket{N}$ to generate the displaced Fock state $D(\gamma=\sqrt{N})\ket{N}$. The displacement amplitude $\gamma$ is chosen in such a way as to maximize the quantum Fisher information, as discussed in section II-D1. Subsequently, a virtual phase operation $U_\phi = e^{-i\phi a^\dagger a}$ is applied to encode the phase $\phi$ on the initial resource state. This virtual operation is implemented by adding a phase offset to the microwave drives of all subsequent pulses on the cavity state and has a duration of zero~\cite{mckay2017}. The phase $\phi$ is finally determined by mapping the displaced parity information to the ancillary qubit state, with the experimental sequence shown in Fig.~\ref{figS6}b.

\subsubsection{Photon-number-resolved quantum metrology experiment}

Although quantum metrology experiments are performed with the Fock states prepared in a probabilistic way, all the photon number states of the initial coherent state can be simultaneously resolved in a single experiment by designing cascaded sinusoidal PNFs~\cite{curtis2021}, and can be used to enhance the estimation precision without postselection.

For both displacement amplitude sensing and phase measurement, a hardware-efficient metrology scheme can be implemented with photon-number-resolved Fock states. By applying a sequence of sinusoidal PNFs, we can simultaneously resolve all the Fock components from the input coherent state $\left|\psi\right\rangle =\sum_{n}c_{n}\left|n\right\rangle$, and all these photon-number-resolved Fock states $\{\left|n\right\rangle\}$ are used as initial resource states for the quantum metrology experiment with probability $\{|c_{n}|^{2}\}$, resulting in classical Fisher information (see Eq.~\ref{QJT_displacement} and Eq.~\ref{QJT_phase} for detailed derivations):
\begin{align}
    F=4\left(2\bar{n}+1\right)
\end{align} 
and
\begin{align}
    F=4\left| \gamma \right|^2\left( 2 \bar{n} + 1 \right)
\end{align}
for displacement amplitude sensing and phase measurement, respectively. Here, $\bar{n}$ is the average photon number of the initial coherent state $\left|\psi\right\rangle$.

\begin{figure}[t]
	\centering
	\includegraphics{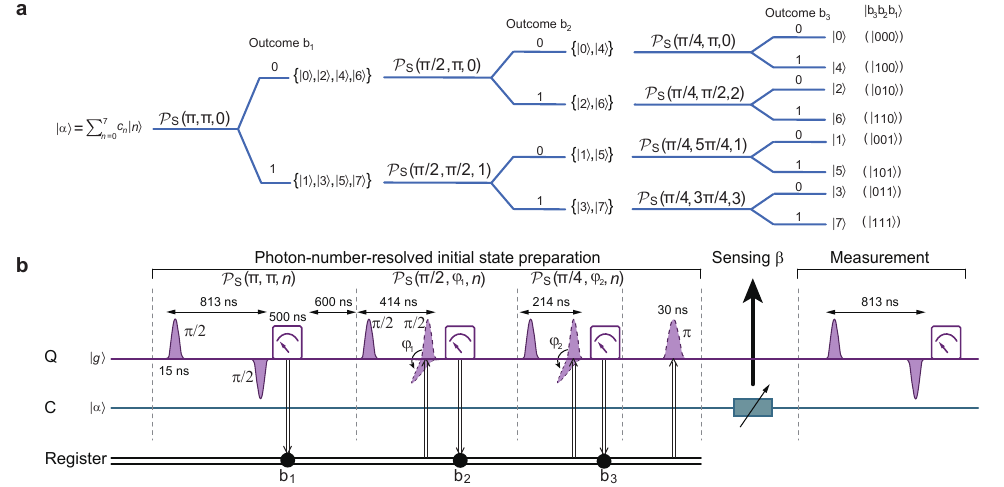}
	\caption{Illustration of the photon-number-resolved Fock state preparation for the quantum metrology experiment (a) and the corresponding experimental sequence (b). }
	\label{figS7}
\end{figure}

We demonstrate this photon-number-resolved metrology experiment by performing a displacement amplitude sensing scheme with engineered sinusoidal PNFs to simultaneously resolve all the photon numbers in the initial coherent state. As an example of the initial coherent state $\ket{\alpha} = \sum_{n=0}^{7}{c_n \ket{n}}$ with $\alpha = \sqrt{3}$, three generalized sinusoidal PNFs $\mathcal{P}_\mathrm{S}(\theta,\varphi)$ (see section I-B1) are engineered to simultaneously resolve eight photon number states for sensing the displacement amplitude, which is illustrated in Fig.~\ref{figS7}a. The phase $\varphi$ of each PNF is chosen adaptively in real-time according to the measurement outcomes of the previous PNFs with fast feedback control. Here $\mathcal{P}_\mathrm{S}(\theta,\varphi) = \sum_n{\sin{\frac{n\theta - \varphi}{2}} \ket{n}\bra{n}}$ or $\sum_n{\cos{\frac{n\theta - \varphi}{2}} \ket{n}\bra{n}}$ depending on the qubit measured in ground or excited state. In this process, the first sinusoidal PNF $\mathcal{P}_\mathrm{S}(\pi,\pi)$ projects the coherent state into two branches of even and odd subspaces depending on the measurement outcomes $b_1=0$ or 1. Specifically, when the qubit is measured in ground and excited states, the cavity output states are $\sum_n{c_n\sin{\frac{n\pi - \pi}{2}} \ket{n}} = \sum_{n=0,2,4,6}{c_n \ket{n}}$ and $\sum_n{c_n\cos{\frac{n\pi - \pi}{2}} \ket{n}} = \sum_{n=1,3,5,7}{c_n \ket{n}}$, respectively, where the phase factors are ignored. In a similar way, the second-layer sinusoidal PNFs $\mathcal{P}_\mathrm{S}(\pi/2,\pi)$ and $\mathcal{P}_\mathrm{S}(\pi/2,\pi/2)$ are dynamically chosen based on the previous outcome $b_1$ to project the even and odd cavity states into $\sum_{n=0,4}{c_n|n\rangle}$, $\sum_{n=2,6}{c_n|n\rangle}$, $\sum_{n=1,5}{c_n|n\rangle}$, and $\sum_{n=3,7}{c_n|n\rangle}$, respectively. Finally, based on the measurement outcomes ($b_1$ and $b_2$) of previous two layer PNFs, another four PNFs $\mathcal{P}_\mathrm{S}(\pi/4,\pi)$, $\mathcal{P}_\mathrm{S}(\pi/4,\pi/2)$, $\mathcal{P}_\mathrm{S}(\pi/4,5\pi/4)$, and $\mathcal{P}_\mathrm{S}(\pi/4,3\pi/4)$ are designed accordingly to simultaneously resolve all the eight photon number states $\ket{N = b_3b_2b_1}$. These photon-number-resolved states are used for the displacement amplitude sensing experiment, with the experimental sequence shown in Fig.~\ref{figS7}b. Without postselection, all the measurement data are collected and classified, giving a total weighted Fisher information surpassing the SQL, as shown in Fig. 4 in the main text.

In addition, sequentially applying these sinusoidal PNFs can also be used to generate multicomponent Schr\"{o}dinger cat states, for example, two PNFs generate the four-leg cat state or compass state, which has important applications in both quantum error correction~\cite{ofek2016,heeres2017} and quantum-enhanced sensing~\cite{toscano2006,dalvit2006}.

\subsubsection{Derivation of the measurement sensitivity}
\begin{figure}[t]
    \includegraphics{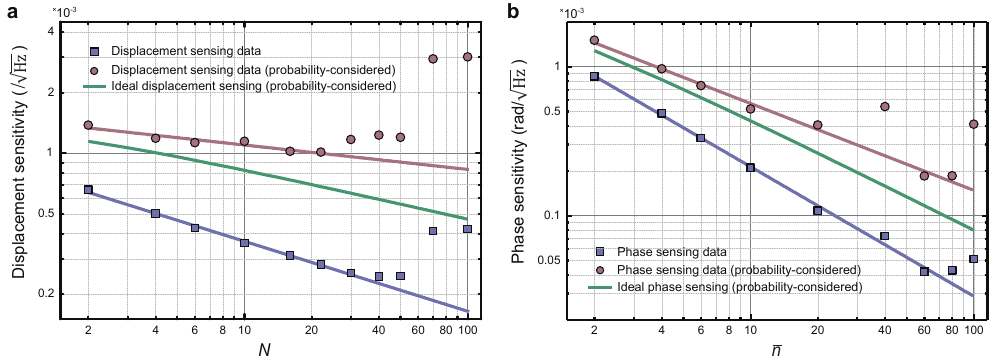}
\caption{Measurement sensitivities for displacement sensing (a) and phase sensing (b) experiments. Blue squares (red circles) represent the experimentally measured sensitivities without (with) considering the success probability of the postselected preparation of initial Fock states. Solid green lines are the ideal cases when considering the success probability. A linear fit in the logarithmic-logarithmic scale reveals a sensitivity scaling of $N^{-0.35}$ ( $N^{-0.12}$ ) for the blue (red) line in the displacement sensing experiment, and a sensitivity scaling of $\bar{n}^{-0.87}$ ( $\bar{n}^{-0.58}$) for the blue (red) line in the phase sensing experiment. }
    \label{figure_Sensitivity}
\end{figure}

For displacement amplitude sensing, the measurement sensitivity of $\beta$ can be calculated as follows:
\begin{equation}
    \sigma_\beta=\delta \beta \times \sqrt{\tau_\mathrm{tot}},
\end{equation}
where $\tau_\mathrm{tot}=7130$~ns is the total duration of the single-shot measurement, including the interrogation time of the displacement operation $\tau_\mathrm{int}=100$~ns, the initialization time of the Fock state $5717$~ns, and the final parity measurement time of $1313$~ns. 
%Take the situation of $N=40$ for example, $\sigma_\beta=0.09\times\sqrt{6630\mathrm{ns}}=2.31\times10^{-4}\mathrm{Hz}^{-1/2}$. 
In Fig.~\ref{figure_Sensitivity}a, we have presented the displacement measurement sensitivities $\sigma_\beta$ as a function of the number of photons $N$ of the initial Fock states. As a result, an optimal displacement sensitivity of $1.01\times10^{-3}\mathrm{Hz}^{-1/2}$ ($2.44\times10^{-4}\mathrm{Hz}^{-1/2}$) is achieved with (without) considering the success probability of the postselected preparation of the initial Fock states. 
In addition, if we are interested in the displacement frequency $E = \beta/\tau_\mathrm{int}$. The sensitivity is given by 
\begin{equation}
    \sigma_E=\delta \beta \times \sqrt{\tau_\mathrm{tot}}/\tau_\mathrm{int}.
\end{equation}
The corresponding optimal sensitivity of displacement frequency is  $10.1~\mathrm{kHz}/\mathrm{Hz}^{1/2}$ ($2.44~\mathrm{kHz}/\mathrm{Hz}^{1/2}$) with (without) the success probability considered.

For phase sensing, the total duration of the single-shot measurement is  $\tau_\mathrm{tot}=7280$, including the duration of two additional displacement operations $250$ ns. Similarly, the measurement sensitivity of $\phi$ can be calculated as follows:
\begin{equation}
    \sigma_{\phi}=\delta \phi\times\sqrt{\tau_\mathrm{tot}}.
\end{equation}
The results are shown in Fig.~\ref{figure_Sensitivity}b, where an optimal phase sensitivity of  $1.84\times10^{-4}\mathrm{rad}/\mathrm{Hz}^{1/2}$ ($4.20\times10^{-5}\mathrm{rad}/\mathrm{Hz}^{1/2}$) is achieved with (without) considering the postselected preparation of the initial Fock states. As a quantitative comparison, the phase measurement sensitivity in our experiment outperforms the sensitivities of $9.5\times10^{-2}\mathrm{rad}/\mathrm{Hz}^{1/2}$~\cite{staudenmaier2021phase}, $1.2\times10^{-3}\mathrm{rad}/\mathrm{Hz}^{1/2}$~\cite{wang2022}, and $4\times10^{-3}\mathrm{rad}/\mathrm{Hz}^{1/2}$~\cite{chen2023quantum} in previous works.

\section{Theoretical  Analysis}
\subsection{Theoretical error analysis}

To understand and assess the influence of different errors on the experiment performance, we employ the perturbation theory to analytically solve the noisy evolution including the initialization of the probe state, the interrogation under the estimated parameter, and the projective measurement.

The evolution of the noisy system can be described by the Lindblad master equation:
\begin{eqnarray}\label{MasterEqu}
\frac{d\rho}{dt}=-i\left[H,\rho\right]+\sum_{m} \mathcal{L}_{m}( \rho ),
\end{eqnarray}
where $H$ is the interaction Hamiltonian and $\mathcal{L}_{m}(\rho) = \kappa_{m}L_{m} \rho L_{m}^{\dagger}- \frac{1}{2} \kappa_{m} L_{m}^{\dagger} L_{m} \rho - \frac{1}{2} \kappa_{m} \rho L_{m}^{\dagger}L_{m}$ is the Lindblad superoperator with jump operators $\{L_m\}$ and the corresponding strength $\{\kappa_m\}$. In this experiment, dominant errors include both decay and dephasing of the probe cavity and ancilla qubit, i.e., $L_{1} = a$, $L_{2} = a^{\dagger}a$, $L_{3}=\ket{g}\bra{e}$ and $L_{4}= \ket{e} \bra{e}$, where $a^\dagger$ ($a$) is the creation (annihilation) operator of the probe mode and $\ket{g}$ ($\ket{e}$) is the ground (excited) state of the qubit. 

The final state $\rho\left(T\right)$ at the final time $T$ can be calculated from perturbation expansion with the assumption that $\kappa_{m}T\ll1$, i.e., $\rho\left(T\right)=\rho_{0}\left(T\right)+\rho_{1}\left(T\right)+\rho_{2}\left(T\right)+...$ with the evolution of the zero-order and first-order terms evolving as 
\begin{eqnarray}
\frac{d\rho_{0}(t)}{dt}=-i\left[H,\rho_{0}(t)\right]
\end{eqnarray}
and 
\begin{eqnarray}
\frac{d\rho_{1}(t)}{dt}=i\left[H,\rho_{1}(t)\right]+\sum_{m}\mathcal{L}_{m}(\rho_{0}(t)) .
\end{eqnarray}
Therefore, the solutions of these two terms are
\begin{eqnarray}\label{0nd order}
\rho_{0}\left(t\right)=e^{-iHt}\rho\left(0\right)e^{iHt}
\end{eqnarray}
and 
\begin{eqnarray}\label{perturbation}
\rho_{1}\left(t\right)=\int^{t}_{0}d\tau \sum_{m} e^{-i H  \left(t-\tau\right)}  \mathcal{L}_{m}(\rho_{0}(\tau)) e^{i H \left(t-\tau\right)} .
\end{eqnarray}
In the following sections, we discuss the dominant error in different processes of the experiment.

\subsubsection{Initialization of the probe state}
During the initialization process, the probe state becomes approximately $\left|N\right\rangle$ after several Gaussian and sinusoidal filters. During the measurement and reset of the ancilla qubit, the probe state is vulnerable to the decay error, i.e., the probe state evolves according to Eq.~\ref{MasterEqu} with $H=I$ and $L_{1} = a$.  Therefore, the final state becomes 
\begin{align}\label{idel_error}
\rho \left( T_i \right) =\left( 1-\kappa _1T_iN \right) \left| \left. N \right> \right. \left< \left. N \right|+\kappa _1T_i \right. N\left| \left. N-1 \right> \right. \left< \left. N-1 \right| \right. ,
\end{align}
where $T_i$ is the measurement and reset duration of the ancilla qubit.

\subsubsection{Projective measurement}
In the displacement amplitude sensing experiment, the initial Fock state becomes $\left|\psi\right\rangle =D\left(\beta\right)\left|N\right\rangle $ after the displacement operation $ D\left(\beta\right)=e^{\beta a^{\dagger}-\beta^{*}a} $. Following the parity measurement with the ancilla qubit initialized in $\left|+\right\rangle=(|g\rangle + |e\rangle)/\sqrt{2}$, the whole system evolves under the cross-Kerr interaction $H=-\chi_\mathrm{qc}a^\dagger a |e\rangle \langle e|$ and the Lindblad operators $L_{1} = a$, $L_{2} = a^{\dagger}a$, $L_{3}=\ket{g}\bra{e}$ and $L_{4}= \ket{e} \bra{e}$ with the parity measurement duration $T_M=\pi/\chi_\mathrm{qc}$. After the cross-Kerr interaction, a $-X/2$ gate and a qubit measurement are implemented on the ancilla qubit, which is equivalent to a projective measurement with $\Pi_{g}=I\otimes\left|+\right\rangle \left\langle +\right|$ and $\Pi_{e}=I\otimes\left|-\right\rangle \left\langle -\right|$.

According to Eqs.~\ref{0nd order} and \ref{perturbation}, the measurement probability is 
\begin{align}\label{MeasurementProb}
P_{g}\left(N\right)=\mathrm{tr}\left[\Pi_{g}\rho\left(T\right)\Pi_{g}\right]=P_{g,0}+P_{g,1}.
\end{align}
Here, $P_{g,0}$ is the zero-order of the measured qubit ground state probability, and can be calculated as
\begin{equation}\label{Prob_0order}
    \begin{aligned}
        P_{g,0}&=\mathrm{tr}\left[ I\otimes \left| + \right. \rangle \left. \langle + \right|\rho _0\left( T_M \right) \right] 
    \\
    &=\mathrm{tr}\left( \frac{I+e^{i\pi a^{\dagger}a}}{2}D\left( \beta \right) \left| N \right. \rangle \left. \langle N \right|D^{-1}\left( \beta \right) \right) 
    \\
    &=\frac{1}{2}+\frac{1}{2}\mathrm{tr}\left( \left| N \right. \rangle \left. \langle N \right|D^{-1}\left( \beta \right) e^{i\pi a^{\dagger}a}D\left( \beta \right) \right) 
    \\
    &=\frac{1}{2}+\frac{1}{2}\left( -1 \right) ^NL_N\left( 4\left| \beta \right|^2 \right) e^{-2\left| \beta \right|^2},
    \end{aligned}
\end{equation}
where 
\begin{equation}\label{WignerFock}
    \begin{aligned}
    &\frac{2}{\pi}\mathrm{tr}\left(\left|N\right\rangle \left\langle N\right|D^{-1}\left(\beta\right)e^{i\pi a^{\dagger}a}D\left(\beta\right)\right)\\
    =& \frac{2}{\pi}\left(-1\right)^{N}L_{N}\left(4\left|\beta\right|^{2}\right)e^{-2\left|\beta\right|^{2}}
    \end{aligned}
\end{equation}
is the Wigner function of the Fock state $\left|N\right\rangle $, and
\begin{align}
    \rho _0\left( T_M \right) =e^{-i\chi _{\mathrm{qc}}T_M a^\dagger a |e\rangle \langle e|}\left( D\left( \beta \right) \left| N \right. \rangle \left. \langle N \right|D^{-1}\left( \beta \right) \otimes \left| + \right. \rangle \left. \langle + \right| \right) e^{i\chi _{\mathrm{qc}}T_M a^\dagger a |e\rangle \langle e|}
\end{align}
is the zero-order term of the final state. The population for different initial states $\left| N \right \rangle$ is fitted as $P_{g}=A e^{-2\left|\beta\right|^{2}} L_{N}\left(4\left|\beta\right|^{2}\right)+B$ with the fitting parameters $A$ and $B$ to account for measurement imperfections such as background errors and fluctuations in the experiment. The first-order term $\rho _1\left( T_M \right)$ can be calculated as Eq.~\ref{perturbation} and the corresponding measurement probability is 
\begin{equation}
    \begin{aligned}
        &P_{g,1}\\
        =&\left(-1\right)^{N}e^{-2\left|\beta\right|^{2}}[\left(-\kappa_{1}T\left(\frac{1}{4}N+\frac{1}{2}\left|\beta\right|^{2}-\frac{1}{4}\right)-\frac{1}{4}(\kappa_{3}+\kappa_{4})T\right) \times\\
        &\quad L_{N}\left(4\left|\beta\right|^{2}\right)-\kappa_{1} T\frac{N+1}{4}L_{N+1}\left(4\left|\beta\right|^{2}\right)-\frac{1}{4}\kappa_{1}TL_{N-1}\left(4\left|\beta\right|^{2}\right)].
    \end{aligned}
\end{equation}

In the phase sensing experiment, the probe state becomes $D^{-1}\left( \gamma \right) e^{-i\phi a^{\dagger}a}D\left( \gamma \right) \left| N \right. \rangle $ before the projective measurement. $D^{-1}\left( \gamma \right) e^{-i\phi a^{\dagger}a}D\left( \gamma \right) =e^{-i\phi \left( a^{\dagger}+\gamma ^* \right) \left( a+\gamma \right)}\simeq e^{-i\phi \gamma ^*\gamma}e^{-i\phi a^{\dagger}a}e^{\left( \alpha a^{\dagger}-\alpha^* a \right)}$, which is equivalent to a displacement operator $D(\alpha)$ with  $\alpha =-i\phi \gamma $ for the Fock state. Therefore, the measurement result is similar to the above with the ideal population $P_{g}=\frac{1}{2}+\frac{1}{2}(-1)^N L_{N}\left(4\left|\phi\gamma\right|^{2}\right)e^{-2\left|\phi \gamma\right|^{2}}$. In the experiment, $\gamma$ is set to be $\sqrt{N}$ to obtain the optimal metrology gain with a fixed average photon number (see Section II-D). In our experiment, the measured qubit population is fitted to the function $P_{g}=A e^{-2N\left|\phi+\phi_0 \right|^{2}} L_{N}\left(4N\left|\phi+\phi_0\right|^{2}\right)+B$  with $A$ and $B$ being the fitting parameters to account for the measurement imperfections and $\phi_0$ to account for the self-Kerr-induced phase of the probe state during the sensing process.

\subsubsection{Displacement}
In the metrology experiment, another important process is the interrogation under the estimated parameter. To estimate the displacement amplitude, a displacement operation $D(\beta)$ is implemented on the probe state, during which one of the dominant errors is the dephasing error of the probe state, i.e., the initial probe state $\left|N\right\rangle$ evolves according to Eq.~\ref{MasterEqu} with $H=i\xi(a^{\dagger}-a)$ and $L_{2} = a^{\dagger}a$. The duration of this displacement operation is $T_D=|\beta|/\xi$. Assuming that the perfect parity measurement is implemented after the displacement, the measured probability can be expressed as $P_g\left( N \right) =\mathrm{tr}\left[ \frac{I+e^{i\pi a^{\dagger}a}}{2}\rho \left( T_D \right) \right] =P_{g,0}+P_{g,1}$, where the zero-order term is the same as Eq.~\ref{Prob_0order}.

The first-order term can be calculated according to the Eq.~\ref{perturbation}. Since $N \kappa_2 T_D \ll 1$ and $N|\beta|^{2} \ll 1$, we neglect the high-order terms and obtain
\begin{align}\label{Displacment_Error}
    P_{g,1} \approx \frac{1}{3} \kappa_2 T_D |\beta|^{2} N^3.
\end{align}

\subsection{Quantum Cramer-Rao bound}

To estimate an unknown parameter $\lambda$ in a quantum metrology experiment, a probe state is first initialized and then evolves under the unknown parameter. After that, the quantum measurement is implemented on the final state $\rho(\lambda)$ with a set of positive operator-valued measure (POVM) operators $\{F_i\}$. The probability distribution $\{P_i\}$ is obtained where $P_i=\mathrm{tr}[\rho(\lambda)F_i]$ and the value of $\lambda$ can be inferred. The achievable precision $\delta \lambda$ is bounded by the quantum Cramer-Rao bound~\cite{braunstein1994}:
\begin{align}
    \delta \lambda \geqslant \frac{1}{\sqrt{F}} \geqslant \frac{1}{\sqrt{Q}}
\end{align}
where 
\begin{align}\label{equ:CFI}
    F=\sum_i{\frac{1}{P_i\left( \lambda \right)}\left[ \frac{\partial P_i\left( \lambda \right)}{\partial \lambda} \right] ^2}
\end{align}
is the classical Fisher information (CFI), and $Q$ is the quantum Fisher information (QFI), which is the supremum of the CFI over all possible POVMs. For a pure final state $\rho (\lambda )=| \psi   \rangle \langle \psi | $, the QFI can be calculated as
\begin{align}\label{QFI_calculation}
Q=4\left( \left. \langle \partial _{\lambda}\psi |\partial _{\lambda}\psi \right. \rangle -\left| \left. \langle \psi |\partial _{\lambda}\psi \right. \rangle \right|^2 \right), 
\end{align}
where$\left|\partial_{\lambda}\psi\right\rangle $ is the partial derivative of $\left|\psi\right\rangle $. For a mixed state, the calculation method of QFI can be found in Ref.~\cite{yuan2017fidelity}.
Drawing inspiration from the quantum jump tracking method presented in Ref.~\cite{wang2022}, if the probe state can be unambiguously initialized differently, the total QFI should be the sum of the individual QFI, i.e., 
\begin{align}\label{QFI_QJT}
Q_\mathrm{QJT}=\sum_i{p_i Q[\rho_i(\lambda) ]},
\end{align}
where $\rho_i(\lambda)$ is the final state of the $i$-th initial state and $p_i$ is the corresponding probability. This equation still holds true for CFI. According to the convexity~\cite{fujiwara2001}, the QFI of this scheme should not be less than that of the scheme without distinguishing the initial state, i.e., $Q_\mathrm{QJT}=\sum_i{p_i Q[\rho_i(\lambda) ]} \geqslant \sum_i{ Q[p_i \rho_i(\lambda) ]}$.

\subsection{Displacement amplitude sensing scheme}

In this part, we will present the theoretical framework and relevant derivations corresponding to the displacement amplitude sensing scheme. The probe state is first initialized to the Fock state $|N\rangle$ in this experiment. Then, the probe state evolves under the displacement $D(|\beta|e^{i\theta})$ with a random angle $\theta$, where $|\beta|$ is the displacement amplitude to be estimated. Finally, a projective parity measurement is implemented to obtain the probability distribution associated with the estimated parameter $|\beta|$.

\subsubsection{Classical Fisher information}

As calculated in Eq.~\ref{Prob_0order}, the probability distribution of the projective measurement is $P_{g}=\frac{1}{2}+\frac{1}{2}\left(-1\right)^{N}L_{N}\left(4\left|\beta\right|^{2}\right)e^{-2\left|\beta\right|^{2}}$. 
 Therefore, the CFI of the displacement amplitude sensing with the Fock state $ \left| N \right. \rangle$ is 
\begin{align}\label{CFI_amplitude}
F
& =\frac{1}{P_{g}\left(1-P_{g}\right)}\left(\partial_{\left|\beta\right|}P_{g}\right)^{2} \notag\\
& =4\left(2N+1\right)-4\left(2N^{2}+2N+5\right)\left|\beta\right|^{2}+O\left(\left|\beta\right|^{3}\right).
\end{align}
When the amplitude is small enough, i.e., $\left|\beta\right|\rightarrow0$, the CFI becomes $\ F\rightarrow4\left(2N+1\right)$. By solving the first zero point from the expression above, the range for the estimated parameter can be obtained approximately as $|\beta|<\sqrt{\frac{2N+1}{2N^{2}+2N+5}}$, i.e., with a scaling of $1/\sqrt{N}$ for large $N$.

\subsubsection{Optimality}

It is worth noting that the scheme presented here is an optimal scheme for displacement amplitude sensing considering an arbitrary initial state and POVM, which can be proven in the following.
If the initial state is $\left|\psi_{0}\right\rangle $, the state after displacement is $\left|\psi\right\rangle =D\left(\beta\right)\left|\psi_{0}\right\rangle  $, and $\left| \partial _{\left| \beta \right|}\psi \right. \rangle =D\left( \beta \right) \left( e^{i\theta}a^{\dagger}-e^{-i\theta}a \right) \left| \psi _0 \right. \rangle 
$. Due to the convexity of the QFI~\cite{fujiwara2001}, 
\begin{align}\label{QFI_amplitude}
& Q\left[ \frac{1}{2\pi}\int_0^{2\pi}{d\theta}D\left( |\beta |e^{i\theta} \right) |\psi _0\rangle \langle \psi _0|D^{\dagger}\left( |\beta |e^{i\theta} \right) \right] \notag\\
& \leq \frac{1}{2\pi}\int_0^{2\pi}{d\theta} Q\left[ D\left( |\beta |e^{i\theta} \right) |\psi _0\rangle \langle \psi _0|D^{\dagger}\left( |\beta |e^{i\theta} \right) \right] \notag\\
 & =\frac{1}{2\pi}\int_{0}^{2\pi}d\theta4\left(\left\langle \partial_{\left|\beta\right|}\psi|\partial_{\left|\beta\right|}\psi\right\rangle -\left|\left\langle \psi|\partial_{\left|\beta\right|}\psi\right\rangle \right|^{2}\right) \notag\\
 & =\frac{1}{2\pi}\int_{0}^{2\pi}d\theta4[\left\langle \psi_{0}\right|\left(e^{-i\theta}a-e^{i\theta}a^{\dagger}\right)\left(e^{i\theta}a^{\dagger}-e^{-i\theta}a\right)\left|\psi_{0}\right\rangle \notag\\
 & \quad-\left|\left\langle \psi_{0}\right|D^{-1}\left(\beta\right)D\left(\beta\right)\left(e^{i\theta}a^{\dagger}-e^{-i\theta}a\right)\left|\psi_{0}\right\rangle \right|^{2}] \notag\\
 & =4\left[1+2\left\langle \psi_{0}\right|a^{\dagger}a\left|\psi_{0}\right\rangle -2\left\langle \psi_{0}\right|a\left|\psi_{0}\right\rangle \left\langle \psi_{0}\right|a^{\dagger}\left|\psi_{0}\right\rangle \right] \notag\\
 & \leq4\left[1+2\left\langle \psi_{0}\right|a^{\dagger}a\left|\psi_{0}\right\rangle \right] \notag\\
 & =4\left[1+2N\right].
\end{align}
As calculated in Eq.~\ref{CFI_amplitude}, the equality is achieved in our scheme with $\left|\psi_{0}\right\rangle =\left|N\right\rangle $, showing that the Fock probe state and parity measurement is optimal in the displacement amplitude sensing.

Although the initial Fock state $|N\rangle$ is generated from the coherent state $D(\sqrt{N})|0\rangle$ in a probabilistic way with a success probability of $e^{-N}N^N/N! \approx 1/\sqrt{2\pi N}$ for $N\gg 1$ according to Stirling's approximation, the postselected quantum Fisher information expressed as
\begin{align}\label{QFIpost}
    Q &= 4(2N+1)e^{-N}N^N/N! \notag\\
      &\approx 4(2N+1) \frac{1}{\sqrt{2\pi N}},
\end{align}
still achieves a scaling enhancement of $N^{1/2}$ compared to the quantum Fisher information $Q_\mathrm{SQL}=4$ for quasiclassical coherent states. Therefore, the scaling behavior of the measurement precision can achieve an enhancement of $N^{1/4}$ compared to the classical limit, thus indicating a quantum metrology advantage even with the nondeterministic preparation of the initial Fock state. Additionally, a $N^{1/2}$-scaling enhancement towards the Heisenberg limit can be further achieved through adaptive control without postselection as demonstrated in the following section.

\subsubsection{Photon-number-resolved scheme}
In addition, the initial Fock states can also be deterministically generated by implementing several sinusoidal PNFs. Without postselection, the system may collapse into arbitrary Fock states for quantum sensing. It can be demonstrated that the scaling of the CFI is similar to that in Eq.~\ref{CFI_amplitude}.
Suppose the input state is $\left|\psi\right\rangle =\sum_{n}c_{n}\left|n\right\rangle$, and all these photon-number-resolved Fock states $\{|n\rangle\}$ are used as initial resource states for the quantum metrology experiment with a probability $\{|c_n|^{2}\}$. According to Eq.~\ref{QFI_QJT}, the resulting CFI is the weighted average of the CFI for each Fock state:
\begin{equation}\label{QJT_displacement}
\begin{aligned}
F   &= \sum_{n}\left|c_{n}\right|^{2}4\left(2n+1\right) \\
	&= 4\left(2\bar{n}+1\right)
\end{aligned}
\end{equation}
where $\bar{n}$ is the average photon number of $\left|\psi\right\rangle$. The projective measurement remains the same for different probe Fock states, which indicates that feedback control is not necessary after the initialization of the probe state.

\subsubsection{Coherent state and squeezed state}

For comparison, the QFI of the coherent state in displacement amplitude sensing is $Q=4$ and can be calculated as follows. The QFI remains unchanged under a unitary operator~\cite{toth2014quantum}; therefore, the QFI is
\begin{align}
    & Q\left( \frac{1}{2\pi}\int_0^{2\pi}{d\theta}D\left( |\beta |e^{i\theta} \right) |\alpha \rangle \langle \alpha |D^{\dagger}\left( |\beta |e^{i\theta} \right) \right) 
\notag\\
=& Q\left( \frac{1}{2\pi}\int_0^{2\pi}{d\theta}D^{\dagger}\left( \alpha \right) D\left( |\beta |e^{i\theta} \right) D\left( \alpha \right) |0\rangle \langle 0|D^{\dagger}\left( \alpha \right) D^{\dagger}\left( |\beta |e^{i\theta} \right) D\left( \alpha \right) \right) 
\notag\\
=& Q\left( \frac{1}{2\pi}\int_0^{2\pi}{d\theta}D\left( |\beta |e^{i\theta} \right) |0\rangle \langle 0|D^{\dagger}\left( |\beta |e^{i\theta} \right) \right),
\end{align}
which is the same as the QFI in Eq.~\ref{QFI_amplitude} when $N=0$. Therefore, the QFI of the coherent state is $Q=4$.

Displaced squeezed state $\left|\psi\right\rangle =D\left(\alpha\right)S\left(\xi\right)\left|0\right\rangle $ is another resource state in displacement sensing with $S\left(\xi\right)=\exp{\left(\frac{1}{2}\xi^{*}a^{2}-\frac{1}{2}\xi a^{\dagger2}\right)}$, $\xi=re^{i\omega}$. Similar to the Fock state, it can achieve the same maximal QFI as that in Eq.~\ref{QFI_amplitude}. However, its potential is currently limited by the limiting squeezing degree with a recent typical parameter $15$ dB, i.e., $r\approx1.73$ and $\bar{n}\approx7.4$~\cite{vahlbruch2016detection}. The QFI of this state is calculated in the following part. Since the QFI remains unchanged under a unitary operator
\begin{align*}
    &Q\left[\frac{1}{2\pi}\int_{0}^{2\pi}d\theta D\left(\left|\beta\right|e^{i\theta}\right)D\left(\alpha\right)S\left(\xi\right)\left|0\right\rangle \left\langle 0\right|S^{\dagger}\left(\xi\right)D^{\dagger}\left(\alpha\right)D^{\dagger}\left(\left|\beta\right|e^{i\theta}\right)\right]\\
    =&Q\left[\frac{1}{2\pi}\int_{0}^{2\pi}d\theta D\left(\left|\beta\right|e^{i\theta}\right)S\left(\xi\right)\left|0\right\rangle \left\langle 0\right|S^{\dagger}\left(\xi\right)D^{\dagger}\left(\left|\beta\right|e^{i\theta}\right)\right].
\end{align*}
The QFI of the displaced squeezed state $\left|\psi\right\rangle =D\left(\alpha\right)S\left(\xi\right)\left|0\right\rangle $ is equal to the squeezed vacuum state with $\alpha=0$. The measurement probability is
\begin{align*}
    P_{g}&=\frac{1}{2\pi}\int_{0}^{2\pi}d\theta\left\langle 0\right|S^{\dagger}\left(\xi\right)D^{\dagger}\left(\left|\beta\right|e^{i\theta}\right)\frac{I+e^{i\pi a^{\dagger}a}}{2}D\left(\left|\beta\right|e^{i\theta}\right)S\left(\xi\right)\left|0\right\rangle \\&=\frac{1}{2}+\frac{1}{2}\frac{1}{2\pi}\int_{0}^{2\pi}d\theta \exp\left(-2\left|\widetilde{\beta}\right|^{2}\right)\\&=1-\frac{1}{2\pi}\int_{0}^{2\pi}d\theta\left|\widetilde{\beta}\right|^{2}+O\left(\left|\beta\right|^{4}\right)\\&=1-\left|\beta\right|^{2}\left[\cosh^{2}\left(r\right)+\sinh^{2}\left(r\right)\right]+O\left(\left|\beta\right|^{4}\right),
\end{align*}
where $\widetilde{\beta}=\left|\beta\right|e^{i\theta}\cosh\left(r\right)-\left|\beta\right|e^{-i\theta}e^{i\omega}\sinh\left(r\right)$. Similar to Eq.~\ref{WignerFock}, here the second equality is obtained by utilizing the Wigner function of the squeezed vacuum state~\cite{biswas2007nonclassicality}. According to Eq.~\ref{equ:CFI}, the CFI is 
\begin{equation}
    \begin{aligned}
        F& \approx 4\left[\cosh^{2}\left(r\right)+\sinh^{2}\left(r\right)\right]+O\left(\left|\beta\right|^{2}\right) \\
        &=4\left[2\sinh^{2}\left(r\right)+1\right] \\
        &=4\left[2\bar{n}+1\right].
    \end{aligned}
\end{equation}
Here, $\bar{n}=\sinh^{2}\left(r\right)$ is the average photon number of the squeezed vacuum state. Asymptotically, this CFI approaches the maximum QFI in Eq.~\ref{QFI_amplitude} .

\subsubsection{Error analysis of the quantum metrology experiment}
In the metrology experiment, the performance of the large Fock state is limited due to imperfections throughout the entire process, including initialization of the probe state, interrogation under the estimated parameter, and the final measurement. In this part, we estimate the error contributions with a perturbation method to qualitatively illustrate the influence of dominant errors on metrology performance. In this error analysis, we only consider the dominant sources of errors in the system, including the decay and dephasing errors of the cavity and the transmon qubits. The errors are analyzed under the assumption of weak noise and small estimated parameters. Nevertheless, the analysis provides a reasonably accurate explanation for the factors that restrict the accuracy of the system.

\begin{figure}[t]
    \includegraphics{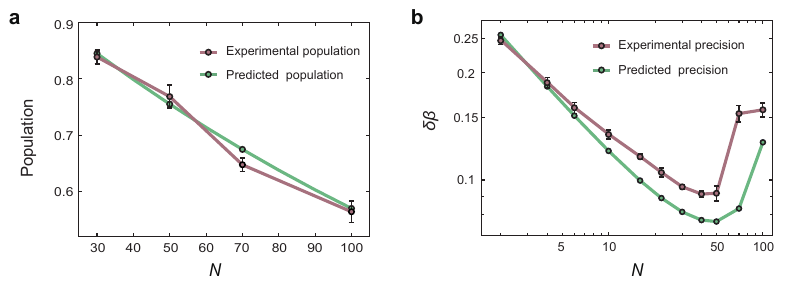}
\caption{(a) Comparison of the generated Fock state population between the experimentally measured (red curve) and prediction from the error analysis (green curve). The error bars of the experimental data are standard errors obtained from the fittings in Fig.~2a in the main text. (b) The predicted precision from the perturbation method (green curve) compared with the measurement result (red curve) in the displacement amplitude sensing experiment. The error bars of the experimental data are standard errors obtained from the error propagation of the fit parameter uncertainties in Fig.~3b in the main text. The total number of postselected measurements is larger than 2,000 for each data point.}
    \label{fig_scaling}
\end{figure}

During initialization, the sinusoidal filter is robust to the dephasing error of the cavity since the error is commuter with the cross Kerr interaction of the cavity and the transmon qubit. One of the dominant errors arises from the decay error of the cavity and considering this error, the probability of the output Fock state $|N\rangle$ decreases to approximately $1-N\kappa _1T_i$ according to Eq.~\ref{idel_error}. Here, $T_i \approx 3~\mu$s is slightly longer than the initialization time in the experiment to include the displacement operation, the $C_\pi$ gate, the qubit measurement, the reset of the ancilla qubit, and the idling time between different pulses. In this model, we focus on the qualitative analysis of the effects caused by error, disregarding the differences in error occurrence in different processes. Considering the duration of the state characterization (approximately $3700$~ns), the population of the target Fock state can be predicted. As shown in Fig.~\ref{fig_scaling}a, the predicted state population for the initialized Fock states exhibits good agreement with the measurement results, indicating that the decoherence errors are the dominant errors for the generated Fock states.

During the displacement operation, the dephasing error becomes obvious with a large Fock state. The influence of the dephasing error is calculated using the perturbation equation as Eq.~\ref{Displacment_Error}. Therefore, an additional term emerges in the measurement probability $\Delta P= \frac{1}{3}\kappa_2 T_D |\beta|^2 N^3$ with $T_D \approx 200$~ns.

Considering the errors above and assuming $N$ is even for simplicity, the measurement probability becomes
\begin{equation}
    \begin{aligned}
P_g
&=(1-N\kappa _1T_i)P_g(N)+N\kappa _1T_iP_g(N-1)+\Delta P
\\
&\approx (1-N\kappa _1T_i)\left[ \left( 1-N\kappa _1T_M-\frac{\kappa _3+\kappa _4}{2}T_M \right) \frac{1}{2}\left( 1-4N\left| \beta \right|^2 \right) +\frac{1}{2} \right]\\
&\quad -N\kappa _1T_i\left[ \frac{1}{2}\left( 1-4N\left| \beta \right|^2 \right) +\frac{1}{2} \right] +\frac{1}{3}\kappa _2T_D|\beta |^2N^3
\\
&\approx \frac{1}{2}+\left( 1-2N\kappa _1T_i-N\kappa _1T_M-\frac{\kappa _3+\kappa _4}{2}T_M \right) \frac{1}{2}\left( 1-4N\left| \beta \right|^2 \right) +\frac{1}{3}\kappa _2T_D|\beta |^2N^3
\\
&\approx1-\lambda _1-2\left( 1-\lambda _2 \right) N\left| \beta \right|^2
    \end{aligned}
\end{equation}
where $\lambda _1=N\kappa _1T_i+\frac{1}{2}N\kappa _1T_M+\frac{\kappa _3+\kappa _4}{4}T_M$ and $\lambda _2=2N\kappa _1T_i+N\kappa _1T_M+\frac{\kappa _3+\kappa _4}{2}T_M+\frac{1}{6}\kappa _2T_D N^2$. Here, $P_g(N)$ refers to the Eq.~\ref{MeasurementProb} considering the decay error of the cavity and both the decay and dephasing error of the ancillary qubit. According to the Eq.~\ref{equ:CFI} and neglecting the high-order term of $N|\beta|^2$ and $N\kappa_m T$ for all $\kappa_m$, the final CFI becomes 
\begin{equation}
    \begin{aligned}
        F&=\frac{\left( 1-\lambda _2 \right) ^2\left( 4N\left| \beta \right| \right) ^2}{\left[ 1-\lambda _1-2\left( 1-\lambda _2 \right) N\left| \beta \right|^2 \right] \left[ \lambda _1+2\left( 1-\lambda _2 \right) N\left| \beta \right|^2 \right]}
\\
&\approx\frac{\left( 1-\lambda _2 \right) ^28N\times 2N\left| \beta \right|^2}{\left[ 1-\lambda _1-2N\left| \beta \right|^2 \right] \left[ \lambda _1+2N\left| \beta \right|^2 \right]}
\\
&\approx\frac{\left( 1-\lambda _2 \right) ^28N}{\lambda _1/2N\left| \beta \right|^2+1}
\\
&\approx\left( 1-\lambda _2 \right) ^28N
\\
&=\left( 1-2N\kappa _1T_i-N\kappa _1T_M-\frac{\kappa _3+\kappa _4}{2}T_M-\frac{1}{6}\kappa _2T_D N^2 \right) ^28N
    \end{aligned}
\end{equation}
when $\lambda_1 \ll  2N|\beta|^2 \ll 1$. Here $T_M \approx 1600$~ns is the measurement duration, including the displacement and projection measurement time. To summarize, the parameters are approximated as follows: $2\kappa_1 T_i \approx 5\times10^{-3}$, $ \frac{1}{6} \kappa_2 T_D \approx 8.3\times 10^{-6}$, $\kappa_1 T_M \approx 1.3\times 10^{-3}$, and $\frac{1}{2}(\kappa_3+\kappa_4)T_M \approx 1.0\times 10^{-2}$.
The corresponding precision is shown in Fig.~\ref{fig_scaling}b and compared with the experimental result. 
Despite the omission of numerous intricate details, we predict the trend and the inflection point close to the experimental result. This indicates that further improvements in experimental precision would require optimizing the dominant errors mentioned above. The error analysis for the phase sensing process is quite similar to the displacement sensing and thus not shown here.

\begin{figure}[t]
    \includegraphics{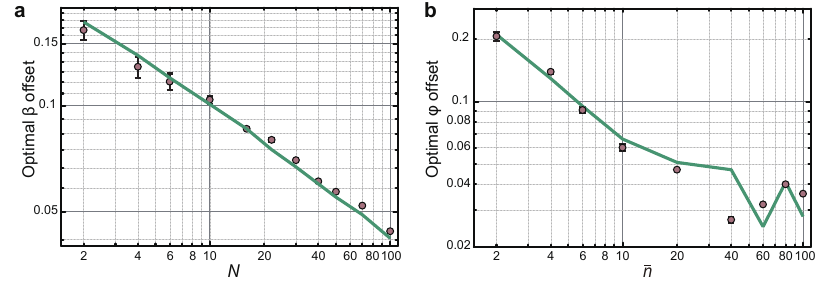}
\caption{The optimal $\beta$ (a) and $\phi$ (b) offsets for achieving the maximal Fisher information in displacement amplitude and phase sensing experiments. Solid lines are from numerical simulations. The error bars of the experimental data are standard errors obtained from the fittings in Fig.~3b and Fig.~3f in the main text. The total number of postselected measurements is larger than 2,000 for each data point.}
    \label{fig_parameteroffsets}
\end{figure}

Another impact of noises is the reduction of the detection range of the estimated parameter.
In the presence of noises, the Fisher information approaches zero when $\left|\beta\right|\rightarrow0$, and the maximal Fisher information is achieved at a specific $\beta$ offset. The optimal $\beta$ offsets for achieving the maximal Fisher information in Fig. 3(d) in the main text are plotted as a function of the number of photons $N$ of the initial Fock states in Fig.~\ref{fig_parameteroffsets}a. To intuitively illustrate the detection range, we can refer to Eq.~\ref{CFI_amplitude}. When noise is present, both $\lambda_1$ and $\lambda_2$ are nonzero. Therefore, the CFI tends towards zero when $\left|\beta\right|\rightarrow0$. In this situation, the CFI becomes sufficiently large when $\left(\partial_{|\beta|} P_g\right)^2/ P_e\approx 1$ , indicating $|\beta|>\frac{\sqrt{\lambda_1}}{4(1-\lambda_2)N}$. Therefore, the detection range can be estimated as $\frac{\sqrt{\lambda_1}}{4(1-\lambda_2)N} < |\beta| < 1/\sqrt{N}$ in the presence of noises.

\subsection{Phase sensing scheme}

\begin{figure}[b]
    \includegraphics{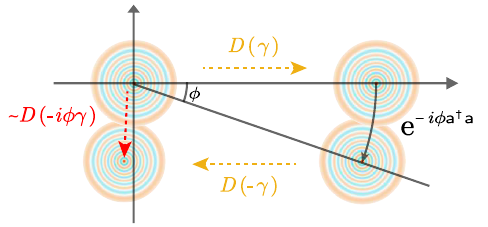}
\caption{The schematic of the phase sensing experiment with the displaced Fock state. The phase rotation with a small angle $\phi$ in phase space is mapped to a displacement operation with an amplitude of $-i\phi\gamma$. }
    \label{fig_phaseMeasurement}
\end{figure}

In the phase sensing experiment, the displaced Fock state $D\left(\gamma\right)\left|N\right\rangle$ serves as the probe state to estimate the unknown phase parameter $\phi$ of a phase rotation $e^{-i \phi a^\dagger a}$. Subsequently, a displacement operation $D\left(-\gamma\right)$ is implemented on the probe state, followed by a parity measurement to obtain the probability distribution associated with the estimated parameter $\phi$.  
The procedure of this scheme is shown in Fig.~\ref{fig_phaseMeasurement}. 
The intuitive idea behind this scheme is to utilize two displacement operations, namely $D(\gamma)$ and $D(-\gamma)$, before and after the phase rotation $e^{-i \phi a^\dagger a}$, respectively. This transforms the phase rotation into a displacement related to $\gamma$ for measurement, i.e., $D(-\gamma)e^{-i \phi a^\dagger a}D(\gamma) \approx D(-i\phi\gamma)e^{-i \phi a^\dagger a}e^{-i \phi \gamma^2}$ for $\phi \ll 1$, thereby overcoming the insensitivity of Fock states to phase rotation.

\subsubsection{Quantum Fisher information}

In the experiment, the final probe state before the parity measurement becomes $\left|\psi\right\rangle =D\left(-\gamma\right)e^{-i\phi a^{\dagger}a}D\left(\gamma\right)\left|N\right\rangle $, and $\left|\partial_{\phi}\psi\right\rangle =D\left(-\gamma\right)\left(-ia^{\dagger}a\right)e^{-i\phi a^{\dagger}a}D\left(\gamma\right)\left|N\right\rangle $. According to Eq.~\ref{QFI_calculation}, the corresponding QFI is:
\begin{equation}
\begin{aligned}\label{QFI_phase}
Q & =4[\left\langle N\right|D^{-1}\left(\gamma\right)e^{i\phi a^{\dagger}a}\left(ia^{\dagger}a\right)\left(-ia^{\dagger}a\right)e^{-i\phi a^{\dagger}a}D\left(\gamma\right)\left|N\right\rangle \\
 & \quad-\left|\left\langle N\right|D^{-1}\left(\gamma\right)e^{i\phi a^{\dagger}a}\left(-ia^{\dagger}a\right)e^{-i\phi a^{\dagger}a}D\left(\gamma\right)\left|N\right\rangle \right|^{2}]\\
 & =4\left|\gamma\right|^{2}\left(2N+1\right).
\end{aligned}
\end{equation}
Here the average photon number is $\bar{n}=N+\left|\gamma\right|^{2}$. For a fixed $\bar{n}$, the optimal QFI $ Q=2\bar{n}\left( \bar{n}+1 \right) +1/2$ can be achieved with the displacement amplitude $\left| \gamma \right|^2=N+1/2$. When $N$ is large enough, the constant term $1/2$ becomes negligible and can be disregarded. Therefore, in our experiment, we choose the optimal parameter $\gamma=\sqrt{N}$, and the average photon number is thus $\bar{n}=N+\left|\gamma\right|^{2}=2N$, with the corresponding QFI $Q=2\bar{n}\left(\bar{n}+1\right)$. 

Similar to the situation in Eq.~\ref{QFIpost}, when considering the success probability $e^{-N}N^N/N! \approx 1/\sqrt{2\pi N}$ ($N\gg 1$) of the generated initial Fock state $|N\rangle$, the postselected quantum Fisher information expressed as
\begin{align}
    Q &= 2\bar{n}(\bar{n}+1)e^{-N}N^N/N! \notag\\
      &\approx \frac{2}{\sqrt{\pi}}(\bar{n}^{3/2}+\bar{n}^{1/2}),
\end{align}
still achieves a scaling enhancement of $\bar{n}^{1/2}$ compared to the quantum Fisher information $Q=4\bar{n}$ for coherent states which will be provided in the following part.

\subsubsection{Classical Fisher information}

The projective parity measurement here can also saturate the QFI and can be proved with perturbation expansion $\phi\ll1$:
\begin{equation}
\begin{aligned}
\left|\psi\right\rangle  & =D\left(-\gamma\right)e^{-i\phi a^{\dagger}a}D\left(\gamma\right)\left|N\right\rangle \\
 & =[I+\left(-i\phi\right)\left(a^{\dagger}+\gamma^*\right)\left(a+\gamma\right)+\frac{1}{2}\left(-i\phi\right)^{2}\left(a^{\dagger}+\gamma^*\right)\\
& \quad \left(a+\gamma\right)\left(a^{\dagger}+\gamma^*\right)\left(a+\gamma\right)+...]\left|N\right\rangle 
\end{aligned}
\end{equation}
Considering perfect measurements, the probability is $P_{g}=1-\phi^{2}\left|\gamma\right|^{2}\left(2N+1\right)+O(\phi^3)$. The CFI of this scheme is
\begin{equation}
    \begin{aligned}
        F&=\frac{1}{P_{g}\left(1-P_{g}\right)}\left(\partial_{\phi}P_{g}\right)^{2}\\
        &=4\left|\gamma\right|^{2}\left(2N+1\right)+O\left(\phi^{2}\right).
    \end{aligned}
\end{equation}
When $\phi\rightarrow0$, the CFI is $\ F\rightarrow4\left|\gamma\right|^{2}\left(2N+1\right)$ which saturates the QFI in Eq.~\ref{QFI_phase}. This shows that a parity measurement combined with a displacement operation constitutes the optimal measurement strategy for displaced Fock states in phase sensing experiments. 

One of the optimal initial states for phase sensing is the maximal variance state $\left| \psi \right. \rangle =\left( \left| 0 \right. \rangle +\left| N \right. \rangle \right) /\sqrt{2}$ with the maximal QFI $Q=4\bar{n}^{2}$ and an average photon number $\bar{n}=N/2$. Another important nonclassical state is the squeezed state, which has been used for high-precision phase measurement in gravitational wave detection~\cite{abadie2011,tse2019}. The displaced squeezed state $\left|\psi\right\rangle =D\left(\alpha\right)S\left(\xi\right)\left|0\right\rangle $ with an average photon number of $\bar{n}=\left|\alpha\right|^{2}+\sinh^{2}r$ and a photon number variance of $\left\langle n^{2}\right\rangle -\left\langle n\right\rangle ^{2}=\left|\alpha \cosh{r}-\alpha^{*}e^{i\omega}\sinh{r}\right|^{2}+2\cosh^{2}r\ \sinh^{2}r$, can achieve a maximum QFI $Q\approx8\bar{n}^{2}$ with $r\gg1$ and $\sinh r \gg |\alpha|$, according to Eq.~\ref{QFI_calculation}. The scaling behavior of the maximum QFI is the same as that of the displaced Fock state, albeit with an additional prefactor of four. Nevertheless, the initialization of the displaced Fock state is less experimentally challenging, as demonstrated in this work. Another important resource state is the GKP state, which is proposed for both displacement and phase sensing~\cite{duivenvoorden2017}. 
The sensing performance of GKP states, in particular the scaling behaviors, are obtained through numerical calculations, with the result shown in Fig.~\ref{fig_summary}. The calculation is implemented with an example of the approximate GKP state~\cite{grimsmo2021quantum} with a finite energy $|\psi\rangle \propto e^{-\Delta^2 a^\dagger a} \sum^{\infty}_{k,l=-\infty}e^{-ikl\pi}|k\sqrt{\pi/2}+i\sqrt{\pi/2}l\rangle$, where $\Delta$ is the squeezing parameter and $|k\sqrt{\pi/2}+i\sqrt{\pi/2}l\rangle$ is a coherent state with an amplitude of $k\sqrt{\pi/2}+i\sqrt{\pi/2}l$. The numerical result indicates that GKP states exhibit the same scaling behavior as Fock states demonstrated in our work but with a different prefactor. However, it is experimentally challenging to prepare high-fidelity GKP states with a large number of photons, and the typical size of the generated GKP states in current superconducting platform is limited, with recent demonstration achieved an average photon number of $4.67$ in Ref.~\cite{sivak2023real}. Therefore, it requires further optimization for GKP states to enhance the applicability to achieve a high metrological gain in quantum sensing.

For phase sensing experiment, another commonly used probe state is the multi-mode state such as NOON state $(|N0\rangle+|0N\rangle)/\sqrt{2}$, with an average photon number $\bar{n}=N$. If the phase is only applied in the first mode, the corresponding QFI $Q=\bar{n}^2$ according to Eq.~\ref{QFI_calculation}. In principle, the displaced Fock states $D\left(\gamma\right)\left|N\right\rangle$ with a QFI of $Q=2\bar{n}(\bar{n}+1)$ can slightly outperform the NOON states with a prefactor enhancement. However, it is important to emphasize that comparisons between single-mode and multi-mode approaches should be made based on the specific application context. In contrast to single-mode sensing, achieving high-precision metrology with multiple bosonic modes often requires establishing high-quality entanglement between each mode, posing a significant challenge with current technology. Therefore, in the metrology scenarios described in the paper, single-mode sensing may be a better choice. However, the multi-mode sensing could provide advantages in certain applications, such as distributed sensing or multi-parameter estimation.

\subsubsection{Photon-number-resolved scheme}

For the phase sensing experiment, the photon-number-resolved scheme can also be implemented with the CFI saturating the QFI. By implementing several sinusoidal PNFs on an initial coherent state $\left|\psi\right\rangle =\sum_{n}c_{n}\left|n\right\rangle$, each Fock component $|n\rangle$ can be resolved with a probability of $|c_n|^2$. According to similar procedure, the CFI can be calculated as:
\begin{equation}
    \begin{aligned}\label{QJT_phase}
F&=\sum_n{\left| c_n \right|^2}4\left| \gamma \right|^2\left( 2n+1 \right) 
\\
&=4\left| \gamma \right|^2\left[ 2\left( \bar{n}-\left| \gamma \right|^2 \right) +1 \right] 
    \end{aligned}
\end{equation}
The average photon number of the ensemble after displacement is $\bar{n}=\left| \gamma \right|^2+\sum_n{\left| c_n \right|^2}n$. Therefore, similar to the postselection scheme, the QFI $F=2\bar{n}\left( \bar{n}+1 \right) +1/2$ can be achieved when $\left| \gamma \right|^2=\sum_n{\left| c_n \right|^2}n+1/2$.

\subsubsection{Coherent state}

For comparison, the coherent state can also be applied as the probe state and the final probe state is $\left|\psi\right\rangle=e^{-i\phi a^{\dagger}a}\left|\alpha\right\rangle $, and $\left| \partial _{\phi}\psi \right. \rangle =e^{-i\phi a^{\dagger}a}\left( -ia^{\dagger}a \right) \left| \alpha \right. \rangle $. Therefore, the QFI is $Q=4|\alpha|^2=4\bar{n}$.

\begin{figure*}
    \includegraphics{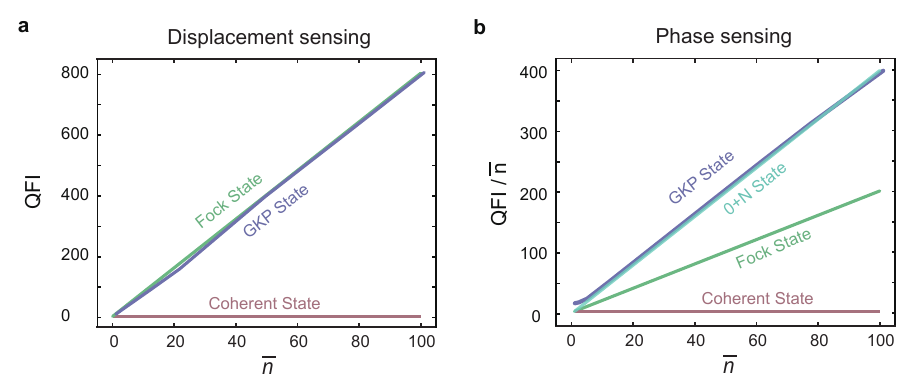}
    \caption{The QFI gain corresponds to different initial states in displacement amplitude sensing (a) and phase sensing (b).The scaling behaviors of GKP states are obtained through numerical calculations.}
    \label{fig_summary}
\end{figure*}

To better illustrate the performance of different schemes in displacement amplitude sensing and phase sensing, the QFI gain $Q/\bar{n}$ for different initial resource states is shown in Fig.~\ref{fig_summary}.

%\bibliographystyle{Zou}
%\bibliography{refs}
%Control: production of eprint (-1) disabled
%